\documentclass{article} 
\usepackage{iclr2026_conference,times}

\usepackage{amsmath,amsfonts,bm}

\def\eqref#1{equation~\ref{#1}}

\def\1{\bm{1}}

\DeclareMathAlphabet{\mathsfit}{\encodingdefault}{\sfdefault}{m}{sl}
\SetMathAlphabet{\mathsfit}{bold}{\encodingdefault}{\sfdefault}{bx}{n}

\PassOptionsToPackage{table}{xcolor}
\usepackage{pifont}
\usepackage{hyperref}
\usepackage{url}
\usepackage{caption}
\usepackage{subcaption}
\usepackage{multirow}
\usepackage{multicol}
\usepackage[table]{xcolor}
\usepackage{booktabs}
\usepackage{wrapfig}
\usepackage{etoc}
\usepackage{enumitem}
\usepackage{graphicx}
\usepackage{colortbl}

\title{EmoPrefer: Can Large Language Models Understand Human Emotion Preferences?}

\author{Zheng Lian$^{1}$, Licai Sun$^{2}$, Lan Chen$^{1}$, Haoyu Chen$^{2}$, Zebang Cheng$^{3,4}$, Fan Zhang$^{5}$, \\
\textbf{Ziyu Jia}$^{1}$, \textbf{Ziyang Ma}$^{6}$, \textbf{Fei Ma}$^{4}$, \textbf{Xiaojiang Peng}$^{7}$, \textbf{Jianhua Tao}$^{8}$\\
$^1$MAIS, Institute of Automation, Chinese Academy of Sciences $^2$University of Oulu \\
$^3$Shenzhen University $^4$Guangdong Laboratory of Artificial Intelligence and Digital Economy (SZ) \\
$^5$The Chinese University of Hong Kong $^6$Shanghai Jiaotong University\\
$^7$Shenzhen Technology University $^8$Department of Automation, BNRist, Tsinghua University \\
\texttt{\{lianzheng2016@ia.ac.cn\}} \\
}

\iclrfinalcopy 
\begin{document}

\maketitle

\begin{abstract}
Descriptive Multimodal Emotion Recognition (DMER) has garnered increasing research attention. Unlike traditional discriminative paradigms that rely on predefined emotion taxonomies, DMER aims to describe human emotional state using free-form natural language, enabling finer-grained and more interpretable emotion representations. However, this free-form prediction paradigm introduces new challenges regarding its evaluation. Previous works depend on ground-truth descriptions, but emotions are inherently tied to diverse human behaviors, and generating a comprehensive and accurate description is inherently demanding. Other researchers reformulate this problem into a more tractable human preference learning task, but pairwise preference annotation involves substantial manual effort. This leads to a question: \emph{can we leverage multimodal LLMs (MLLMs) to achieve more cost-efficient preference annotation?} To answer this, we propose \textbf{EmoPrefer}, a pioneering work exploring the potential of LLMs in decoding human emotion preferences. Specifically, we construct the first emotion preference dataset, \textbf{EmoPrefer-Data}, featuring high-quality preference annotations from experts. Additionally, we introduce \textbf{EmoPrefer-Bench}, which evaluates the performance of various MLLMs and prompting techniques in preference prediction, while also revealing new strategies to enhance their performance. To the best of our knowledge, this is the first work exploring the capabilities of LLMs in understanding human emotion preferences. Our work advances the field of DMER and lays the foundation for more intelligent human-computer interaction.
\end{abstract}

\section{Introduction}
\label{sec:introduction}
\begin{figure}[h]
	\centering
	\includegraphics[width=\linewidth]{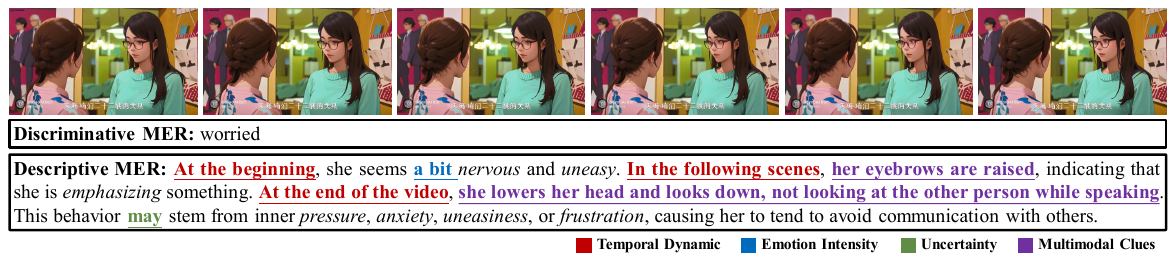}
	\caption{\textbf{Task comparison.} \emph{Discriminative MER} assigns a label from a predefined taxonomy, whereas \emph{Descriptive MER} provides a more detailed description by incorporating emotional temporal dynamics, intensity, uncertainty, and the existence and reasonableness of multimodal clues.}
	\label{fig:task_comparison}
\end{figure}

Descriptive Multimodal Emotion Recognition (DMER) \citep{lian2023explainable,lian2025affectgpt} aims to use free-form natural language to describe emotional states. Unlike traditional \emph{discriminative} methods that depend on predefined emotion taxonomies \citep{el2011survey, lian2024merbench}, DMER adopts a \emph{generative} paradigm, offering greater flexibility in emotion representation. This enables fine-grained and interpretable emotional expression, presenting significant opportunities for advancing emotion-intelligent human-computer interaction technologies \citep{brave2007emotion}. Recent advancements in Multimodal Large Language Models (MLLMs) \citep{zhao2023survey,yin2024survey}, with their rich vocabularies and multimodal understanding capabilities, have made this task feasible. Figure \ref{fig:task_comparison} demonstrates the distinctions between descriptive and discriminative MER.

\begin{figure}[h]
	\centering
	\includegraphics[width=\linewidth]{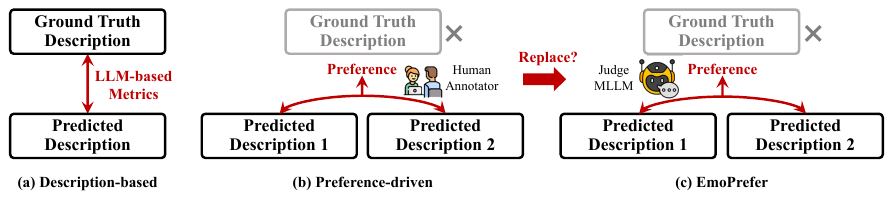}
	\caption{\textbf{Evaluation strategies for DMER.} (a) \emph{Description-based evaluation}, which relies on costly and often incomplete human-annotated descriptions; (b) \emph{Preference-driven evaluation}, where pairwise preference annotations require substantial human effort; (c) \emph{EmoPrefer}, a pioneering approach that explores whether MLLMs can decode human emotion preferences, aiming to replace expensive human annotations with MLLM-based judges.}
	\label{fig:evaluation_strategy}
\end{figure}

However, evaluating the quality of such open-ended descriptions remains a non-trivial task. Figure \ref{fig:evaluation_strategy} summarizes the existing evaluation strategies. The first approach leverages ground-truth emotion descriptions and employs LLMs, such as OpenAI GPT \citep{openai24gpt4o} or Google Gemini \citep{team2023gemini}, to measure the similarity between predicted and ground-truth descriptions \citep{lian2023explainable}. Nevertheless, emotions are inherently tied to diverse human behaviors, including facial expressions, (micro-)gestures, head movements, hand actions, and vocal tones \citep{noroozi2018survey, chen2023smg, li2020deep}. As a result, generating a comprehensive and accurate description of a person’s emotional state is inherently challenging, and unreliable ground-truth descriptions can lead to inaccurate evaluations. To address this, researchers propose abandoning costly and often incomplete ground-truth descriptions \citep{lian2025mer}. Instead, they reformulate the complex problem of designing indicators to measure semantic similarity with ground-truth descriptions into a more tractable problem of learning human preferences. However, this approach requires preference annotations for each model pair across multiple samples, entailing substantial human effort. For example, given $M$ models and $N$ samples, the total number of required comparisons is $C(M, 2) \times N$.

Given the high cost of manual preference annotation, a heuristic idea emerges: \emph{Can we leverage MLLMs to achieve automatic emotion preference decoding, thereby providing a more cost-efficient evaluation strategy for DMER?} To answer this, we propose \textbf{EmoPrefer}, the first work exploring the potential of MLLMs in emotion preference prediction. (1) We introduce \textbf{EmoPrefer-Data}, the first multimodal preference dataset centered on human emotions. In this dataset, we provide pairwise emotion descriptions for videos and recruit multiple expert annotators to label preferences. Only samples with unanimous agreement among all annotators are retained, ensuring high-quality preference annotations. (2) We establish \textbf{EmoPrefer-Bench}, the first benchmark for emotion preference prediction. In this benchmark, we conduct a comprehensive evaluation of different MLLMs and prompting techniques, further exploring strategies to enhance their alignment with human preferences. This paper presents pioneering work that reveals the capabilities of MLLMs in human emotion preference. Beyond evaluation, the dataset and insights from our benchmark will support the training of emotion-aware reward models, thereby enhancing the consistency of MLLMs with human emotional understanding. The main contributions of this paper are summarized as follows:
\begin{itemize}[leftmargin=20pt]

    \item \textbf{(EmoPrefer)} This is a pioneering work that explores the potential of MLLMs in emotion preference. Beyond evaluation, the insights from this paper will contribute to training reward models capable of understanding human emotions, paving the way for emotion-intelligent models.
    
    \item \textbf{(EmoPrefer-Data)} This paper constructs the first preference dataset centered on emotions. Our work makes a substantial contribution to current research on LLMs-as-judges by not only expanding the modality to full multimodal scenarios (encompassing audio, video, and text) but also extending preference learning to the domain of human emotions.
    
    \item \textbf{(EmoPrefer-Bench)} This paper proposes comprehensive evaluation metrics and explores diverse solutions, demonstrating the feasibility of automatic preference prediction for human emotions. Additionally, we introduce additional strategies to enhance prediction performance.
    
\end{itemize}

\begin{figure}[t]
	\centering
	\includegraphics[width=\linewidth]{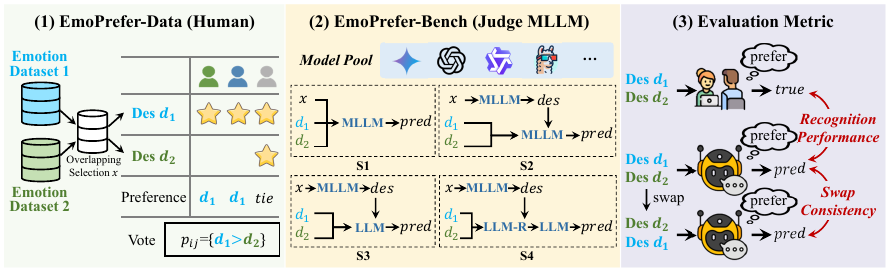}
	\caption{\textbf{Pipeline of EmoPrefer.} (1) EmoPrefer-Data. We select overlapping videos across two descriptive emotion datasets, then conduct human preference annotations and adopt their majority-voted results as final labels. (2) EmoPrefer-Bench. We systematically evaluate various MLLMs and prompting techniques, revealing their potential in emotion preference decoding. (3) Evaluation metric. We introduce two metrics for performance evaluation: \emph{recognition performance} measures alignment between MLLM-based judges and human annotations; \emph{swap consistency} evaluates the MLLM-based judges' robustness to swapping the order of description pairs.}
	\label{fig:emoprefer_pipeline}
\end{figure}

\section{EmoPrefer-Data}
\label{sec:preference_dataset}
Figure \ref{fig:emoprefer_pipeline} illustrates the overall pipeline of \textbf{EmoPrefer}, which is the first work to explore the potential of MLLMs for emotion preference decoding. To achieve this, we require a high-quality, human-annotated emotion preference dataset. In this section, we detail the construction process of our \textbf{EmoPrefer-Data} from four aspects: (1) sample selection, (2) description quality analysis, (3) preference annotation, and (4) inter-annotator agreement analysis.

\paragraph{Sample Selection.}
For preference annotation, we provide pairwise emotion descriptions for each video and ask human annotators to select the better one. However, if the quality of two descriptions varies significantly, the preference prediction task will become trivial, potentially diminishing the advantages of superior solutions. To ensure high-quality descriptions, we leverage two benchmark datasets for descriptive emotion: MERR-Fine \citep{cheng2024emotion} and MER-Caption+ \citep{lian2025affectgpt}. The raw videos in these datasets are sourced from the MER2024 dataset \citep{lian2024mer}. We only select videos that appear in both datasets, resulting in 1,368 samples, each with two descriptions (one from each dataset). Appendix \ref{appendix:dataset_data_visualization} provides visualizations of the selected video samples. These videos primarily feature a single front-facing character with complete audio content, ensuring the full expression of emotions for the target speaker.

\begin{figure}[h]
    \captionsetup[subfigure]{font=scriptsize} 
    \begin{center}
        \begin{subcaptionbox}{\scriptsize{MERR-Fine (Emotion)}\label{fig:emotion_clouds-1}}[0.24\linewidth]
            {\includegraphics[width=\linewidth]{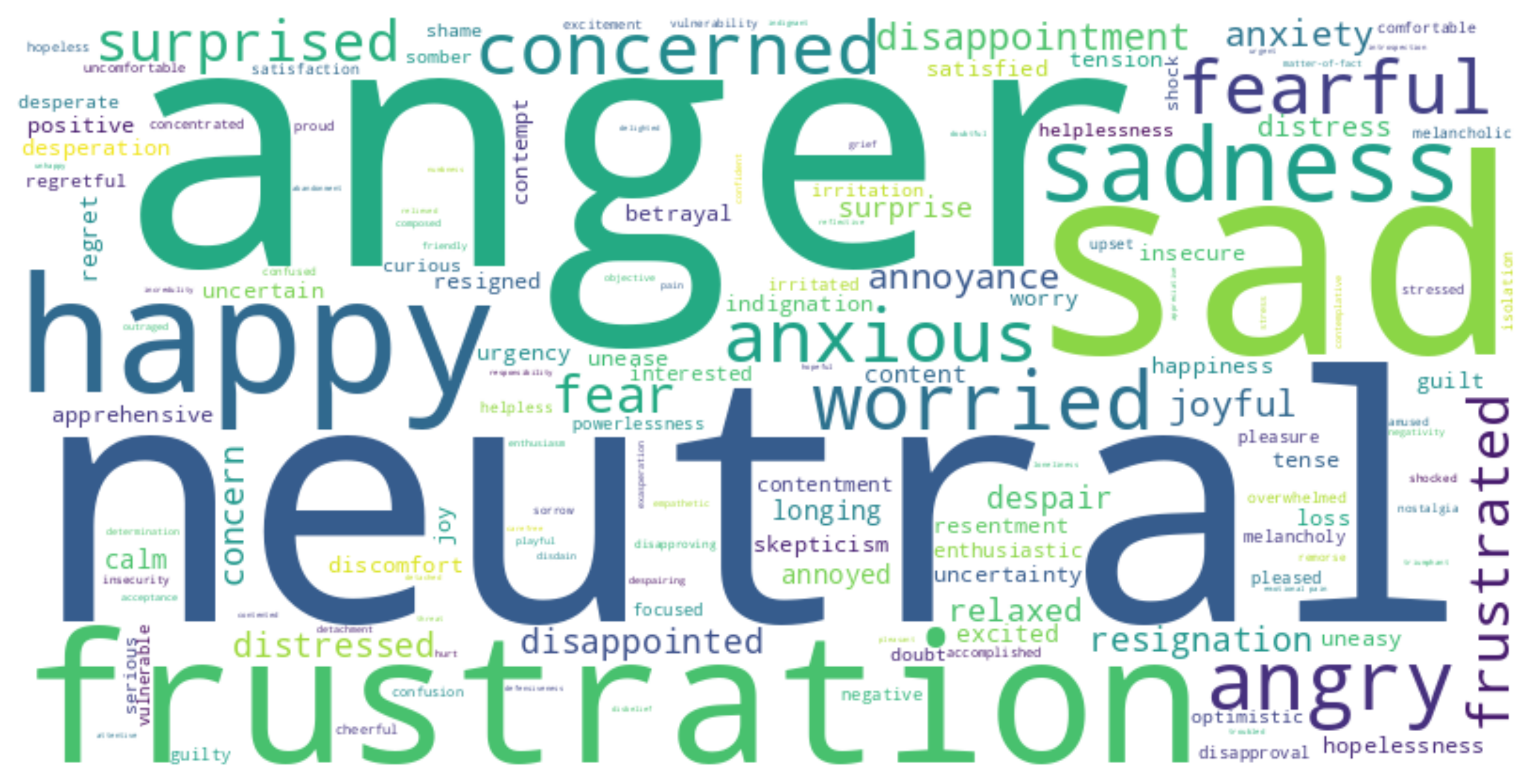}}
        \end{subcaptionbox}
        \begin{subcaptionbox}{\scriptsize{MER-Caption+ (Emotion)}\label{fig:emotion_clouds-2}}[0.24\linewidth]
            {\includegraphics[width=\linewidth]{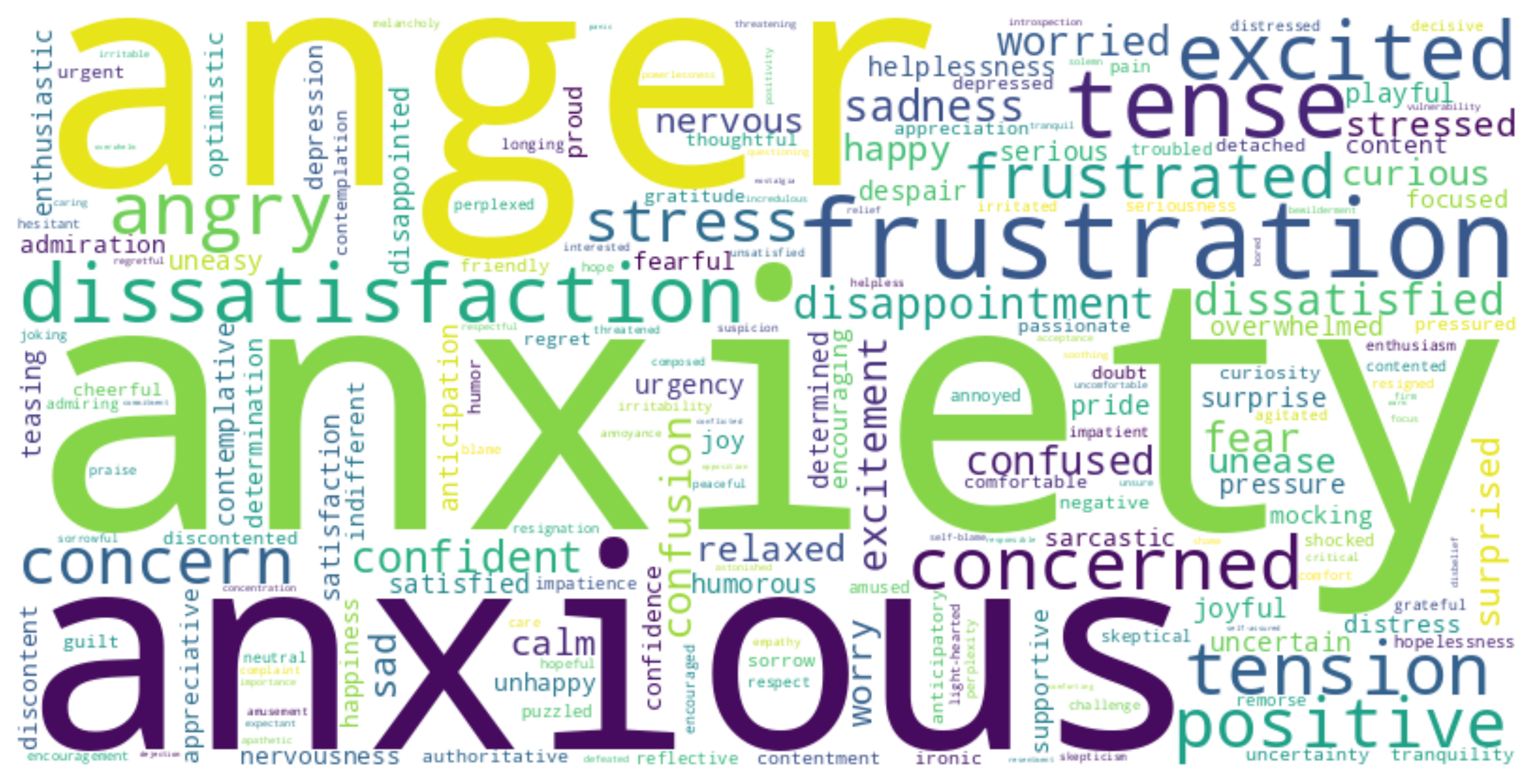}}
        \end{subcaptionbox}
        \begin{subcaptionbox}{\scriptsize{MERR-Fine (Noun)}\label{fig:noun_clouds-1}}[0.24\linewidth]
            {\includegraphics[width=\linewidth]{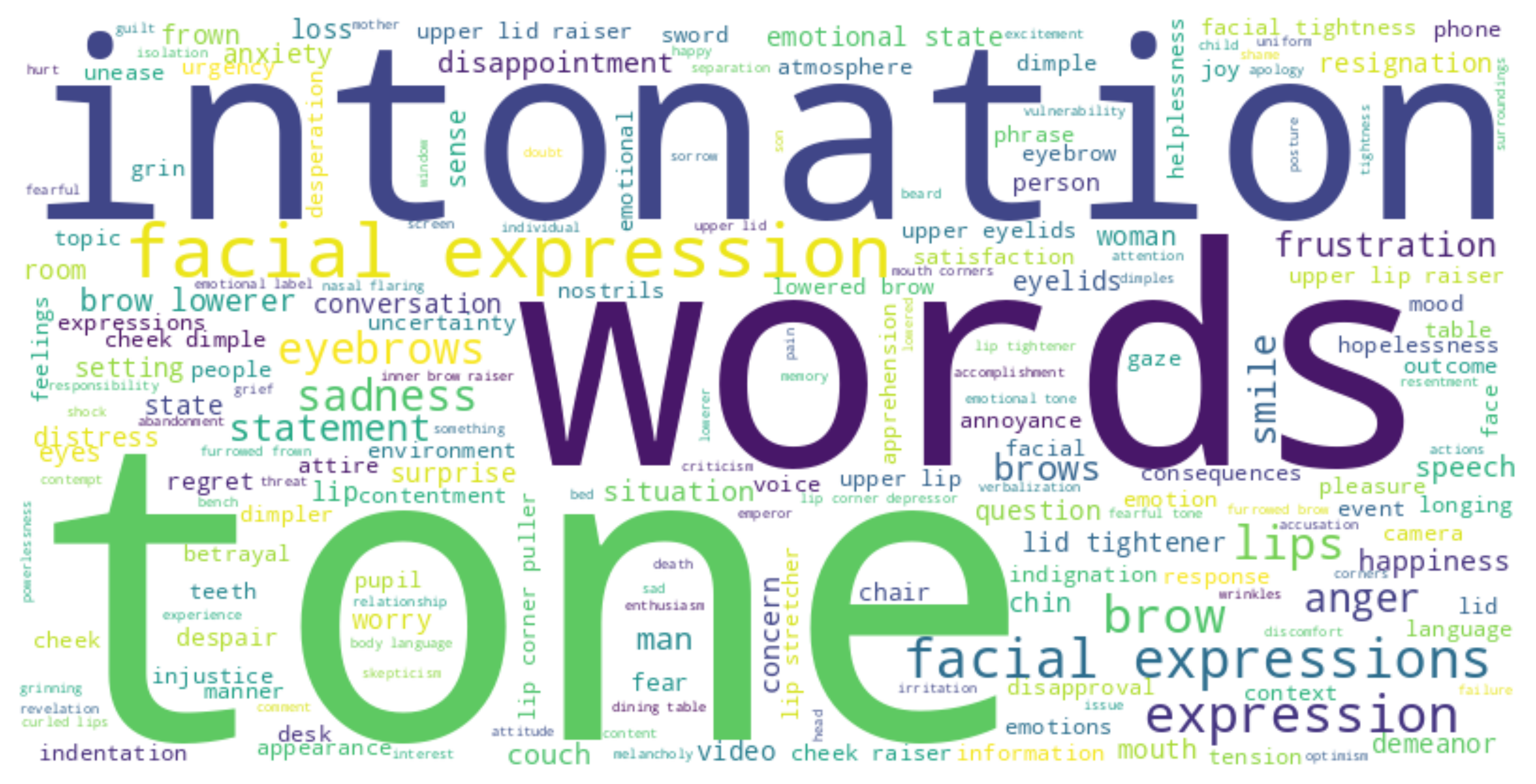}}
        \end{subcaptionbox}
        \begin{subcaptionbox}{\scriptsize{MER-Caption+ (Noun)}\label{fig:noun_clouds-2}}[0.24\linewidth]
            {\includegraphics[width=\linewidth]{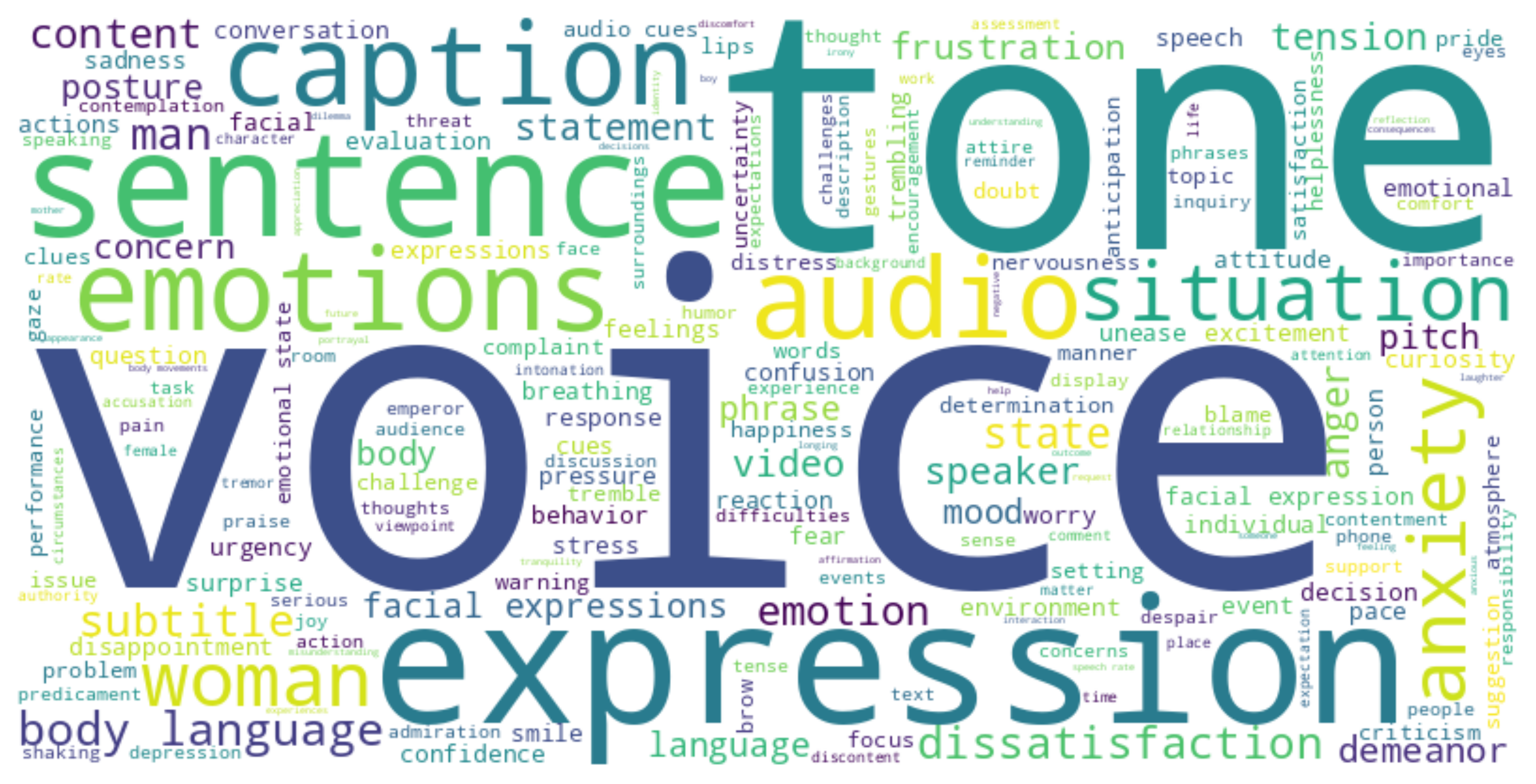}}
        \end{subcaptionbox}
    \end{center}
    \caption{\textbf{Word cloud visualization.} We extract emotion words and nouns from the descriptions and generate corresponding word clouds for both datasets. Appendix \ref{appendix:dataset_description_statistic} details the extraction process.}
    \label{fig:word_clouds}
\end{figure}

\paragraph{Description Quality Analysis.}
This section analyzes the descriptions provided by MERR-Fine and MER-Caption+. These datasets focus on capturing the character's multimodal cues in the video, including facial expressions, body language, vocal tone, contextual events, and environmental factors, offering a comprehensive depiction of the person's emotional state. Figure \ref{fig:word_clouds} presents word clouds of emotion words and nouns extracted from these descriptions. The detailed extraction process is described in Appendix \ref{appendix:dataset_description_statistic}. Our analysis reveals that these descriptions contain a rich diversity of emotional vocabulary and multimodal cues related to emotions, confirming their high quality.

\paragraph{Preference Annotation.}
During the annotation process, we recruit master's students from our lab as candidate annotators. Given their research focus on affective computing, these individuals are already familiar with emotion definitions. To ensure their understanding aligns with that of the general population, we first conduct a preliminary test. Specifically, we select 12 samples that have reached consensus among eight external annotators (i.e., all eight have assigned identical emotion labels). The candidates are then asked to annotate these 12 samples, and only those who achieve at least 75\% accuracy are retained, resulting in three qualified annotators. The selected annotators are then tasked with determining which of two descriptions more accurately reflects the character's emotional state. Appendix \ref{appendix:dataset_annotation_platform} illustrates the layout of our annotation platform. For each comparison, annotators can choose one of three options: \emph{the first description is better}, \emph{the second description is better}, or \emph{tie}. We retain only samples where all three annotators reach consensus. This strict selection strategy ensures the reliability of the preference labels. 

\begin{wrapfigure}{r}{0.46\linewidth}
    \centering
    \begin{subcaptionbox}{w/o tie\label{fig:annotator_consistency-1}}[0.476\linewidth]
            {\includegraphics[width=\linewidth, trim=50 0 50 0, clip]{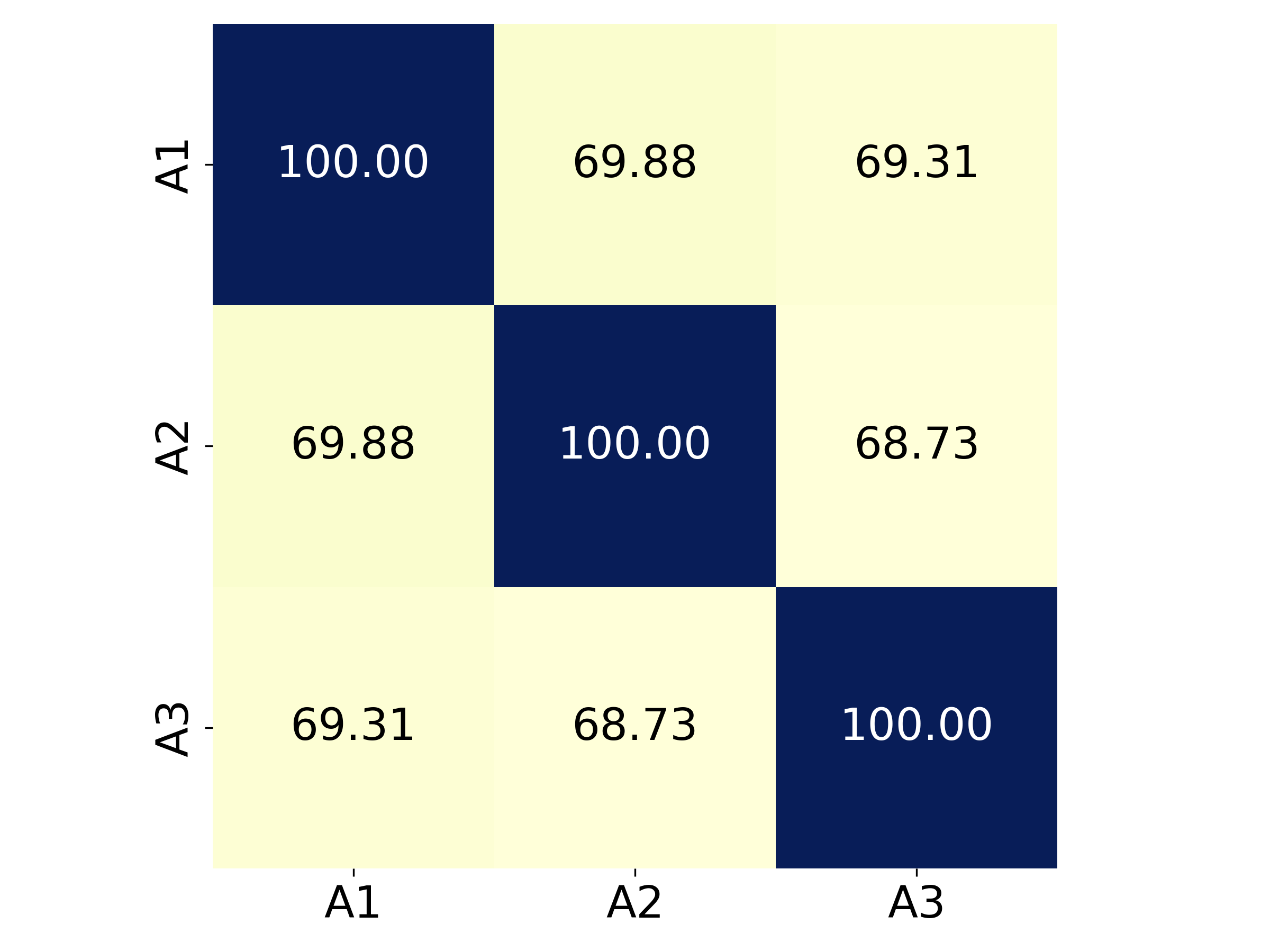}}
        \end{subcaptionbox}
        \begin{subcaptionbox}{w/ tie\label{fig:annotator_consistency-2}}[0.476\linewidth]
            {\includegraphics[width=\linewidth, trim=50 0 50 0, clip]{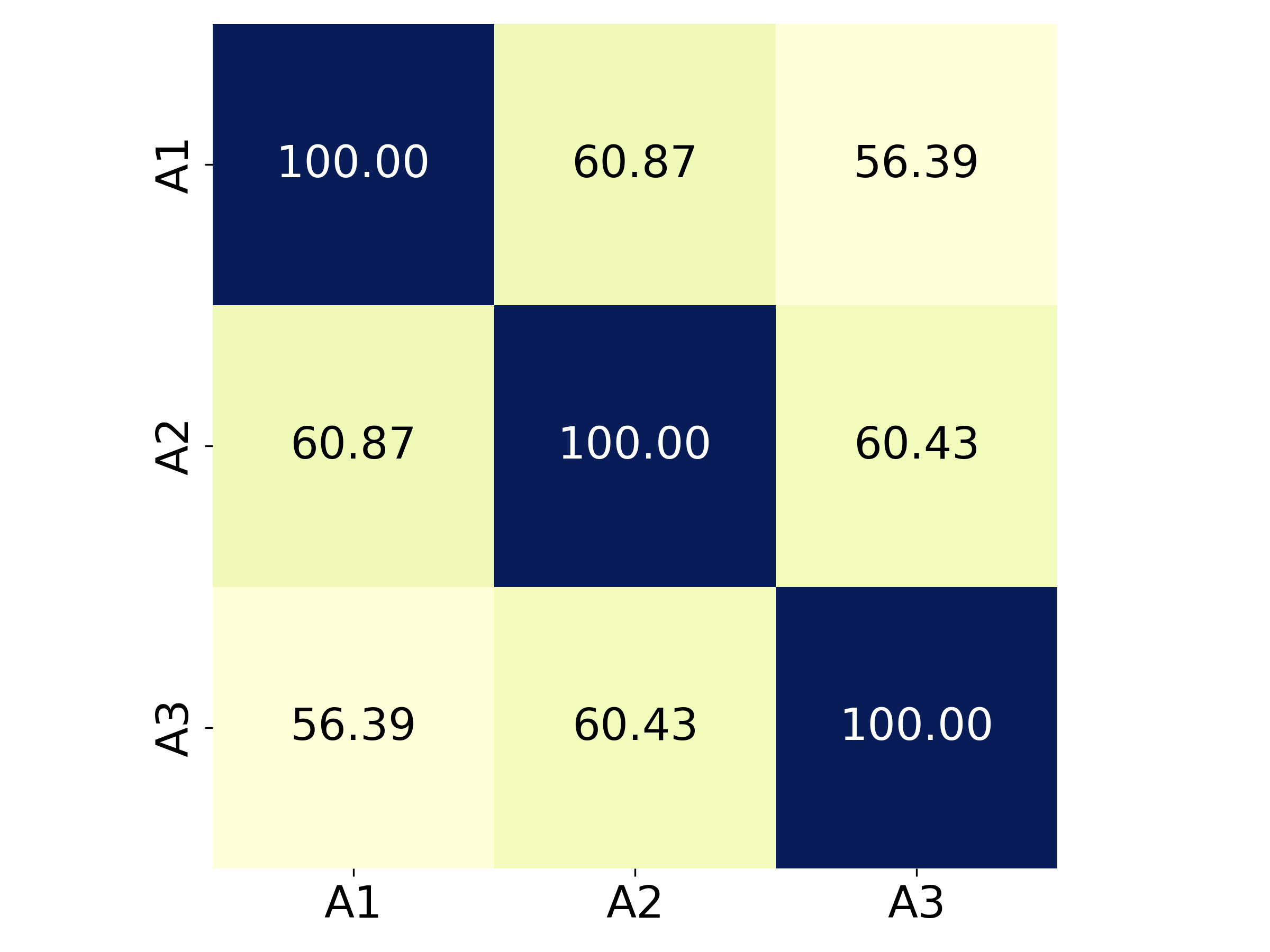}}
        \end{subcaptionbox}
    \caption{\textbf{Inter-annotator agreement analysis.} We present the inter-annotator agreement under two conditions: w/o ties and w/ ties.}
    \label{fig:annotator_consistency}
\end{wrapfigure}
\paragraph{Inter-annotation Agreement Analysis.}
After annotation, we evaluate the inter-annotator agreement for emotion preference recognition. Figure \ref{fig:annotator_consistency-1} presents the agreement scores excluding ties, with an average of $\mathrm{avg}(69.88, 69.31, 68.73)=69.31\%$. This reflects the upper bound of human agreement and confirms that different annotators can reach a reasonable consensus on this task. Figure \ref{fig:annotator_consistency-2} shows the agreement scores including ties, where the average drops to $\mathrm{avg}(60.87, 56.39, 60.43)=59.23\%$, a 10\% decrease compared to Figure \ref{fig:annotator_consistency-1}. This suggests that annotators struggle more to reach consensus on ties, indicating that ties represent higher ambiguity than clear preferences (e.g., when one description is distinctly better than another). Table \ref{tab:dataset_compare} compares EmoPrefer-Data with existing datasets. Our dataset not only covers more modalities but also extends the preference task to human emotions. To the best of our knowledge, this is the first multimodal preference dataset centered on human emotions, providing valuable resources for future research.
\begin{table*}[h]
	\centering
	\caption{\textbf{Dataset comparison.} The second column specifies the modalities used for preference tasks. Since MLLM-as-a-Judge \citep{chen2024mllm} relies on image sequences rather than videos, we use the symbol \ding{52}\rotatebox[origin=c]{-9.2}{\kern-0.7em\ding{55}} to indicate this. Our EmoPrefer-Data not only expands the range of modalities but also extends the preference task to human emotions.}
	\label{tab:dataset_compare}
	\scalebox{0.8}{
		\begin{tabular}{l|cccc|l}
			\toprule
			\multirow{2}{*}{\textbf{Dataset}} & \multicolumn{4}{c|}{\textbf{Modality}} & \multirow{2}{*}{\textbf{Task Description}} \\
                & \textbf{T} & \textbf{I} & \textbf{A} & \textbf{V} & \\
			\midrule
MT-Bench \citep{zheng2023judging} & \ding{52} & \ding{55} & \ding{55} & \ding{55} & writing, math, general knowledge, etc. \\
LLMEval \citep{zhang2023wider} & \ding{52} & \ding{55} & \ding{55}  & \ding{55} & storytelling, summarization, dialogue, etc.\\
FairEval \citep{wang2024large}  & \ding{52} & \ding{55} & \ding{55} & \ding{55} & writing, role play, etc. \\
PandaLM \citep{wang2024pandalm} & \ding{52} & \ding{55} & \ding{55} & \ding{55} & law, biomedical data, etc. \\
JudgeBench \citep{tan2024judgebench} & \ding{52} & \ding{55} & \ding{55}  & \ding{55} & general knowledge, reasoning, math, coding \\
CALM \citep{ye2025justice} & \ding{52} & \ding{55} & \ding{55}  & \ding{55} & commonsense, math, science, etc. \\
MLLM-as-a-Judge \citep{chen2024mllm} & \ding{52} & \ding{52} & \ding{55} & \ding{52}\rotatebox[origin=c]{-9.2}{\kern-0.7em\ding{55}} & captioning, reasoning, etc. \\
VL-RewardBench \citep{li2025vl} & \ding{52} & \ding{52} & \ding{55} & \ding{55} & general visual perception, reasoning, etc. \\
MM-RLHF \citep{zhang2025mm} & \ding{52} & \ding{52} & \ding{55} & \ding{52} & captions, safety, reasoning, etc. \\
Multimodal RewardBench \citep{yasunaga2025multimodal} & \ding{52} & \ding{52} & \ding{55} & \ding{55} & knowledge, safety, reasoning, etc. \\
\textbf{EmoPrefer-Data (Ours)} & \ding{52} & \ding{52} & \ding{52} & \ding{52} & human emotion understanding \\
			\bottomrule
		\end{tabular}
	}
\end{table*}

\section{Evaluation Metrics}
\label{sec:evaluation_metric}
In this section, we propose two metrics to evaluate the performance of MLLM-based judges: \emph{recognition performance} and \emph{swap consistency}. Figure \ref{fig:emoprefer_pipeline} illustrates the detailed calculation pipeline.

\paragraph{Recognition Performance.}
This metric evaluates the alignment between model predictions and human annotations. In EmoPrefer-Data, the labels fall into three categories: (1) \emph{the first description is better}, (2) \emph{the second description is better}, or (3) \emph{tie}. Since human annotators often struggle to reach consensus on ties (as shown in Figure \ref{fig:annotator_consistency}), we report both the weighted average F1-score (WAF) and accuracy (ACC) under two settings: two-class (excluding ties) and three-class (including ties) setups. Given the inherent class imbalance in the dataset, we default to the two-class WAF when reporting a single metric in the following experiments.

\paragraph{Swap Consistency.}
Ideally, swapping the order of description pairs should not affect the model’s preferred description. However, in practice, we observe that order swaps sometimes lead to different outcomes. This discrepancy may arise because the model’s preferences are determined by the order of inputs rather than the actual content of the descriptions. Therefore, we introduce \emph{swap consistency}, a metric that evaluates a model’s robustness to order swapping and ensures predictions are based on content rather than ordering. Specifically, given $N$ examples $\{(x^n, d_1^n, d_2^n)\}_{i=1}^N$, where $x^n$ is a video and $d_1^n$ and $d_2^n$ are two corresponding descriptions, we use MLLM to predict preferences for both the normal order input $(x^n, d_1^n, d_2^n)$ and the swapped order input $(x^n, d_2^n, d_1^n)$. This yields two outcomes: $o_{\text{normal}}^n$ and $o_{\text{swapped}}^n$. We compute swap consistency as follows:
\begin{equation}
    \text{Swap Consistency}=\frac{\sum_{i=1}^{N}\mathbb{I}(o_{\text{normal}}^n = o_{\text{swapped}}^n)}{N},
\end{equation}
where $\mathbb{I}(\cdot)$ is the indicator function, and higher scores indicate greater robustness to input order variations. Appendix \ref{appendix:metric_swap_consistency} provides the prompt template used in the calculation process.

\section{EmoPrefer-Bench}
\label{sec:emoprefer_bench}
This is the first benchmark designed for emotion preference recognition. We evaluate various MLLMs and explore the effectiveness of model-based crowdsourcing. The paper reports the zero-shot performance of these MLLMs, with all experiments conducted on A100 GPUs. To reduce randomness, each experiment is run twice, and both the mean and standard deviation are reported.

\subsection{Judge MLLM}
\label{sec:solutions}
Given that humans express emotions through multiple modalities, we require MLLMs to support at least audio or video inputs. To enable preference prediction, we propose four prompting strategies (see Figure \ref{fig:emoprefer_pipeline}). \textbf{Strategy 1 (S1):} We input the video along with two descriptions into the MLLM and ask the model to determine the better one based on the character's emotional state. \textbf{Strategy 2 (S2):} We decompose S1 into two steps. First, we instruct the MLLM to generate a detailed description of the video. Then, using the output from the first step as a reference, we prompt the model to determine which of the two descriptions better aligns with it. \textbf{Strategy 3 (S3):} The second step in S2 relies solely on text input, meaning it can be implemented using an external LLM. The rationale for this alternative is that current MLLM training typically begins with a pretrained LLM, followed by additional multimodal fine-tuning. While this enables the model to process other modalities, such training may compromise its text understanding capabilities. Thus, we opt to use an external LLM for the second step. \textbf{Strategy 4 (S4):} We further break down the second step in S3 into two sub-steps, requiring the model to perform additional reasoning before making a decision. This allows us to investigate whether the reasoning process is necessary for accurate preference prediction. We adopt Qwen2.5-7B \citep{qwen2025qwen25technicalreport} as the external LLM for S3/S4, and the rationale for this choice is discussed in Section \ref{sec:main_results}. Appendix \ref{appendix:prompts} provides the prompt templates for all strategies.

\subsection{Main Results}
\label{sec:main_results}
\begin{figure}[h]
\begingroup  
    \captionsetup[subfigure]{font=tiny} 
    \begin{center}
       \begin{subcaptionbox}{\tiny{Video-LLaVA}\label{fig:solution_result-1}}[0.156\linewidth]
    {\includegraphics[width=\linewidth]{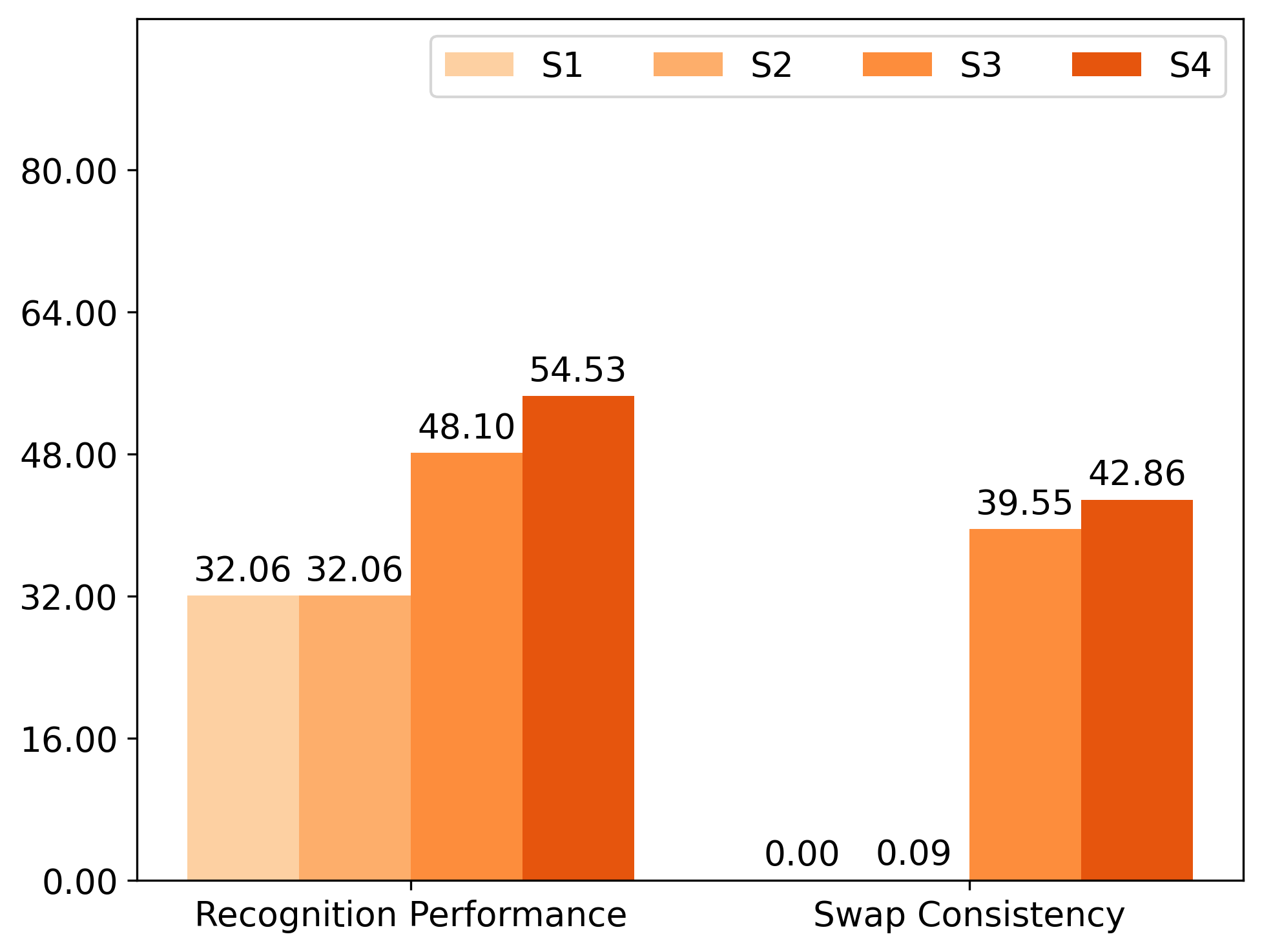}}
\end{subcaptionbox}
\begin{subcaptionbox}{\tiny{Qwen2-Audio}\label{fig:solution_result-2}}[0.156\linewidth]
    {\includegraphics[width=\linewidth]{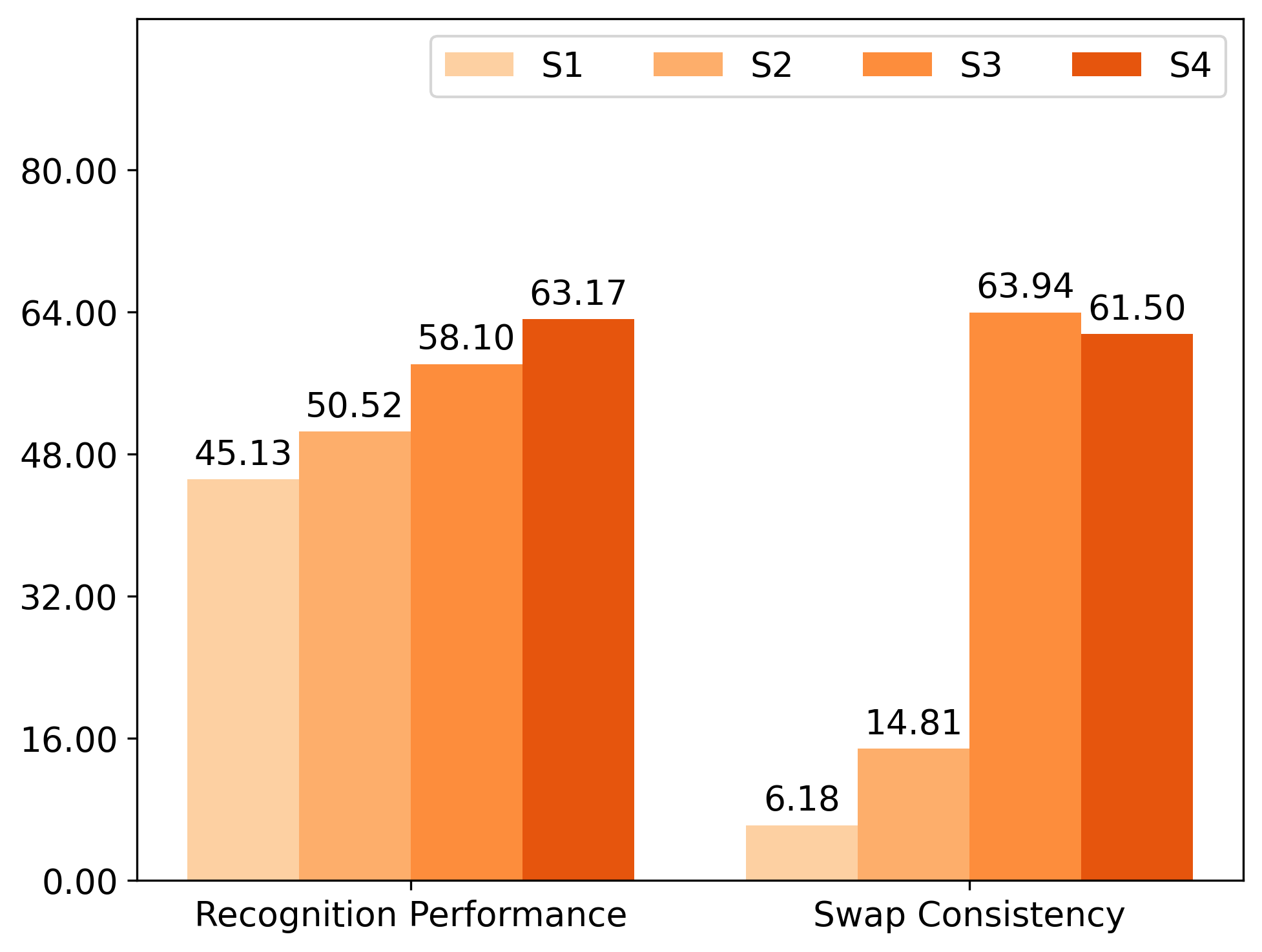}}
\end{subcaptionbox}
\begin{subcaptionbox}{\tiny{LLaMA-VID}\label{fig:solution_result-3}}[0.156\linewidth]
    {\includegraphics[width=\linewidth]{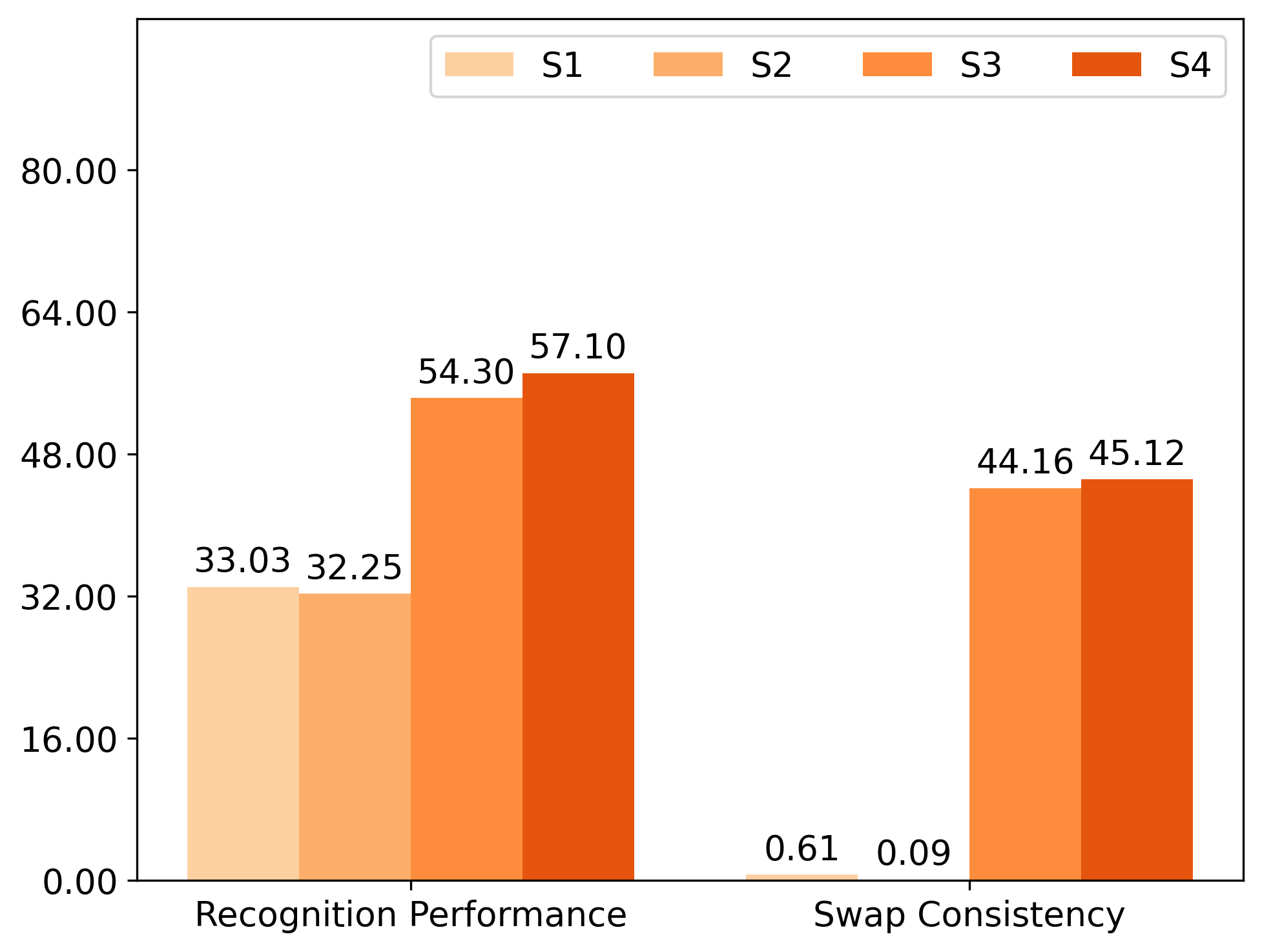}}
\end{subcaptionbox}
\begin{subcaptionbox}{\tiny{Chat-UniVi}\label{fig:solution_result-4}}[0.156\linewidth]
    {\includegraphics[width=\linewidth]{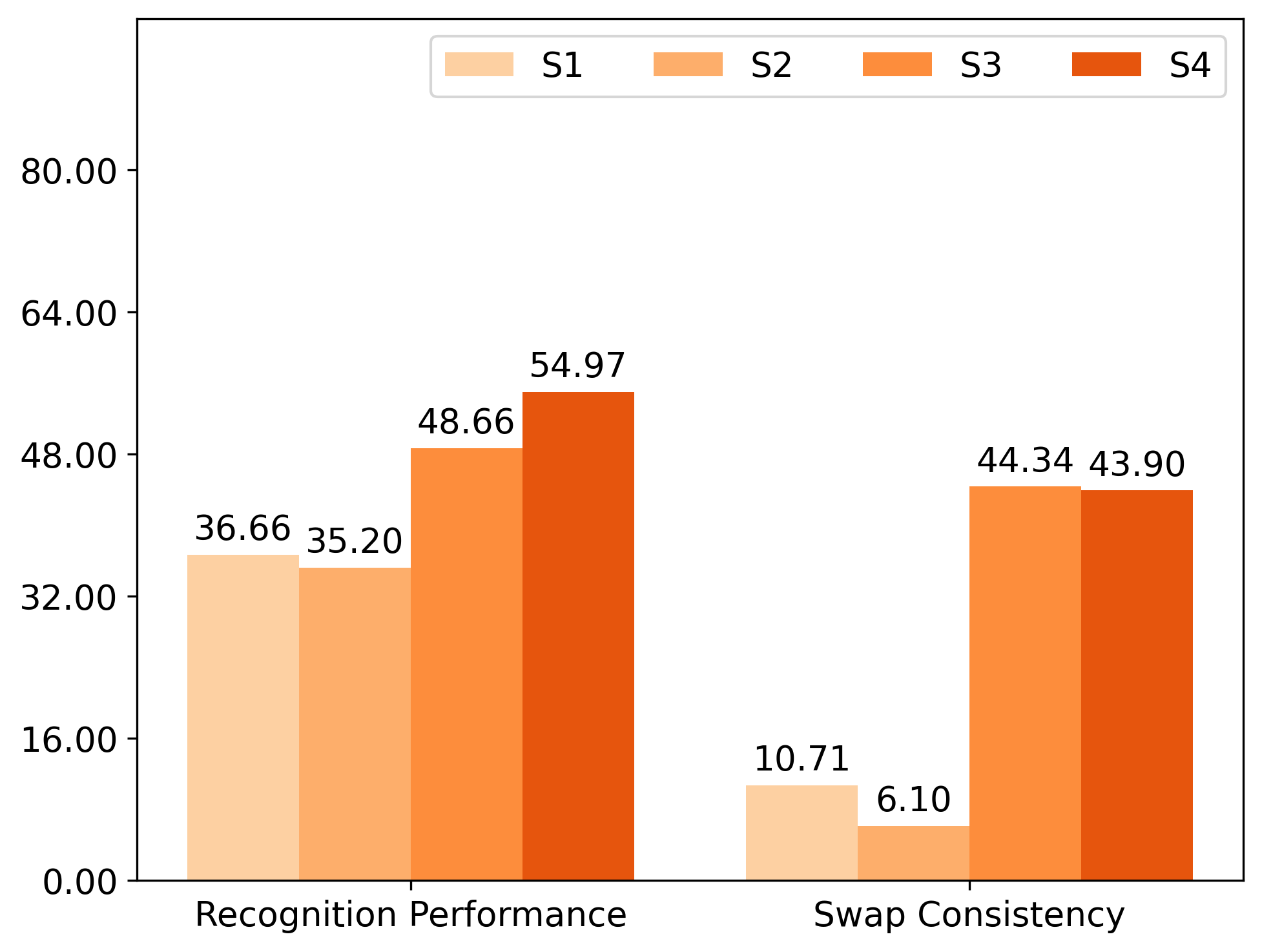}}
\end{subcaptionbox}
\begin{subcaptionbox}{\tiny{mPLUG-Owl}\label{fig:solution_result-5}}[0.156\linewidth]
    {\includegraphics[width=\linewidth]{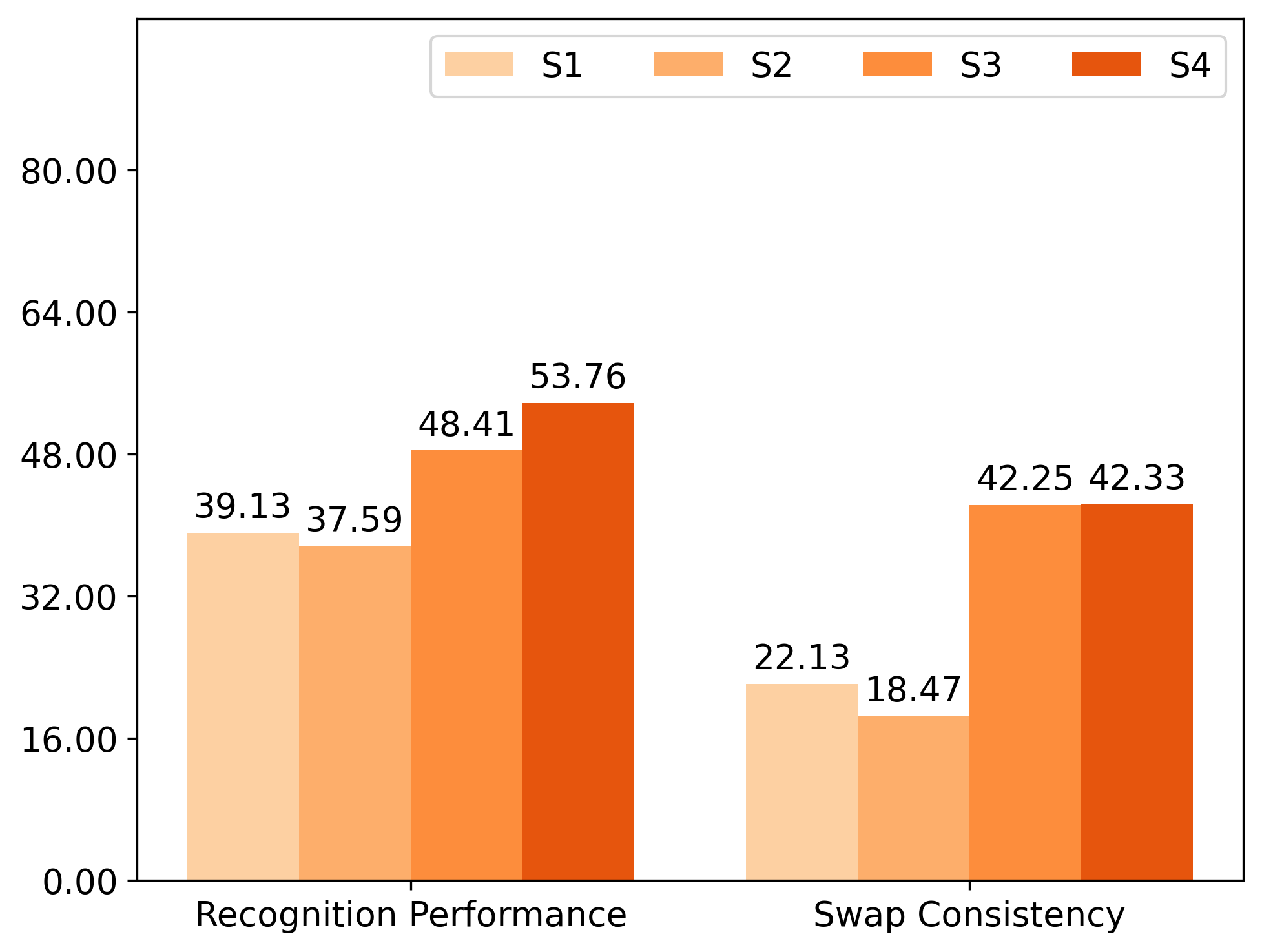}}
\end{subcaptionbox}
\begin{subcaptionbox}{\tiny{VideoChat}\label{fig:solution_result-6}}[0.156\linewidth]
    {\includegraphics[width=\linewidth]{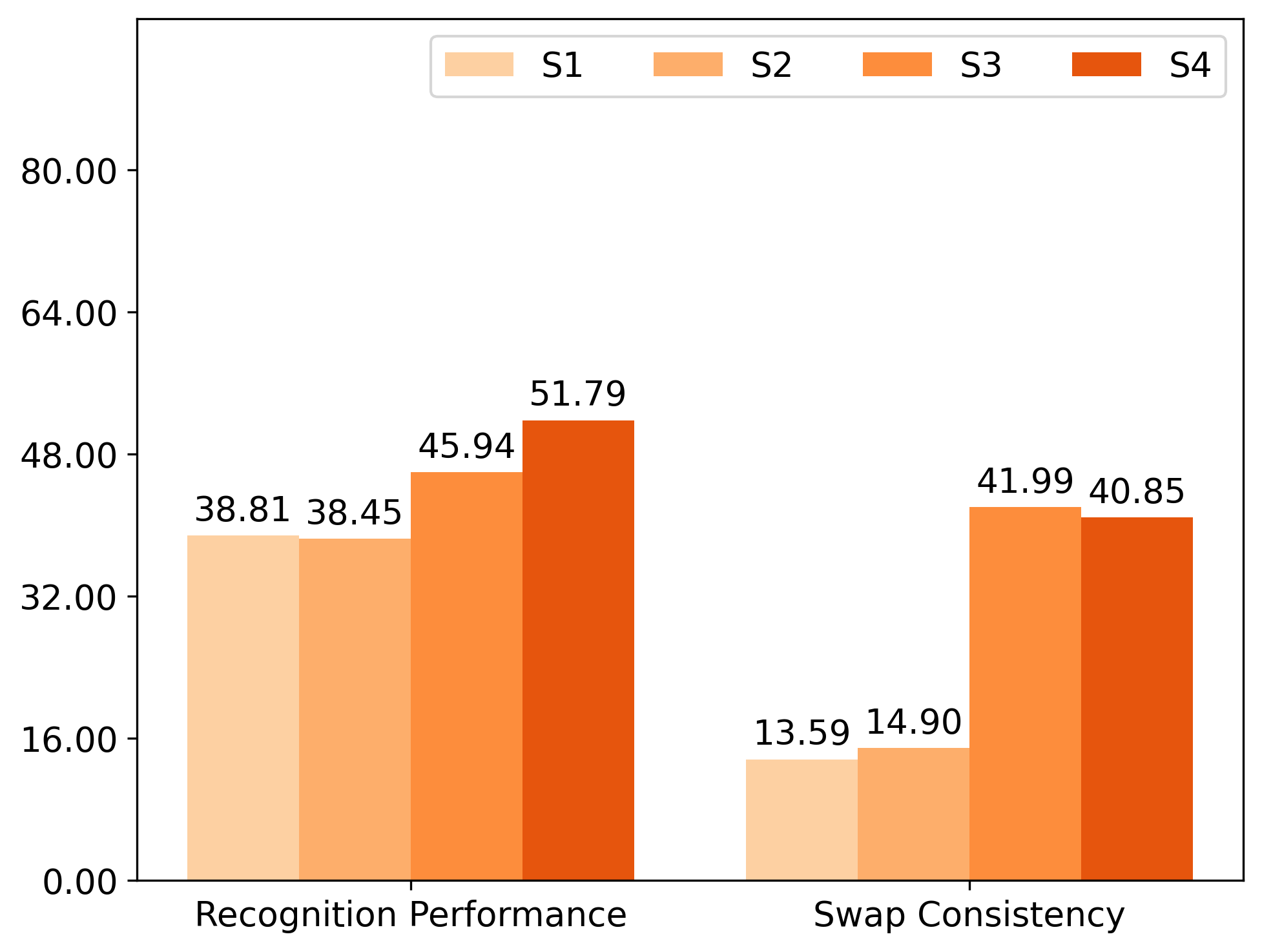}}
\end{subcaptionbox}
\begin{subcaptionbox}{\tiny{VideoChat2}\label{fig:solution_result-7}}[0.156\linewidth]
    {\includegraphics[width=\linewidth]{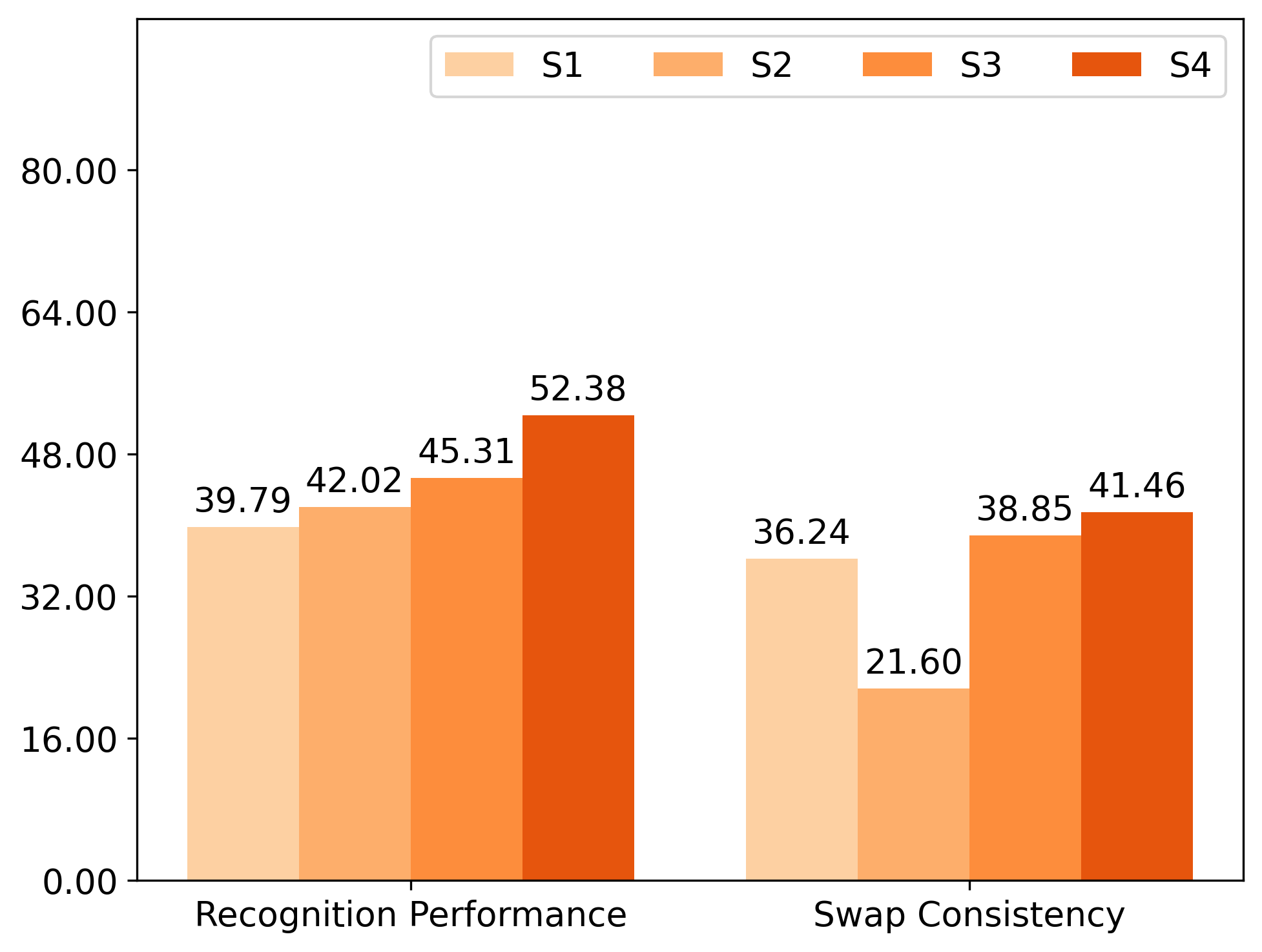}}
\end{subcaptionbox}
\begin{subcaptionbox}{\tiny{LLaVA-Next-Video}\label{fig:solution_result-8}}[0.156\linewidth]
    {\includegraphics[width=\linewidth]{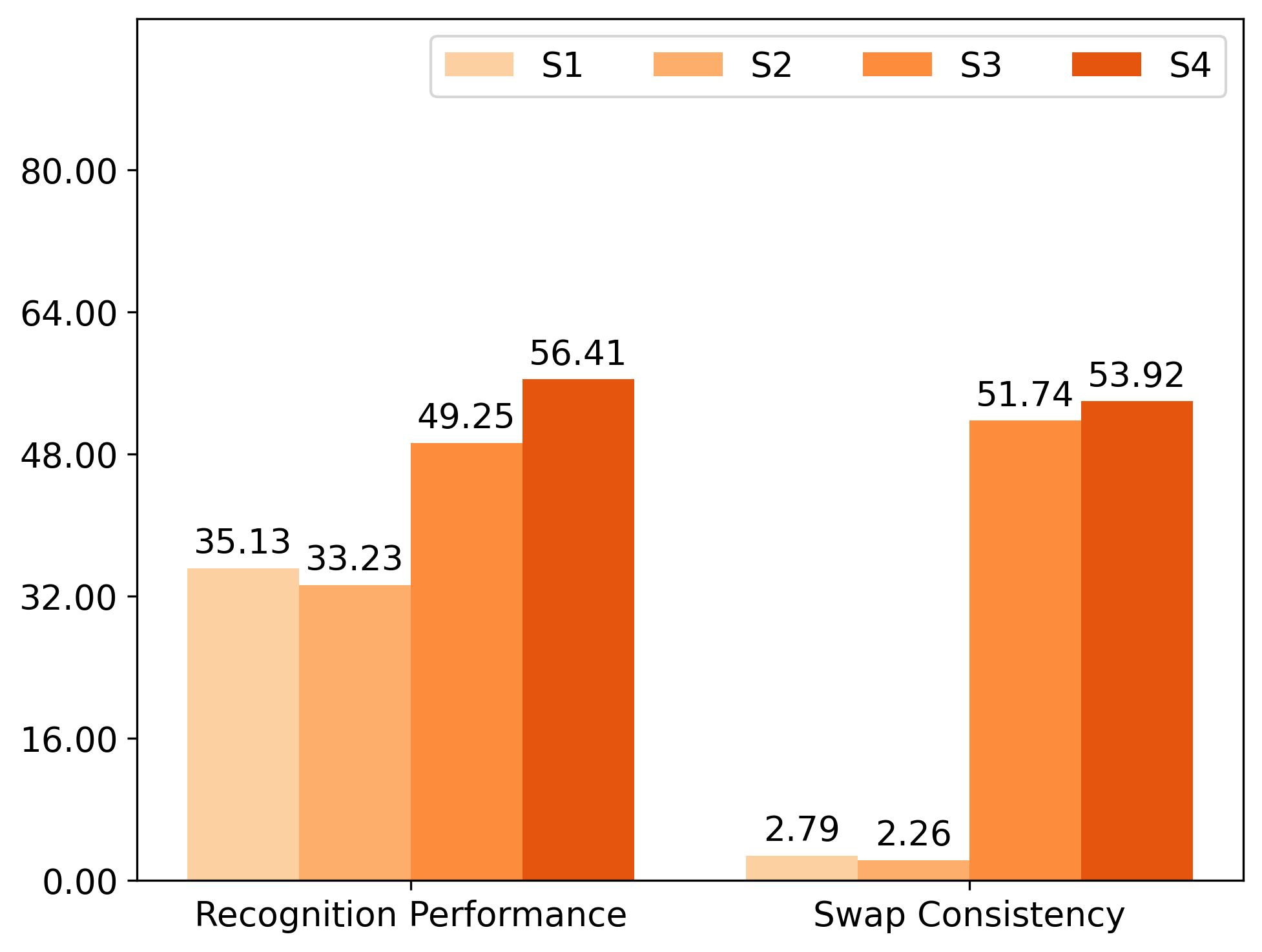}}
\end{subcaptionbox}
\begin{subcaptionbox}{\tiny{VITA-1.5}\label{fig:solution_result-9}}[0.156\linewidth]
    {\includegraphics[width=\linewidth]{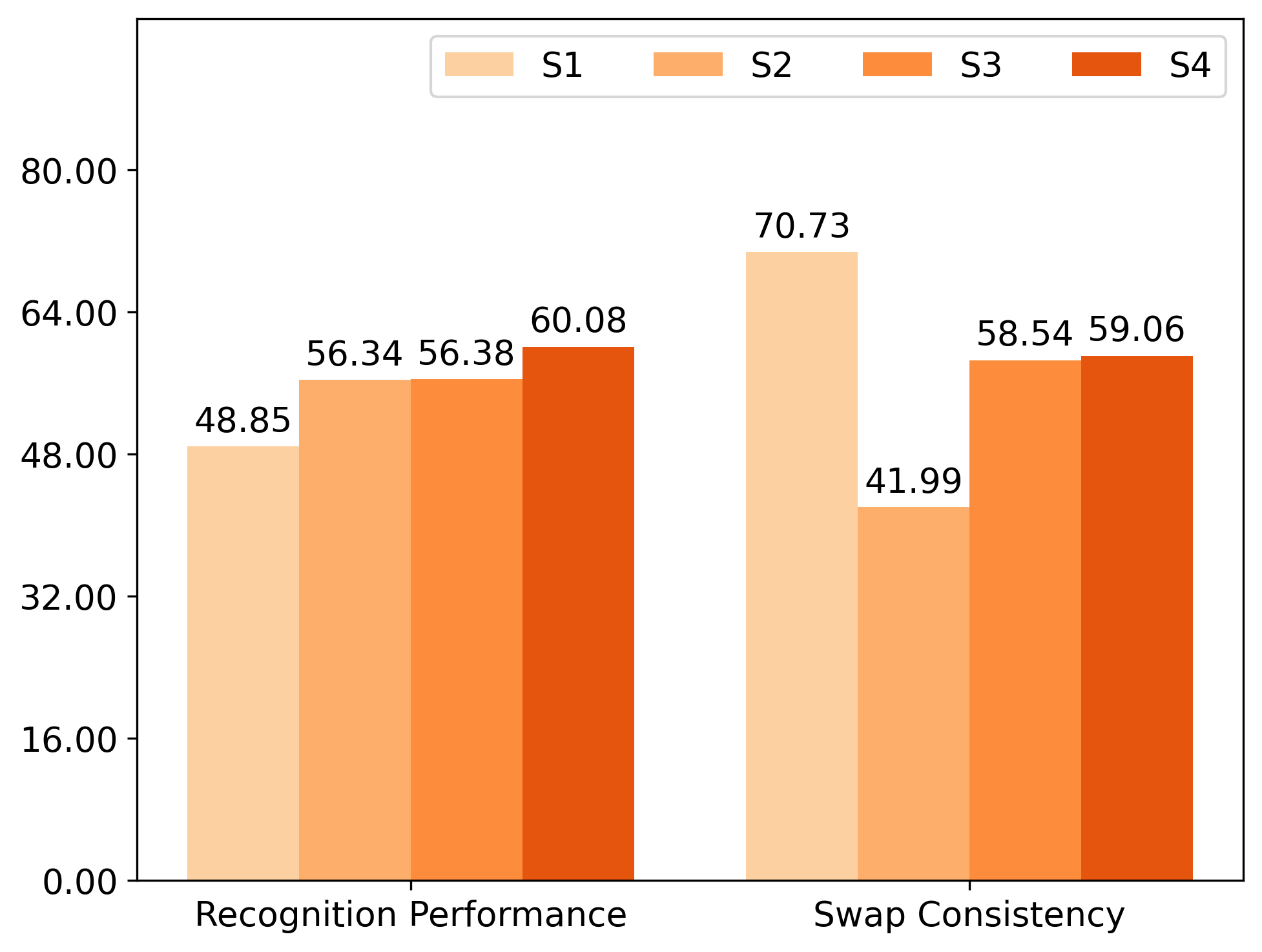}}
\end{subcaptionbox}
\begin{subcaptionbox}{\tiny{PLLAVA}\label{fig:solution_result-10}}[0.156\linewidth]
    {\includegraphics[width=\linewidth]{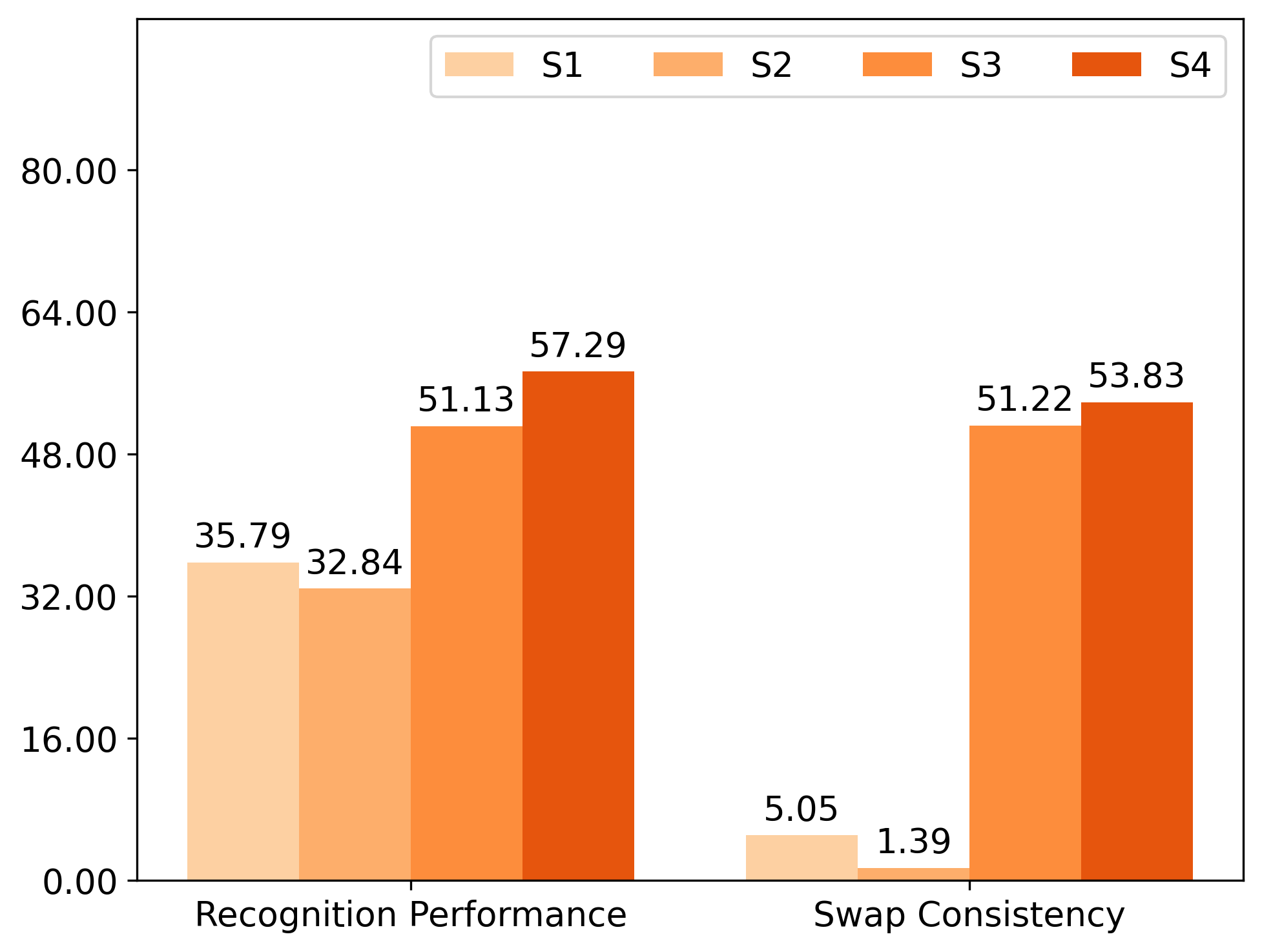}}
\end{subcaptionbox}
\begin{subcaptionbox}{\tiny{Qwen2.5-VL}\label{fig:solution_result-11}}[0.156\linewidth]
    {\includegraphics[width=\linewidth]{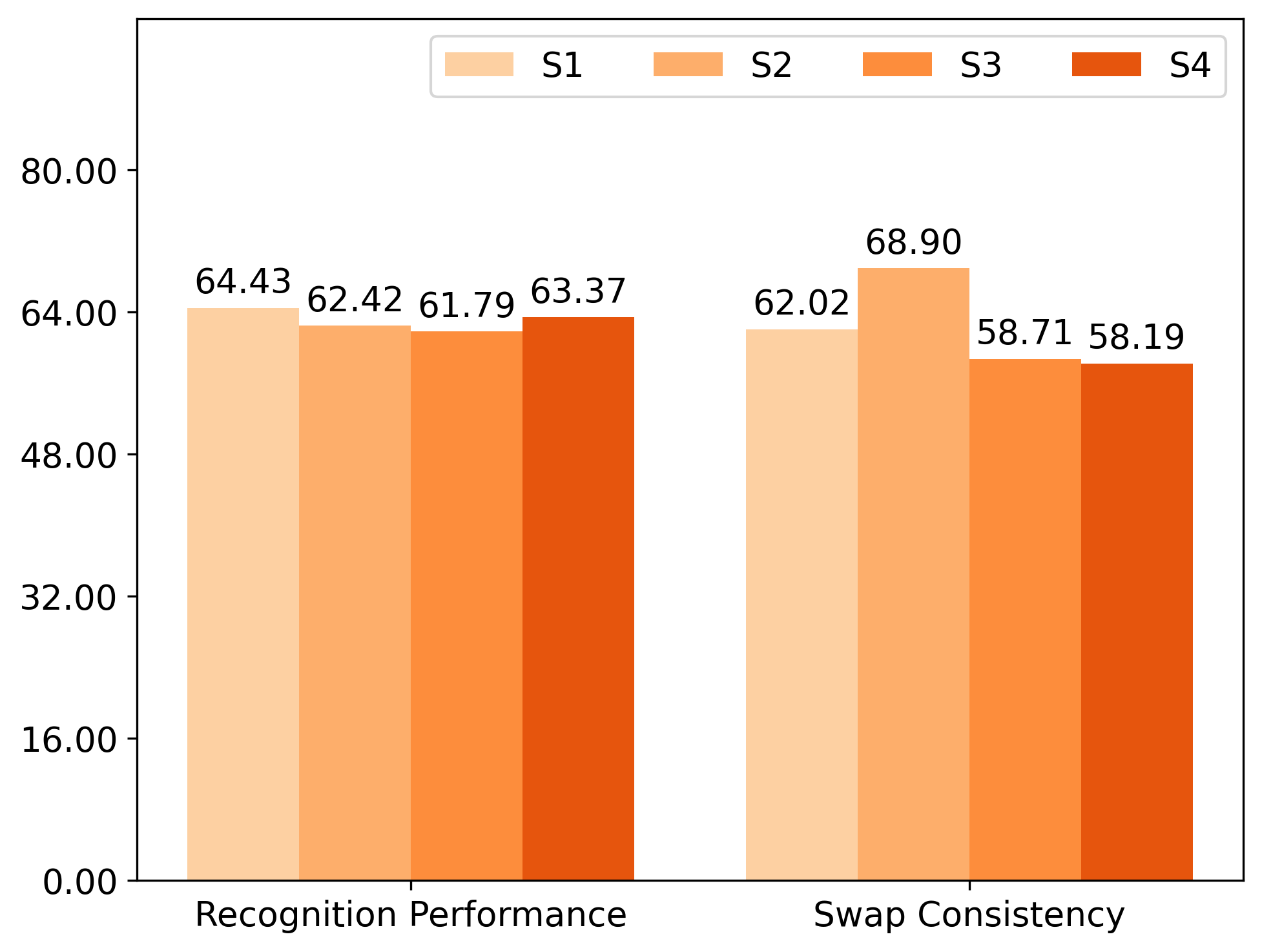}}
\end{subcaptionbox}
\begin{subcaptionbox}{\tiny{Qwen2.5-Omni}\label{fig:solution_result-12}}[0.156\linewidth]
    {\includegraphics[width=\linewidth]{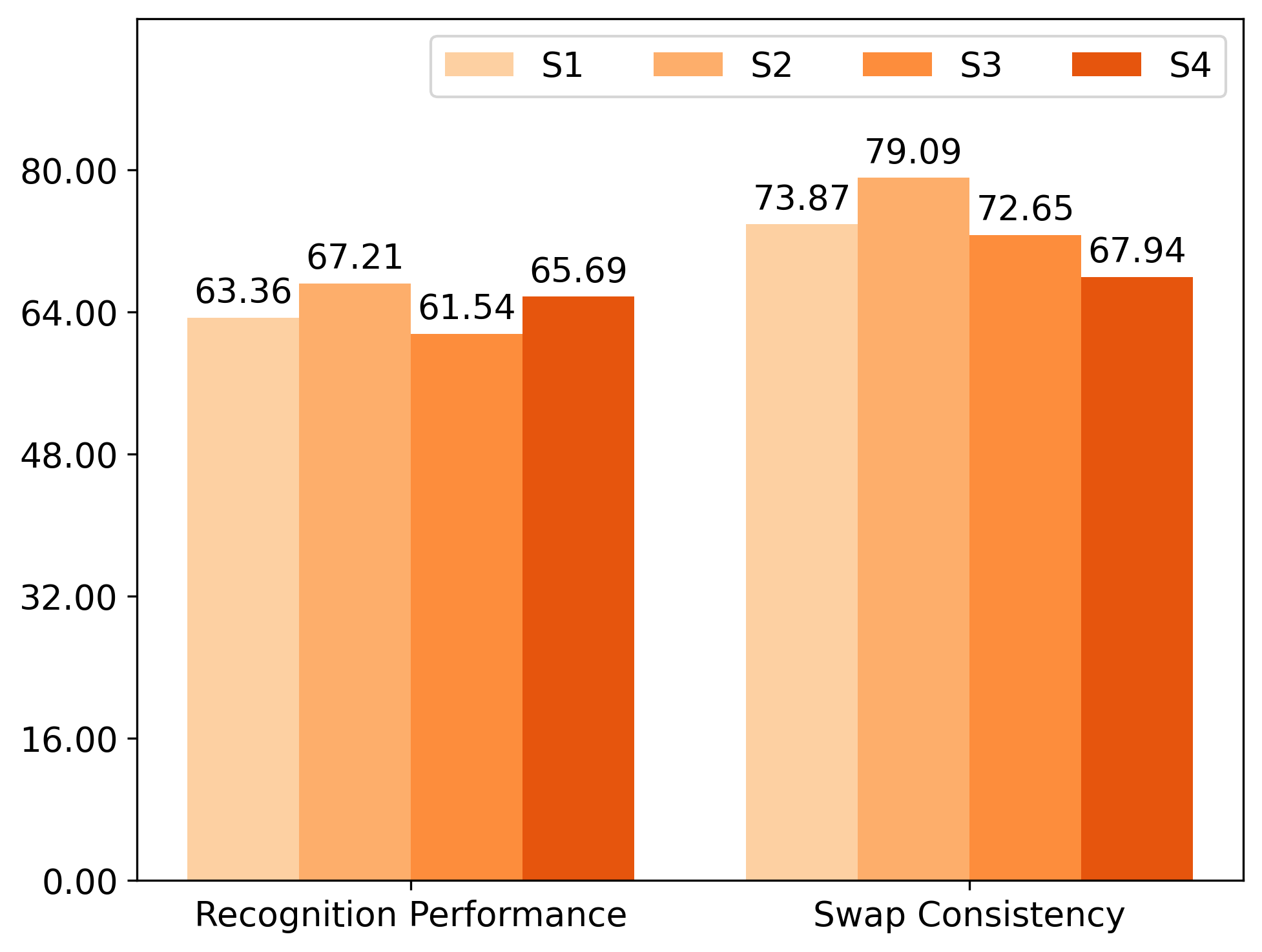}}
\end{subcaptionbox}
    \end{center}
    \endgroup
    \caption{\textbf{Impact of prompting strategies.} In these figures, we compare the performance of S1$\sim$S4 across various MLLMs. All experiments are conducted on the EmoPrefer-Data dataset.}
    \label{fig:solution_result}
\end{figure}

\paragraph{Impact of Prompting Strategy.}
Section \ref{sec:solutions} proposes four prompting strategies, and their impacts on different MLLMs are summarized in Figure \ref{fig:solution_result}. For most models, S3 and S4 outperform S1 and S2. The key distinction between these strategies lies in whether they leverage an external LLM. These findings highlight the advantages of integrating an external LLM, as MLLMs may have limitations in language understanding. However, some exceptions exist. For instance, Qwen2.5-VL \citep{bai2025qwen2} and Qwen2.5-Omni \citep{xu2025qwen2} perform better with S1 and S2 than with S3 and S4. This is because these models retain strong language processing capabilities due to their advanced training methodologies, enabling them to interpret complex prompts effectively. In such cases, the longer inference chains in S3 and S4 may introduce error accumulation and degrade performance. Thus, a trade-off emerges: \emph{while using an external LLM can compensate for the limited language processing abilities of some MLLMs, longer inference chains may also increase error accumulation.}
\vspace{-0.6cm}

\begin{figure}[h]
	\centering
	\includegraphics[width=\linewidth]{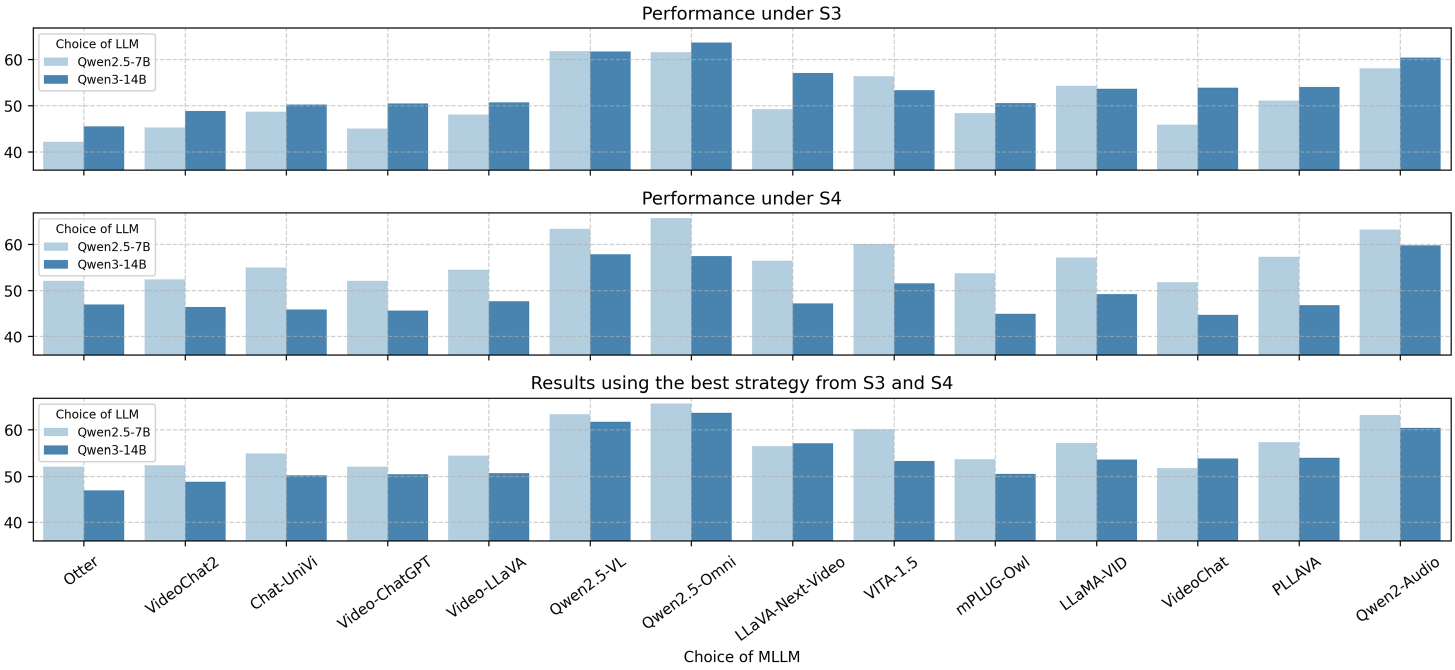}
	\caption{\textbf{Role of external LLMs.} From top to bottom, we show the performance of two LLMs under three configurations: \emph{the S3 strategy}, \emph{the S4 strategy}, and \emph{the best of the S3 and S4 strategies}.}
    \label{fig:llm_impact}
\end{figure}

\paragraph{Role of External LLMs.}
S3 and S4 rely on external LLMs. Figure~\ref{fig:llm_impact} examines the impact of LLMs on emotion preference prediction. We observe that Qwen3-14B \citep{yang2025qwen3} performs better under the S3 strategy, while Qwen2.5-7B performs better under the S4 strategy. These results reveal an interesting phenomenon: \emph{different LLMs exhibit varying preferences for different prompting strategies}. To further investigate, we conduct a statistical analysis of the two LLMs using their respective best strategies. Experimental results demonstrate that Qwen2.5-7B generally outperforms Qwen3-14B. These findings suggest that a larger LLM does not necessarily guarantee better alignment with human preferences. In this paper, we adopt Qwen2.5-7B as the default external LLM.
\vspace{-0.6cm}

\paragraph{Performance on Different MLLMs.}
Table \ref{tab:main} presents the performance of MLLMs under their best strategies. Full results for four strategies are provided in Appendix \ref{appendix:main_raw}. Experimental results demonstrate that open-source models do not show a noticeable performance gap compared to closed-source models; some open-source models even surpass closed-source ones. Among all models, Qwen2.5-Omni achieves promising results, attaining 67.21\% on the two-class WAF and 79.09\% on swap consistency, highlighting its strong capability in multimodal information processing and human emotion understanding. Additionally, most models outperform random guessing (i.e., 50\%), yet even the best-performing one falls short of the upper bound. This indicates the challenge of emotion preference decoding and suggests that further exploration is needed in the future.
\begin{table*}[h]
	\centering
	\renewcommand\arraystretch{0.9}
	\caption{\textbf{Main results.} The second column indicates the optimal strategy for each MLLM.}
	\label{tab:main}
        \scalebox{0.8}{
		\begin{tabular}{l|c|cc|cc|c}
            \toprule
            \multirow{2}{*}{\textbf{Model}} & \multirow{2}{*}{\textbf{S.}} & \multicolumn{2}{c|}{\textbf{Rec. (2-Class)}} & \multicolumn{2}{c|}{\textbf{Rec. (3-Class)}} & \multirow{2}{*}{\begin{tabular}[c]{@{}c@{}}\textbf{Swap} \\ \textbf{Cons. ($\uparrow$)}\end{tabular}} \\
            & & \textbf{WAF($\uparrow$)} & \textbf{ACC($\uparrow$)} & \textbf{WAF($\uparrow$)} & \textbf{ACC($\uparrow$)} \\
            \midrule
            \rowcolor{lightgray}
            \multicolumn{7}{c}{\emph{Open-source MLLMs}} \\
            \midrule

VideoChat \citep{li2023videochat}& S4& 51.79$\pm$1.21& 52.31$\pm$0.98& 44.88$\pm$1.54& 40.77$\pm$1.57& 40.85$\pm$1.83\\
Video-ChatGPT \citep{maaz2024video}& S4& 52.10$\pm$0.94& 52.13$\pm$0.98& 43.97$\pm$0.44& 40.42$\pm$0.35& 45.91$\pm$1.13\\
Otter \citep{li2023otter}& S4& 52.12$\pm$1.10& 52.49$\pm$0.98& 43.41$\pm$1.67& 38.15$\pm$1.74& 46.08$\pm$0.96\\
VideoChat2 \citep{li2024mvbench}& S4& 52.38$\pm$0.10& 52.49$\pm$0.09& 45.88$\pm$0.96& 43.82$\pm$0.44& 41.46$\pm$1.74\\
mPLUG-Owl \citep{ye2023mplugowl}& S4& 53.76$\pm$0.17& 53.82$\pm$0.18& 46.90$\pm$0.24& 44.34$\pm$0.44& 42.33$\pm$1.39\\
Video-LLaVA \citep{lin2024video}& S4& 54.53$\pm$0.81& 54.62$\pm$0.80& 43.61$\pm$0.82& 38.94$\pm$0.44& 42.86$\pm$3.31\\
Chat-UniVi \citep{jin2024chat}& S4& 54.97$\pm$1.64& 55.06$\pm$1.60& 47.03$\pm$2.32& 44.69$\pm$2.00& 43.90$\pm$0.52\\
LLaVA-Next-Video \citep{li2024llava}& S4& 56.41$\pm$0.98& 56.57$\pm$0.98& 52.84$\pm$1.25& 50.78$\pm$1.31& 53.92$\pm$2.00\\
LLaMA-VID \citep{li2024llama}& S4& 57.10$\pm$0.63& 57.10$\pm$0.62& 50.42$\pm$1.17& 47.13$\pm$1.66& 45.12$\pm$0.70\\
PLLAVA \citep{xu2024pllava}& S4& 57.29$\pm$0.18& 57.55$\pm$0.18& 54.02$\pm$0.05& 52.61$\pm$0.00& 53.83$\pm$1.22\\
VITA-1.5 \citep{fu2025vita}& S4& 60.08$\pm$0.61& 60.12$\pm$0.62& 57.08$\pm$0.16& 56.01$\pm$0.09& 59.06$\pm$1.22\\
Qwen2-Audio \citep{chu2024qwen2}& S4& 63.17$\pm$0.19& 63.23$\pm$0.18& 60.15$\pm$0.76& 59.32$\pm$0.78& 61.50$\pm$0.17\\
Qwen2.5-VL \citep{bai2025qwen2}& S1& 64.43$\pm$0.87& 65.28$\pm$0.80& 62.60$\pm$0.84& 64.02$\pm$0.78& 62.02$\pm$0.35\\
Qwen2.5-Omni \citep{xu2025qwen2}& S2& \textbf{67.21}$\pm$0.00& \textbf{67.32}$\pm$0.00& \textbf{65.29}$\pm$0.00& \textbf{66.03}$\pm$0.00& 79.09$\pm$0.00\\

\midrule
\rowcolor{lightgray}
\multicolumn{7}{c}{\emph{Closed-source MLLMs}} \\
\midrule 

GPT-4o \citep{openai24gpt4o}& S1& 59.28$\pm$0.08& 59.41$\pm$0.09& 56.57$\pm$0.04& 53.75$\pm$0.09& 64.55$\pm$0.96\\
Gemini-2.0-Flash \citep{gemini20flash}& S1& 59.80$\pm$0.88& 60.39$\pm$0.89& 58.14$\pm$0.65& 55.66$\pm$0.61& 57.14$\pm$2.26\\
Gemini-2.5-Flash \citep{gemini25flash}& S1& 60.60$\pm$0.64& 61.19$\pm$0.62& 58.77$\pm$0.55& 57.93$\pm$0.78& 64.98$\pm$1.39\\
GPT-4.1 \citep{achiam2023gpt}& S1& 60.75$\pm$0.18& 60.75$\pm$0.18& 59.23$\pm$0.35& 59.67$\pm$0.26& \textbf{80.84}$\pm$0.52\\
Gemini-1.5-Pro \citep{team2024gemini}& S1& 60.79$\pm$0.68& 61.55$\pm$0.62& 59.06$\pm$0.66& 60.37$\pm$0.61& 62.37$\pm$0.35\\
Gemini-1.5-Flash \citep{team2024gemini}& S1& 64.64$\pm$0.12& 65.19$\pm$0.18& 62.55$\pm$0.03& 63.50$\pm$0.09& 72.04$\pm$0.44\\

\bottomrule
		\end{tabular}
        }
\end{table*}

\begin{wrapfigure}{r}{0.5\linewidth}
  \centering
  \captionsetup[subfigure]{font=scriptsize} 
  \begin{subcaptionbox}{\scriptsize{all strategies}\label{fig:metric_correlation-1}}[0.43\linewidth]
        {\includegraphics[width=\linewidth]{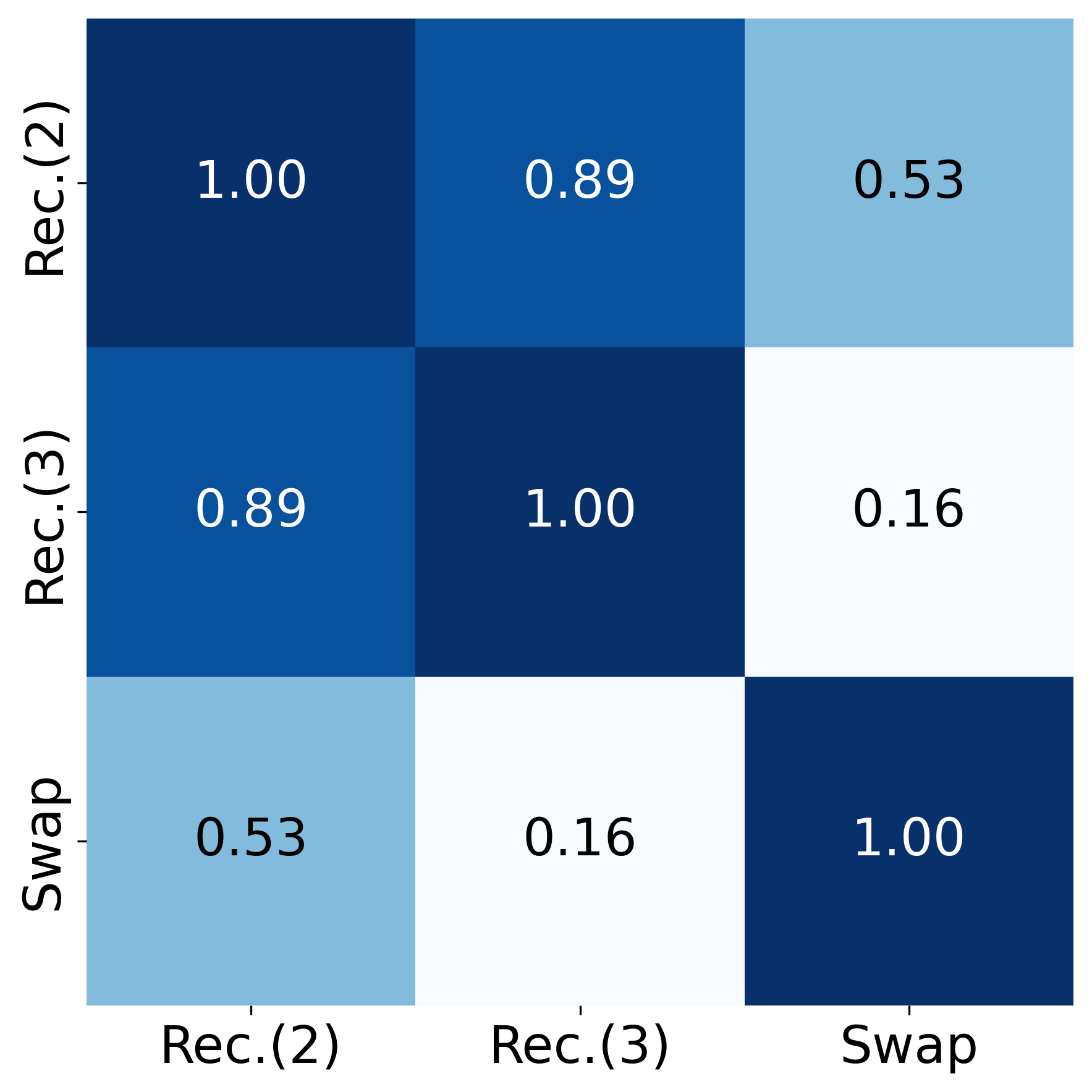}}
    \end{subcaptionbox}
    \begin{subcaptionbox}{\scriptsize{best strategy}\label{fig:metric_correlation-2}}[0.43\linewidth]
        {\includegraphics[width=\linewidth]{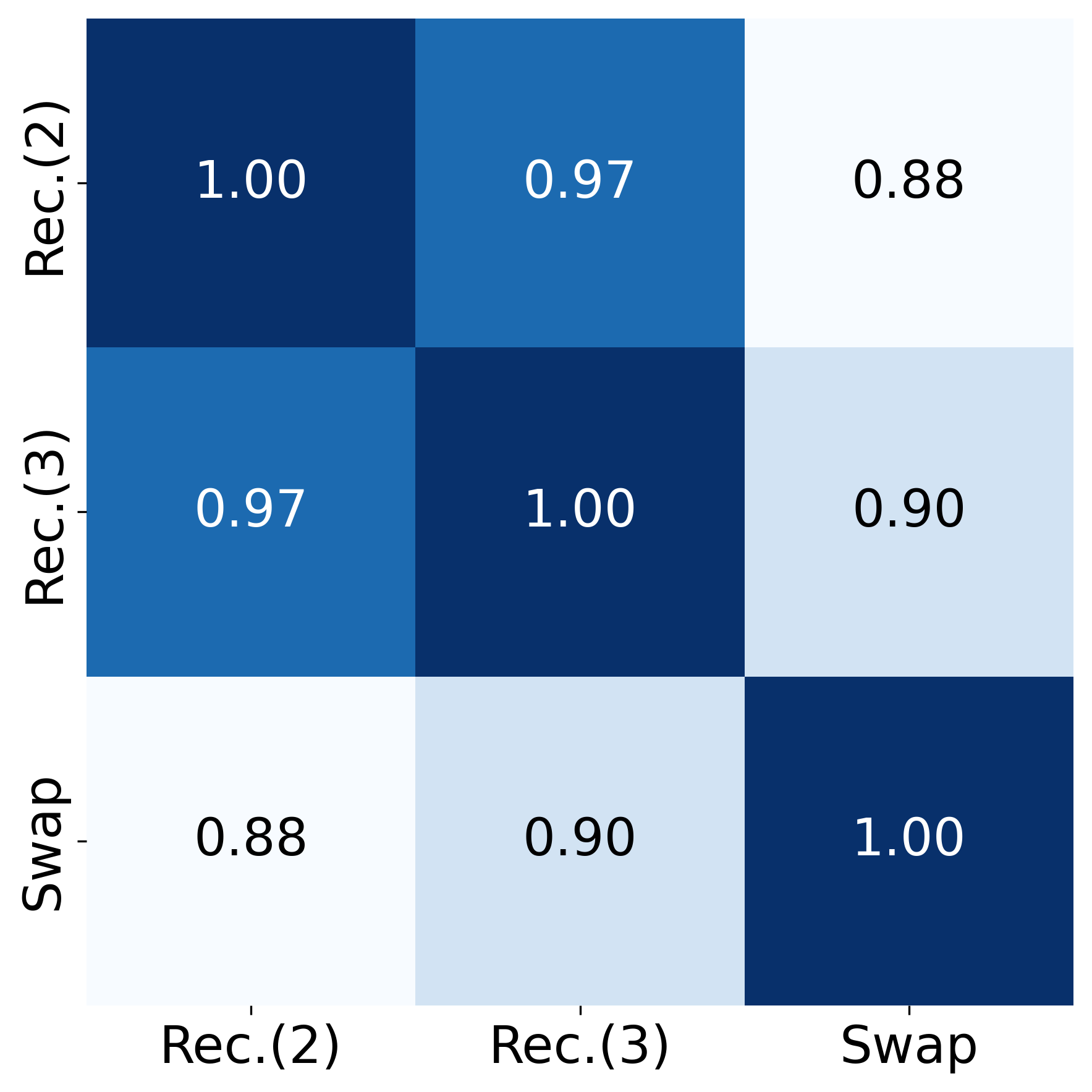}}
    \end{subcaptionbox}
  \caption{Metric correlation analysis.}
  \label{fig:metric_correlation}
\end{wrapfigure}
\paragraph{Metric Correlation.} Figure \ref{fig:metric_correlation-1} calculates the Pearson correlation coefficients (PCC) using scores from all strategies (see Appendix \ref{appendix:main_raw}), whereas Figure \ref{fig:metric_correlation-2} uses scores from the optimal strategies (see Table \ref{tab:main}). In Figure \ref{fig:metric_correlation-1}, we observe a strong correlation between two-class and three-class \emph{recognition performance}, whereas these metrics show only a relatively weak correlation with \emph{swap consistency}. These results suggest that the two metrics capture distinct aspects of model capability: one reflects preference prediction accuracy, while the other reflects model robustness. A good model should perform well on both metrics. In Figure \ref{fig:metric_correlation-2}, the correlation between these metrics is higher than in Figure \ref{fig:metric_correlation-1}. This indicates that models with poor \emph{recognition performance} may exhibit more varied \emph{swap consistency}. But for well-performing models, these metrics remain related.

\begin{figure}[h]
	\centering
	\includegraphics[width=\linewidth]{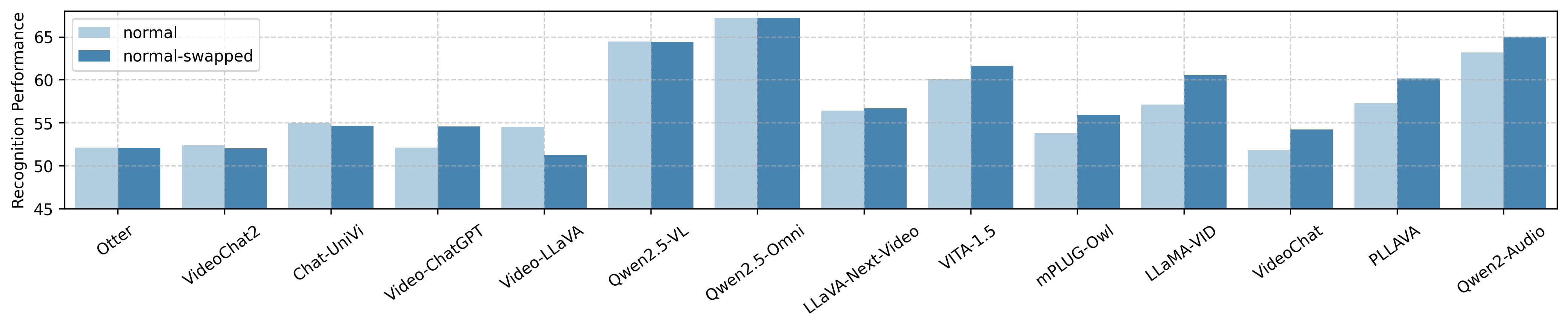}
	\caption{Normal-only vs. Normal-swapped combinations.}
	\label{fig:normal_swapped}
\end{figure}

\paragraph{Effect of Normal-Swapped Combination.}
Table \ref{tab:main} reveals that swapping input orders may lead to distinct preference prediction results. This observation naturally raises the hypothesis: \emph{Can we enhance the reliability of the results by aggregating the outputs of normal and swapped orders?} To verify this, we execute each model twice in normal order and twice in swapped order, generating four outcomes. We then apply majority voting to determine the final prediction. Figure \ref{fig:normal_swapped} demonstrates the impact of this strategy across different MLLMs. We observe that most MLLMs benefit from this approach. However, some exceptions exist, such as Chat-UniVi \citep{jin2024chat}, indicating that the effectiveness of this strategy is model-dependent.

\begin{figure}[h]
    \captionsetup[subfigure]{font=tiny} 
    \begin{center}

        \begin{subcaptionbox}{\tiny{All (w/o NS)}\label{fig:crowdsourcing_recognition-1}}[0.468\linewidth]
            {\includegraphics[width=\linewidth]{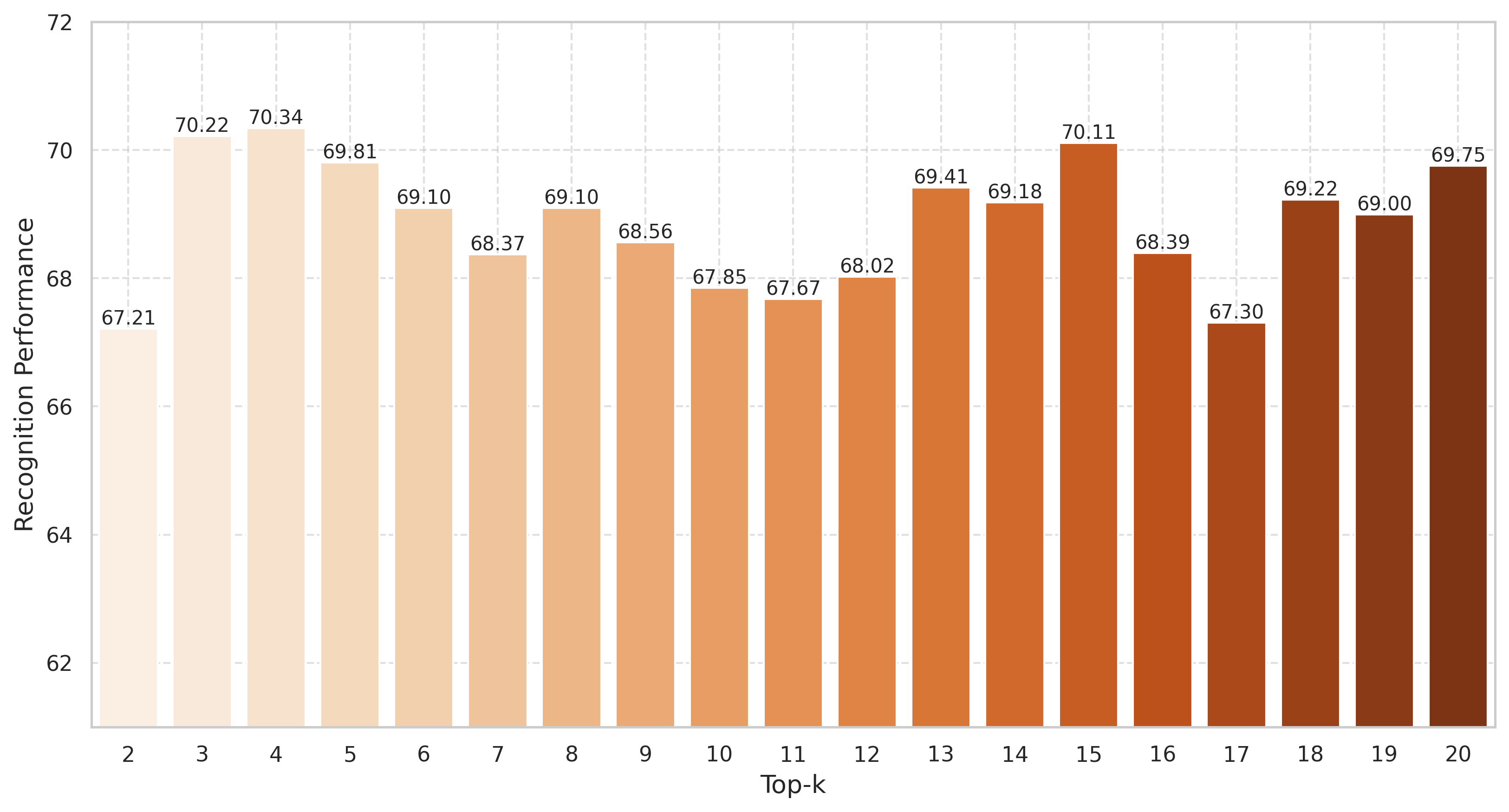}}
        \end{subcaptionbox}
        \begin{subcaptionbox}{\tiny{Open-source (w/o NS)}\label{fig:crowdsourcing_recognition-2}}[0.338\linewidth]
            {\includegraphics[width=\linewidth]{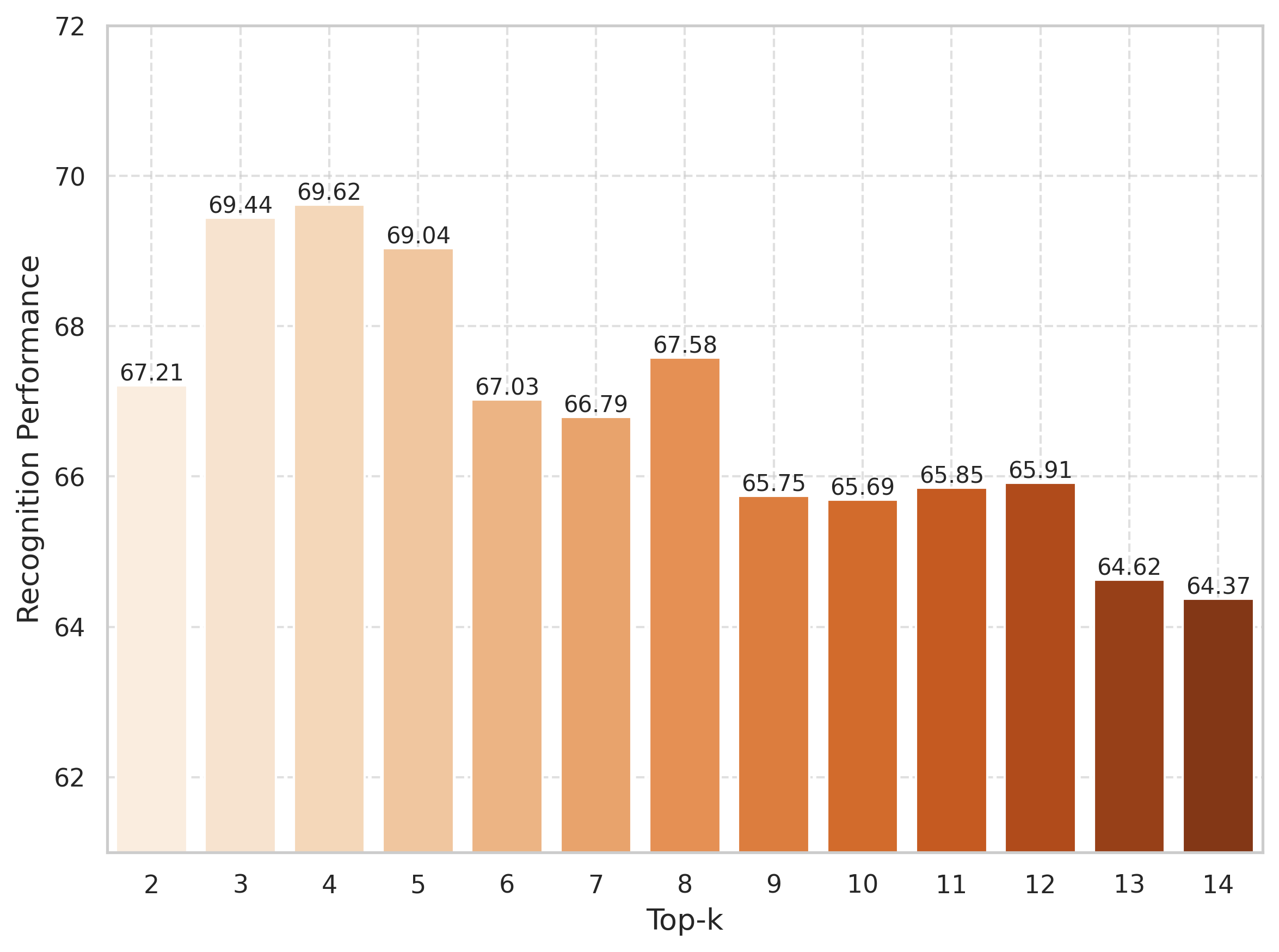}}
        \end{subcaptionbox}
        \begin{subcaptionbox}{\tiny{Closed-source (w/o NS)}\label{fig:crowdsourcing_recognition-3}}[0.165\linewidth]
            {\includegraphics[width=\linewidth]{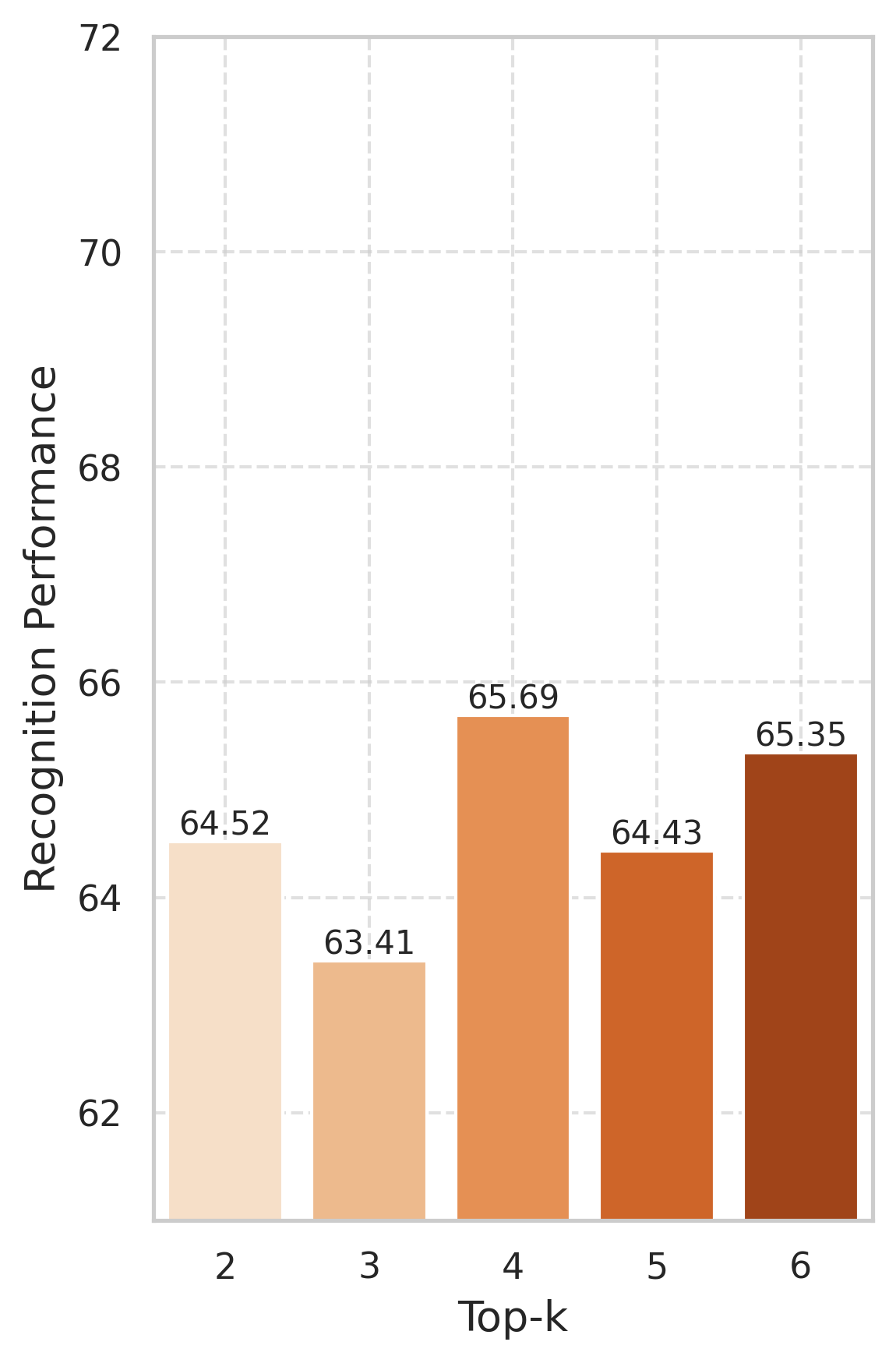}}
        \end{subcaptionbox}

        \begin{subcaptionbox}{\tiny{All (w/ NS)}\label{fig:crowdsourcing_recognition-4}}[0.468\linewidth]
            {\includegraphics[width=\linewidth]{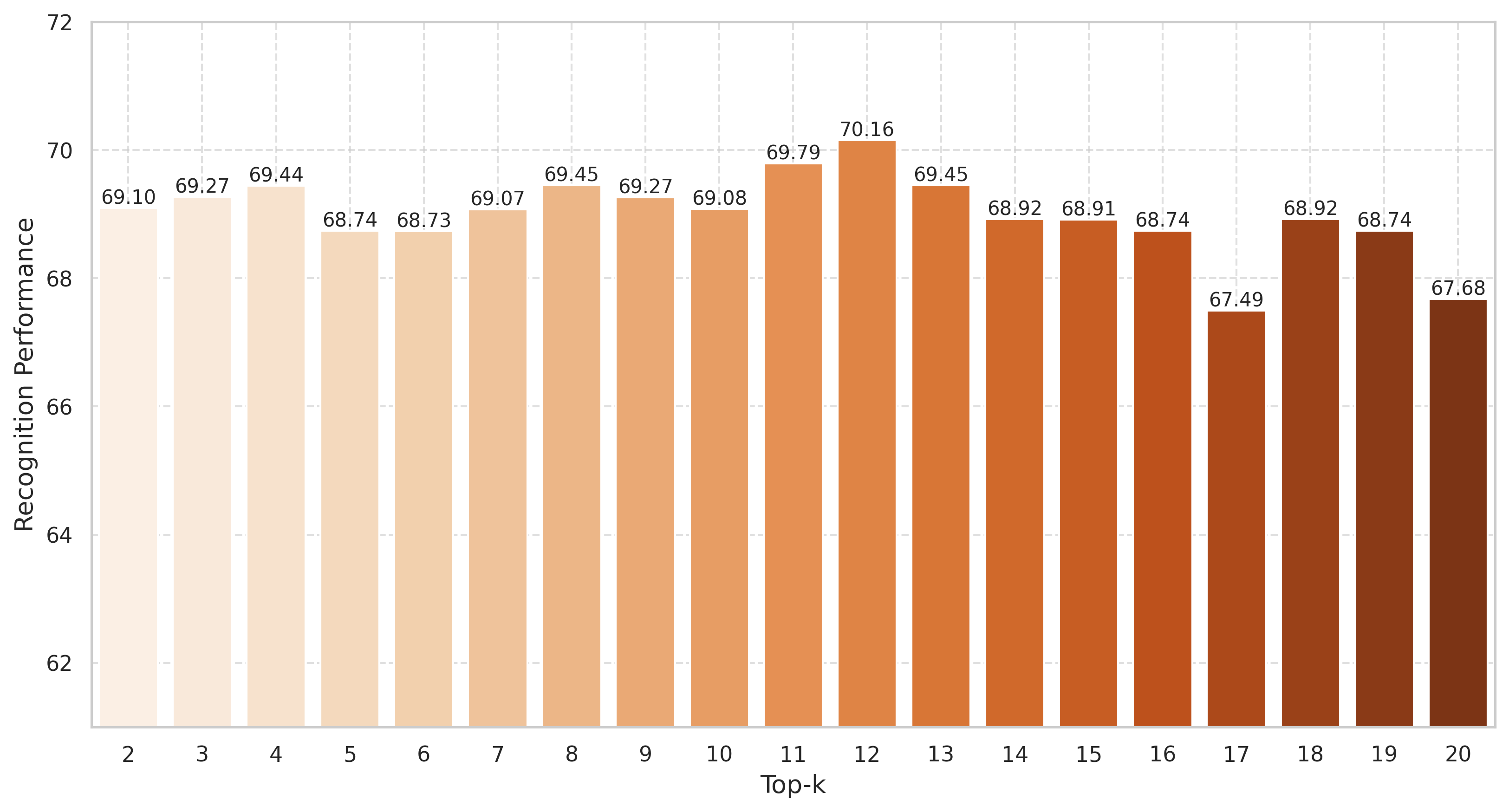}}
        \end{subcaptionbox}
        \begin{subcaptionbox}{\tiny{Open-source (w/ NS)}\label{fig:crowdsourcing_recognition-5}}[0.338\linewidth]
            {\includegraphics[width=\linewidth]{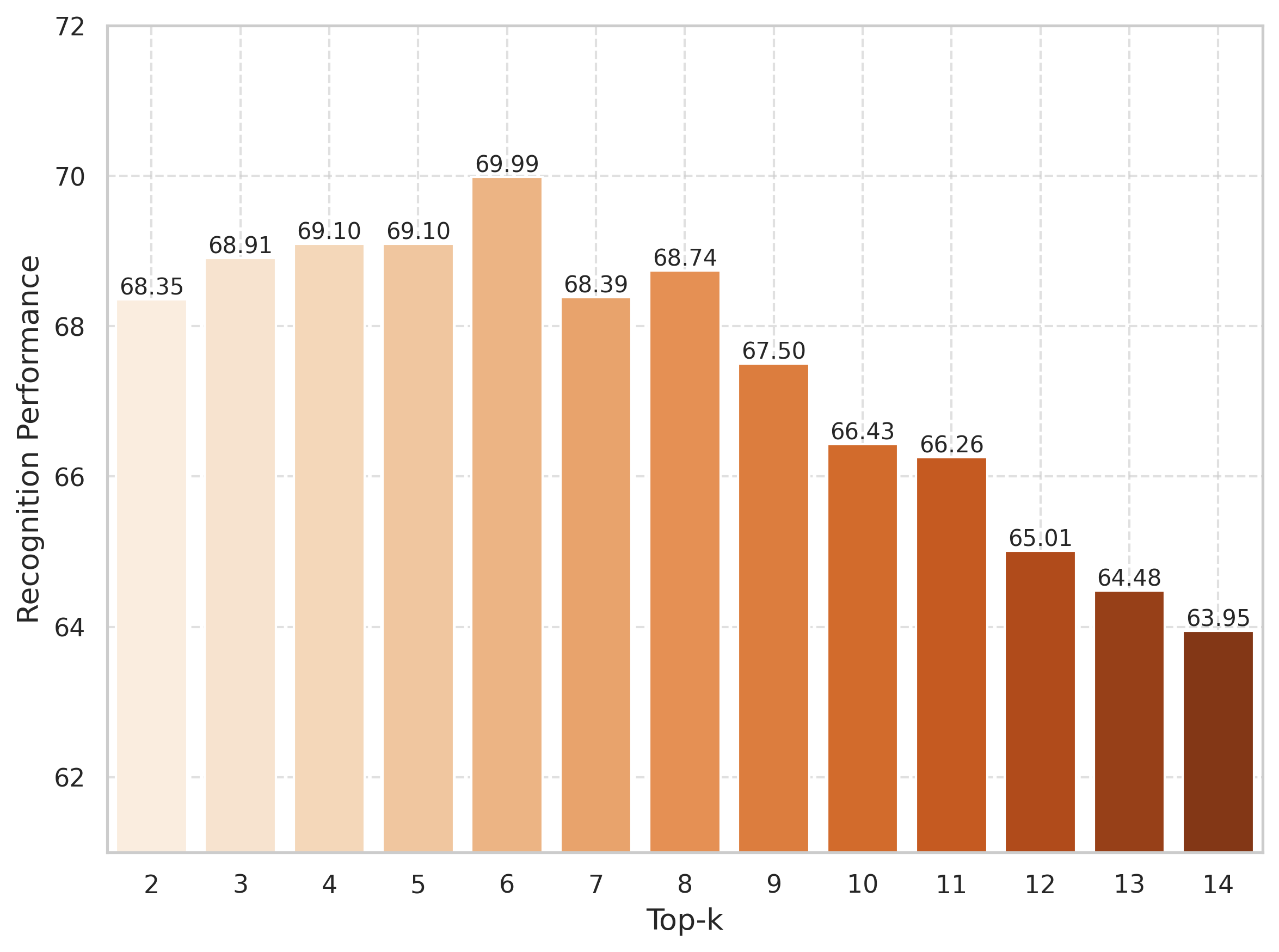}}
        \end{subcaptionbox}
        \begin{subcaptionbox}{\tiny{Closed-source (w/ NS)}\label{fig:crowdsourcing_recognition-6}}[0.165\linewidth]
            {\includegraphics[width=\linewidth]{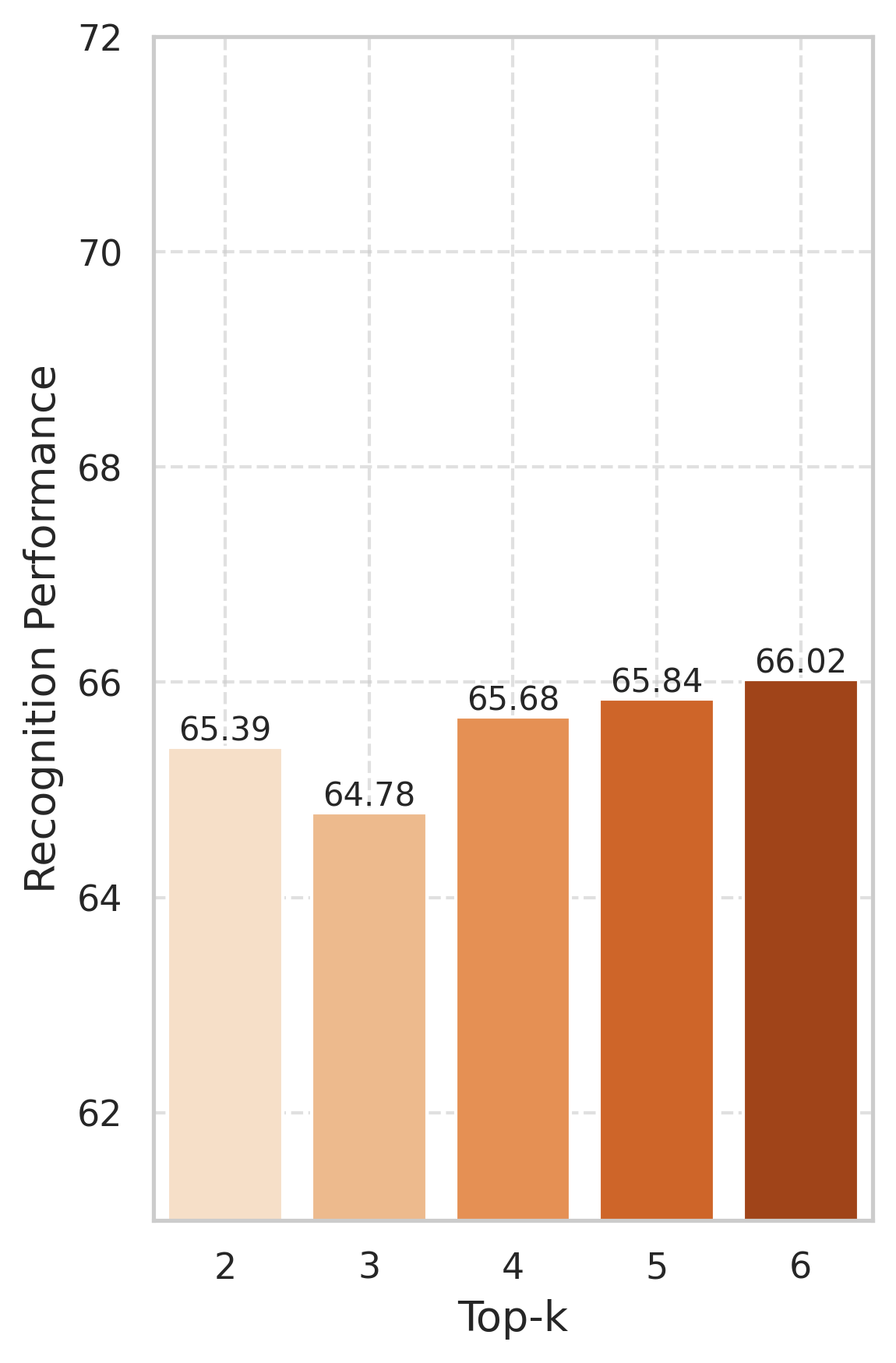}}
        \end{subcaptionbox}

    \end{center}
    \caption{\textbf{Impact of model-based crowdsourcing on \emph{recognition performance}.} In these figures, ``w/o NS'' and ``w/ NS'' denote whether normal-swapped combinations are used, while ``all'', ``open-source'', and ``closed-source'' indicate whether the model scope is restricted.}
    \label{fig:crowdsourcing_recognition}
\end{figure}

\paragraph{Model-Based Crowdsourcing.}
This section investigates the impact of model-based crowdsourcing. Specifically, we rank models based on their recognition performance and aggregate the predictions of the top-$k$ models. As shown in Figure \ref{fig:crowdsourcing_recognition}, crowdsourcing generally improves performance, but the extent of the gain depends on $k$: a large $k$ introduces noise, whereas a small $k$ limits effectiveness. We then conduct an analysis from two perspectives: \emph{(1) Necessity of normal-swapped combinations.} We examine the effectiveness of normal-swapped combinations in crowdsourcing. Unlike the findings for single models (see Figure \ref{fig:normal_swapped}), normal-swapped combinations yield only marginal improvements and may even degrade performance in some cases (e.g., Figures~\ref{fig:crowdsourcing_recognition-1} and \ref{fig:crowdsourcing_recognition-4}). Therefore, we do not use normal-swapped combinations by default in crowdsourcing. \emph{(2) Restricting the model scope.} Figures~\ref{fig:crowdsourcing_recognition-1}$\sim$\ref{fig:crowdsourcing_recognition-3} show performance when restricting the model scope, with the scope ranging from \emph{all}, \emph{open-source}, to \emph{closed-source} models, respectively. Compared to no restrictions, restricting the model scope to closed-source models leads to a noticeable performance degradation, whereas limiting it to open-source models still yields promising results. This suggests that restricting evaluations to open-source models could reduce costs without significant performance loss.

\begin{figure}[h]
    \captionsetup[subfigure]{font=tiny} 
    \begin{center}
        \begin{subcaptionbox}{\tiny{All (w/o NS)}\label{fig:crowdsourcing_swap-1}}[0.468\linewidth]
            {\includegraphics[width=\linewidth]{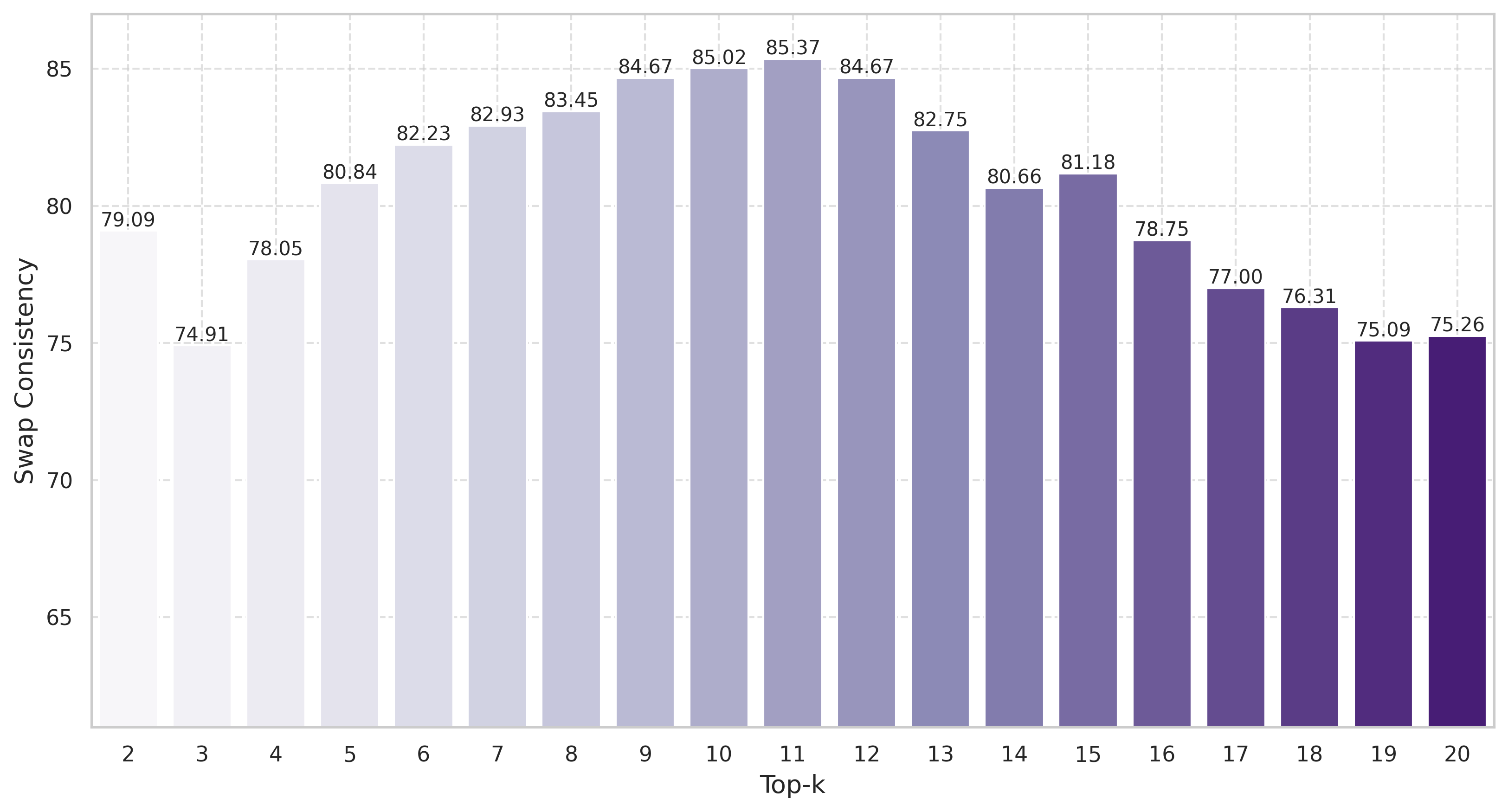}}
        \end{subcaptionbox}
        \begin{subcaptionbox}{\tiny{Open-source (w/o NS)}\label{fig:crowdsourcing_swap-2}}[0.338\linewidth]
            {\includegraphics[width=\linewidth]{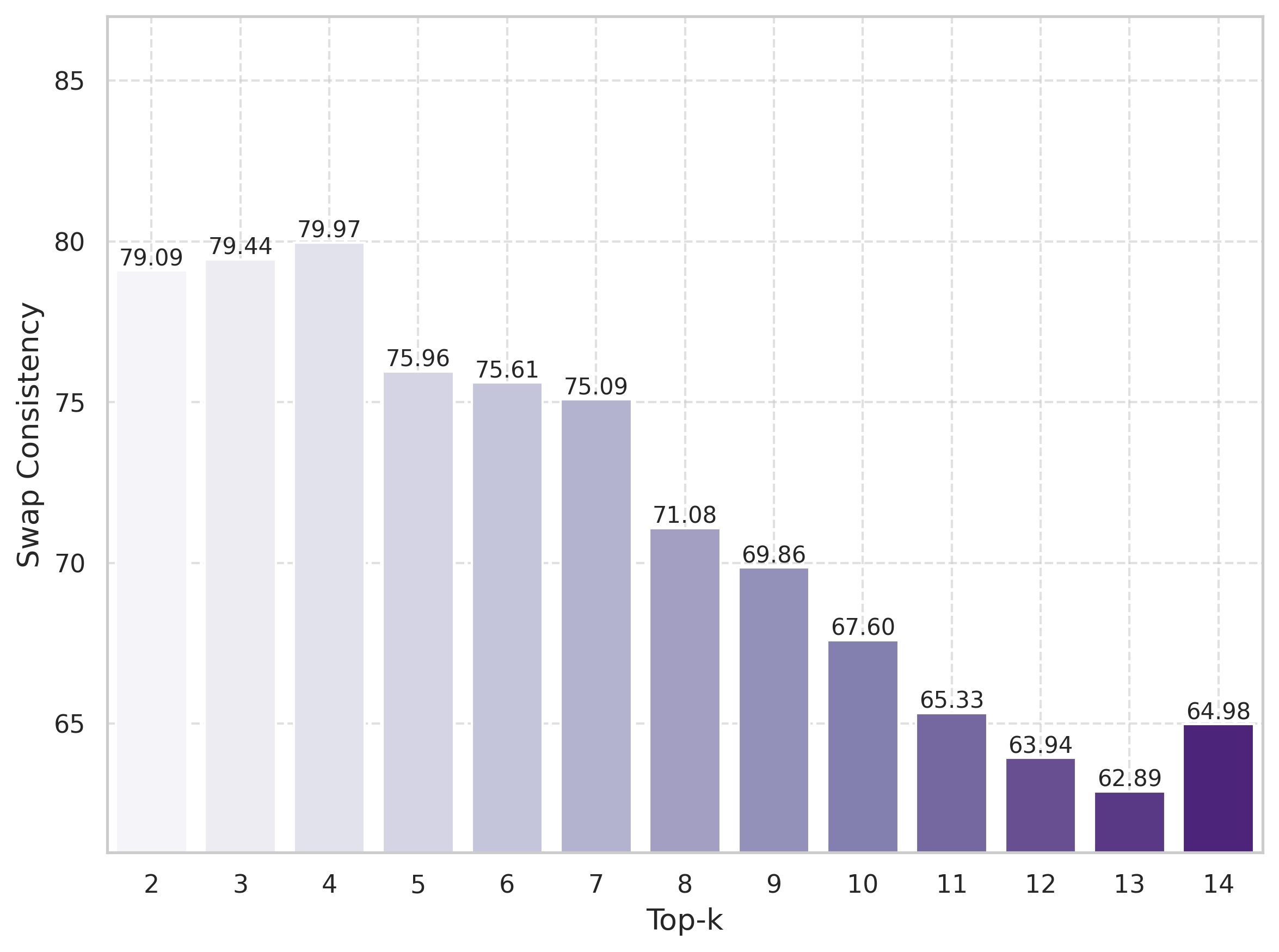}}
        \end{subcaptionbox}
        \begin{subcaptionbox}{\tiny{Closed-source (w/o NS)}\label{fig:crowdsourcing_swap-3}}[0.165\linewidth]
            {\includegraphics[width=\linewidth]{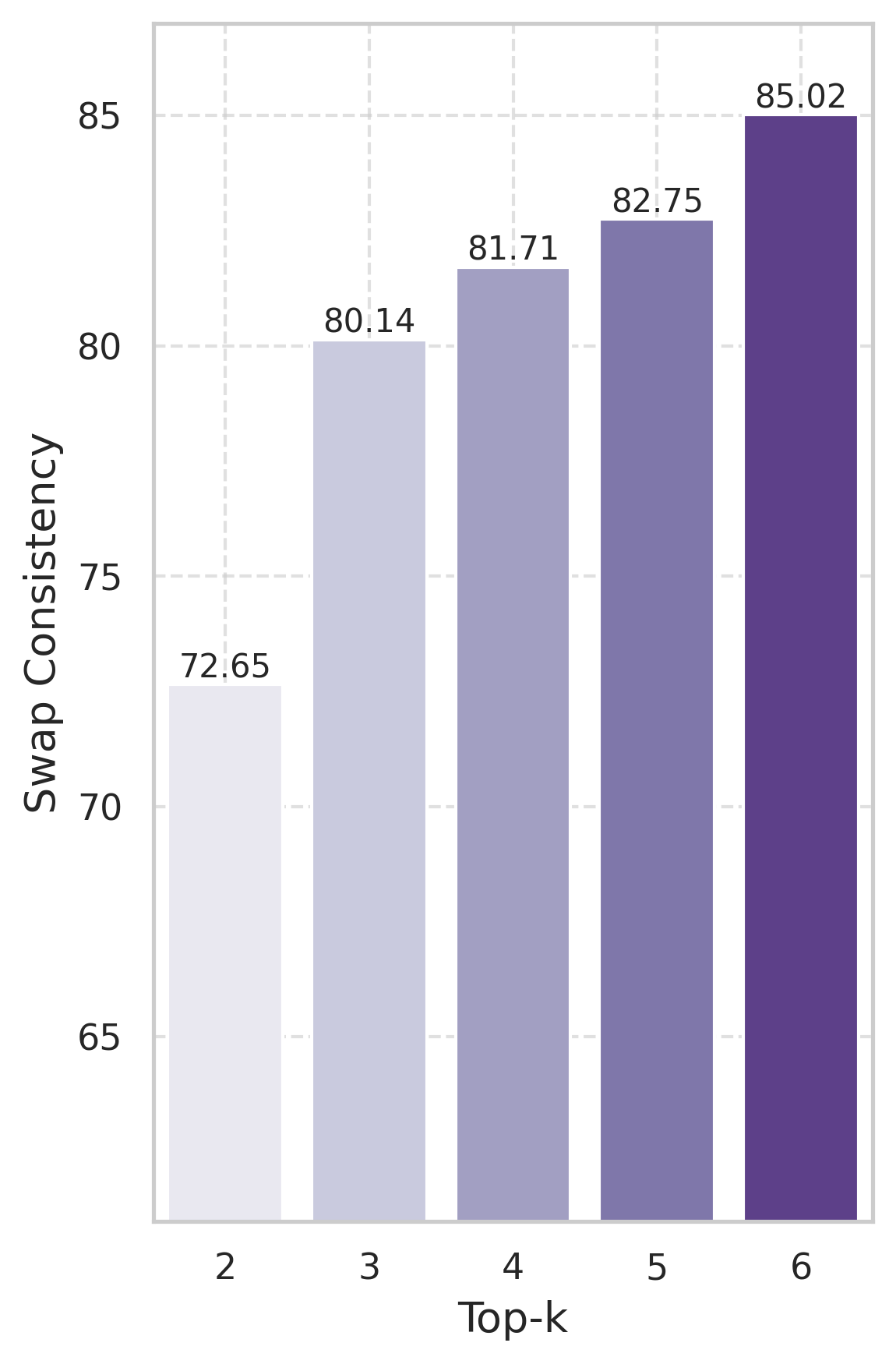}}
        \end{subcaptionbox}
    \end{center}
    \caption{Impact of model-based crowdsourcing on \emph{swap consistency}.}
    \label{fig:crowdsourcing_swap}
\end{figure}

Besides its impact on \emph{recognition performance}, Figure \ref{fig:crowdsourcing_swap} further reveals its effect on \emph{swap consistency}. We observe that selecting an appropriate $k$ also improves \emph{swap consistency}, indicating greater robustness to input order changes with crowdsourcing. Interestingly, the optimal $k$ differs for \emph{swap consistency} and \emph{recognition performance} (e.g., Figures \ref{fig:crowdsourcing_recognition-1} and \ref{fig:crowdsourcing_swap-1}). Since a good model should perform well on both metrics, we should consider both when selecting $k$. For example, $k=3 \;\text{or}\; 4$ is a good choice when restricting the model scope to open-source models.

\section{Practical Application}
\begin{figure}[h]
	\centering
	\includegraphics[width=\linewidth]{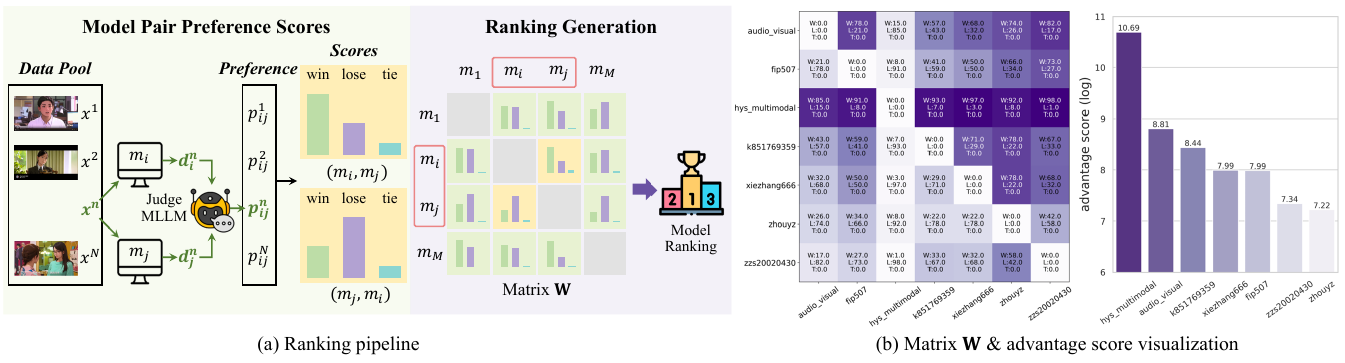}
	\caption{\textbf{Practical application.} (a) Ranking pipeline. We first employ pairwise models $m_i$ and $m_j$ to generate descriptions for $x^n$, obtaining $d_i^n$ and $d_j^n$. Next,  we use MLLM-based judges to derive the preference prediction $p_{ij}^n$. By repeating this process, we collect the counts of wins, losses, and ties for each model pair comparison. These results are then used to compute the model rankings. (b) We visualize the matrix $\mathbf{W}$ and advantage scores derived from a concrete example.}
	\label{fig:implementation}
\end{figure}

Figure \ref{fig:implementation}(a) integrates preference prediction with the Bradley-Terry algorithm \citep{hunter2004mm} for model ranking. Specifically, suppose there are $M$ models and $N$ samples, where each sample is denoted as $x^n$ and each model as $m_i$. Taking the comparison between models $m_i$ and $m_j$ as an example, we first obtain their emotion descriptions for $x^n$, yielding $d^n_i$ and $d^n_j$. We then use judge MLLMs to predict preference. Inspired by the results of EmoPrefer-Bench, we use model-based crowdsourcing, restrict the model scope to open-source models, and set top-$k$=3. The predicted preference is denoted as $p^n_{ij}$. By repeating this process across multiple samples, we obtain the numbers of wins, losses, and ties between $m_i$ and $m_j$, denoted as $\text{win}_{ij}$, $\text{lose}_{ij}$, and $\text{tie}_{ij}$, respectively. These scores are stored in the matrix $\mathbf{W}$, and then the Bradley-Terry algorithm is used to estimate the advantages. Appendix \ref{appendix:bradley_terry} provides the detailed calculation process of the ranking algorithm. 

Figure \ref{fig:implementation} provides a concrete example. Specifically, MER2025 \citep{lian2025mer} is a prominent challenge in affective computing, featuring a track focused on descriptive emotions. With permission from the challenge organizers, we use participants' submissions for model ranking. Figure \ref{fig:implementation}(b) visualizes the matrix W, and Figure \ref{fig:implementation}(c) presents the estimated advantage scores.

\section{Conclusion}
This paper proposes \emph{EmoPrefer}, a pioneering work that explores the capabilities of MLLMs in recognizing human emotion preferences. To this end, we introduce \emph{EmoPrefer-Data} and establish \emph{EmoPrefer-Bench}, the first dataset and benchmark focused on human emotion preference. Extensive experiments show that different MLLMs favor different prompting strategies, and model-based crowdsourcing can enhance both recognition performance and model robustness. Meanwhile, we observe that current MLLMs still struggle with accurate emotion preference prediction. This highlights both the limitations of current MLLMs and the inherent challenges in emotion understanding.

\textbf{Limitations and Future Work.} This paper reports the zero-shot performance of MLLMs, without exploring more effective architectures or training paradigms. Such exploration is left for our future work. Beyond evaluation, we will leverage our dataset and the insights drawn from our benchmark to train reward models and apply reinforcement learning to maximize these rewards. This approach will enhance MLLMs' understanding of human emotions and, in turn, improve their emotional intelligence. Additionally, this paper focuses on emotion preference in descriptive emotion. Beyond descriptive emotions, other forms of emotion representation exist, including categorical emotions (e.g., using discrete emotion words) and dimensional emotions (e.g., using float-point values). In the future, we will extend our preference task to other emotion representations.

\section*{Ethics Statement}
We do not collect new data but instead use the raw samples from MER2024 with the permission of the owners. To construct EmoPrefer-Data, we hire annotators. They are generously compensated for their work. During the annotation process, we provide a clear task description and utilize a well-designed annotation platform. Furthermore, we restrict the use of EmoPrefer-Datato to non-commercial purposes under the CC BY-NC 4.0 license, which explicitly outlines the appropriate and responsible use of our dataset. As a result, no ethical concerns are raised in this paper.

\section*{Reproducibility Statement}
To ensure reproducibility, we provide the dataset, source code, and intermediate outputs in the supplementary materials. After paper acceptance, we will open-source them on GitHub. For dataset annotation, we provide a detailed introduction in Appendix \ref{appendix:dataset_details}. In summary, we have made every effort to ensure the reproducibility of this paper.

\bibliography{iclr2026_conference}

\begin{thebibliography}{60}
\providecommand{\natexlab}[1]{#1}
\providecommand{\url}[1]{\texttt{#1}}
\expandafter\ifx\csname urlstyle\endcsname\relax
  \providecommand{\doi}[1]{doi: #1}\else
  \providecommand{\doi}{doi: \begingroup \urlstyle{rm}\Url}\fi

\bibitem[Achiam et~al.(2023)Achiam, Adler, Agarwal, Ahmad, Akkaya, Aleman,
  Almeida, Altenschmidt, Altman, Anadkat, et~al.]{achiam2023gpt}
Josh Achiam, Steven Adler, Sandhini Agarwal, Lama Ahmad, Ilge Akkaya,
  Florencia~Leoni Aleman, Diogo Almeida, Janko Altenschmidt, Sam Altman,
  Shyamal Anadkat, et~al.
\newblock Gpt-4 technical report.
\newblock \emph{arXiv preprint arXiv:2303.08774}, 2023.

\bibitem[Bai et~al.(2025)Bai, Chen, Liu, Wang, Ge, Song, Dang, Wang, Wang,
  Tang, et~al.]{bai2025qwen2}
Shuai Bai, Keqin Chen, Xuejing Liu, Jialin Wang, Wenbin Ge, Sibo Song, Kai
  Dang, Peng Wang, Shijie Wang, Jun Tang, et~al.
\newblock Qwen2. 5-vl technical report.
\newblock \emph{arXiv preprint arXiv:2502.13923}, 2025.

\bibitem[Brave \& Nass(2007)Brave and Nass]{brave2007emotion}
Scott Brave and Cliff Nass.
\newblock Emotion in human-computer interaction.
\newblock In \emph{Human-Computer Interaction Handbook}, pp.\  103--118. CRC
  Press, 2007.

\bibitem[Cacioppo \& Gardner(1999)Cacioppo and Gardner]{cacioppo1999emotion}
John~T Cacioppo and Wendi~L Gardner.
\newblock Emotion.
\newblock \emph{Annual Review of Psychology}, 50\penalty0 (1):\penalty0
  191--214, 1999.

\bibitem[Chen et~al.(2024)Chen, Chen, Zhang, Wang, Liu, Zhou, Zhang, Wan, Zhou,
  and Sun]{chen2024mllm}
Dongping Chen, Ruoxi Chen, Shilin Zhang, Yaochen Wang, Yinuo Liu, Huichi Zhou,
  Qihui Zhang, Yao Wan, Pan Zhou, and Lichao Sun.
\newblock Mllm-as-a-judge: assessing multimodal llm-as-a-judge with
  vision-language benchmark.
\newblock In \emph{Proceedings of the 41st International Conference on Machine
  Learning}, pp.\  6562--6595, 2024.

\bibitem[Chen et~al.(2023)Chen, Shi, Liu, Li, and Zhao]{chen2023smg}
Haoyu Chen, Henglin Shi, Xin Liu, Xiaobai Li, and Guoying Zhao.
\newblock Smg: A micro-gesture dataset towards spontaneous body gestures for
  emotional stress state analysis.
\newblock \emph{International Journal of Computer Vision}, 131\penalty0
  (6):\penalty0 1346--1366, 2023.

\bibitem[Cheng et~al.(2024)Cheng, Cheng, He, Wang, Lin, Lian, Peng, and
  Hauptmann]{cheng2024emotion}
Zebang Cheng, Zhi-Qi Cheng, Jun-Yan He, Kai Wang, Yuxiang Lin, Zheng Lian,
  Xiaojiang Peng, and Alexander Hauptmann.
\newblock Emotion-llama: Multimodal emotion recognition and reasoning with
  instruction tuning.
\newblock \emph{Proceedings of the Advances in Neural Information Processing
  Systems}, 37:\penalty0 110805--110853, 2024.

\bibitem[Christiano et~al.(2017)Christiano, Leike, Brown, Martic, Legg, and
  Amodei]{christiano2017deep}
Paul~F Christiano, Jan Leike, Tom~B Brown, Miljan Martic, Shane Legg, and Dario
  Amodei.
\newblock Deep reinforcement learning from human preferences.
\newblock In \emph{Proceedings of the 31st International Conference on Neural
  Information Processing Systems}, pp.\  4302--4310, 2017.

\bibitem[Chu et~al.(2024)Chu, Xu, Yang, Wei, Wei, Guo, Leng, Lv, He, Lin,
  et~al.]{chu2024qwen2}
Yunfei Chu, Jin Xu, Qian Yang, Haojie Wei, Xipin Wei, Zhifang Guo, Yichong
  Leng, Yuanjun Lv, Jinzheng He, Junyang Lin, et~al.
\newblock Qwen2-audio technical report.
\newblock \emph{arXiv preprint arXiv:2407.10759}, 2024.

\bibitem[Cobbe et~al.(2021)Cobbe, Kosaraju, Bavarian, Chen, Jun, Kaiser,
  Plappert, Tworek, Hilton, Nakano, et~al.]{cobbe2021training}
Karl Cobbe, Vineet Kosaraju, Mohammad Bavarian, Mark Chen, Heewoo Jun, Lukasz
  Kaiser, Matthias Plappert, Jerry Tworek, Jacob Hilton, Reiichiro Nakano,
  et~al.
\newblock Training verifiers to solve math word problems.
\newblock \emph{arXiv preprint arXiv:2110.14168}, 2021.

\bibitem[Dolan(2002)]{dolan2002emotion}
Raymond~J Dolan.
\newblock Emotion, cognition, and behavior.
\newblock \emph{science}, 298\penalty0 (5596):\penalty0 1191--1194, 2002.

\bibitem[Ekman \& Keltner(1970)Ekman and Keltner]{ekman1970universal}
Paul Ekman and Dacher Keltner.
\newblock Universal facial expressions of emotion.
\newblock \emph{California mental health research digest}, 8\penalty0
  (4):\penalty0 151--158, 1970.

\bibitem[El~Ayadi et~al.(2011)El~Ayadi, Kamel, and Karray]{el2011survey}
Moataz El~Ayadi, Mohamed~S Kamel, and Fakhri Karray.
\newblock Survey on speech emotion recognition: Features, classification
  schemes, and databases.
\newblock \emph{Pattern Recognition}, 44\penalty0 (3):\penalty0 572--587, 2011.

\bibitem[Fu et~al.(2025)Fu, Lin, Wang, Zhang, Shen, Liu, Cao, Long, Gao, Li,
  et~al.]{fu2025vita}
Chaoyou Fu, Haojia Lin, Xiong Wang, Yi-Fan Zhang, Yunhang Shen, Xiaoyu Liu,
  Haoyu Cao, Zuwei Long, Heting Gao, Ke~Li, et~al.
\newblock Vita-1.5: Towards gpt-4o level real-time vision and speech
  interaction.
\newblock \emph{arXiv preprint arXiv:2501.01957}, 2025.

\bibitem[Google(2025{\natexlab{a}})]{gemini20flash}
Google.
\newblock Gemini 2.0 flash model card, 2025{\natexlab{a}}.
\newblock URL
  \url{https://storage.googleapis.com/model-cards/documents/gemini-2-flash.pdf}.

\bibitem[Google(2025{\natexlab{b}})]{gemini25flash}
Google.
\newblock Gemini 2.5 flash preview model card, 2025{\natexlab{b}}.
\newblock URL
  \url{https://storage.googleapis.com/model-cards/documents/gemini-2.5-flash-preview.pdf}.

\bibitem[Gu et~al.(2024)Gu, Jiang, Shi, Tan, Zhai, Xu, Li, Shen, Ma, Liu,
  et~al.]{gu2024survey}
Jiawei Gu, Xuhui Jiang, Zhichao Shi, Hexiang Tan, Xuehao Zhai, Chengjin Xu, Wei
  Li, Yinghan Shen, Shengjie Ma, Honghao Liu, et~al.
\newblock A survey on llm-as-a-judge.
\newblock \emph{arXiv preprint arXiv:2411.15594}, 2024.

\bibitem[Gunes et~al.(2011)Gunes, Schuller, Pantic, and
  Cowie]{gunes2011emotion}
Hatice Gunes, Bj{\"o}rn Schuller, Maja Pantic, and Roddy Cowie.
\newblock Emotion representation, analysis and synthesis in continuous space: A
  survey.
\newblock In \emph{2011 IEEE international conference on automatic face \&
  gesture recognition (FG)}, pp.\  827--834. IEEE, 2011.

\bibitem[Hunter(2004)]{hunter2004mm}
David~R Hunter.
\newblock Mm algorithms for generalized bradley-terry models.
\newblock \emph{Annals of Statistics}, 32\penalty0 (1):\penalty0 384--406,
  2004.

\bibitem[Jin et~al.(2024)Jin, Takanobu, Zhang, Cao, and Yuan]{jin2024chat}
Peng Jin, Ryuichi Takanobu, Wancai Zhang, Xiaochun Cao, and Li~Yuan.
\newblock Chat-univi: Unified visual representation empowers large language
  models with image and video understanding.
\newblock In \emph{Proceedings of the IEEE/CVF Conference on Computer Vision
  and Pattern Recognition}, pp.\  13700--13710, 2024.

\bibitem[Lerner et~al.(2015)Lerner, Li, Valdesolo, and
  Kassam]{lerner2015emotion}
Jennifer~S Lerner, Ye~Li, Piercarlo Valdesolo, and Karim~S Kassam.
\newblock Emotion and decision making.
\newblock \emph{Annual review of psychology}, 66\penalty0 (1):\penalty0
  799--823, 2015.

\bibitem[Li et~al.(2023{\natexlab{a}})Li, Zhang, Chen, Wang, Yang, and
  Liu]{li2023otter}
Bo~Li, Yuanhan Zhang, Liangyu Chen, Jinghao Wang, Jingkang Yang, and Ziwei Liu.
\newblock Otter: A multi-modal model with in-context instruction tuning.
\newblock \emph{arXiv preprint arXiv:2305.03726}, 2023{\natexlab{a}}.

\bibitem[Li et~al.(2024{\natexlab{a}})Li, Zhang, Zhang, Zhang, Li, Li, Ma, and
  Li]{li2024llava}
Feng Li, Renrui Zhang, Hao Zhang, Yuanhan Zhang, Bo~Li, Wei Li, Zejun Ma, and
  Chunyuan Li.
\newblock Llava-next-interleave: Tackling multi-image, video, and 3d in large
  multimodal models.
\newblock \emph{arXiv preprint arXiv:2407.07895}, 2024{\natexlab{a}}.

\bibitem[Li et~al.(2023{\natexlab{b}})Li, He, Wang, Li, Wang, Luo, Wang, Wang,
  and Qiao]{li2023videochat}
KunChang Li, Yinan He, Yi~Wang, Yizhuo Li, Wenhai Wang, Ping Luo, Yali Wang,
  Limin Wang, and Yu~Qiao.
\newblock Videochat: Chat-centric video understanding.
\newblock \emph{arXiv preprint arXiv:2305.06355}, 2023{\natexlab{b}}.

\bibitem[Li et~al.(2024{\natexlab{b}})Li, Wang, He, Li, Wang, Liu, Wang, Xu,
  Chen, Luo, Wang, and Qiao]{li2024mvbench}
Kunchang Li, Yali Wang, Yinan He, Yizhuo Li, Yi~Wang, Yi~Liu, Zun Wang, Jilan
  Xu, Guo Chen, Ping Luo, Limin Wang, and Yu~Qiao.
\newblock Mvbench: A comprehensive multi-modal video understanding benchmark.
\newblock In \emph{Proceedings of the IEEE/CVF Conference on Computer Vision
  and Pattern Recognition}, 2024{\natexlab{b}}.

\bibitem[Li et~al.(2025)Li, Wei, Xie, Yang, Song, Wang, An, Liu, Li, Lin,
  et~al.]{li2025vl}
Lei Li, Yuancheng Wei, Zhihui Xie, Xuqing Yang, Yifan Song, Peiyi Wang, Chenxin
  An, Tianyu Liu, Sujian Li, Bill~Yuchen Lin, et~al.
\newblock Vl-rewardbench: A challenging benchmark for vision-language
  generative reward models.
\newblock In \emph{Proceedings of the Computer Vision and Pattern Recognition
  Conference}, pp.\  24657--24668, 2025.

\bibitem[Li \& Deng(2020)Li and Deng]{li2020deep}
Shan Li and Weihong Deng.
\newblock Deep facial expression recognition: A survey.
\newblock \emph{IEEE Transactions on Affective Computing}, 13\penalty0
  (3):\penalty0 1195--1215, 2020.

\bibitem[Li et~al.(2024{\natexlab{c}})Li, Wang, and Jia]{li2024llama}
Yanwei Li, Chengyao Wang, and Jiaya Jia.
\newblock Llama-vid: An image is worth 2 tokens in large language models.
\newblock In \emph{European Conference on Computer Vision}, pp.\  323--340.
  Springer, 2024{\natexlab{c}}.

\bibitem[Lian et~al.(2023)Lian, Sun, Xu, Sun, Xu, Wen, Chen, Liu, and
  Tao]{lian2023explainable}
Zheng Lian, Licai Sun, Mingyu Xu, Haiyang Sun, Ke~Xu, Zhuofan Wen, Shun Chen,
  Bin Liu, and Jianhua Tao.
\newblock Explainable multimodal emotion reasoning.
\newblock \emph{arXiv preprint arXiv:2306.15401}, 2023.

\bibitem[Lian et~al.(2024{\natexlab{a}})Lian, Sun, Sun, Wen, Zhang, Chen, Gu,
  Zhao, Ma, Chen, et~al.]{lian2024mer}
Zheng Lian, Haiyang Sun, Licai Sun, Zhuofan Wen, Siyuan Zhang, Shun Chen, Hao
  Gu, Jinming Zhao, Ziyang Ma, Xie Chen, et~al.
\newblock Mer 2024: Semi-supervised learning, noise robustness, and
  open-vocabulary multimodal emotion recognition.
\newblock In \emph{Proceedings of the 2nd International Workshop on Multimodal
  and Responsible Affective Computing}, pp.\  41--48, 2024{\natexlab{a}}.

\bibitem[Lian et~al.(2024{\natexlab{b}})Lian, Sun, Ren, Gu, Sun, Chen, Liu, and
  Tao]{lian2024merbench}
Zheng Lian, Licai Sun, Yong Ren, Hao Gu, Haiyang Sun, Lan Chen, Bin Liu, and
  Jianhua Tao.
\newblock Merbench: A unified evaluation benchmark for multimodal emotion
  recognition.
\newblock \emph{arXiv preprint arXiv:2401.03429}, 2024{\natexlab{b}}.

\bibitem[Lian et~al.(2025{\natexlab{a}})Lian, Chen, Chen, Sun, Sun, Ren, Cheng,
  Liu, Liu, Peng, et~al.]{lian2025affectgpt}
Zheng Lian, Haoyu Chen, Lan Chen, Haiyang Sun, Licai Sun, Yong Ren, Zebang
  Cheng, Bin Liu, Rui Liu, Xiaojiang Peng, et~al.
\newblock Affectgpt: A new dataset, model, and benchmark for emotion
  understanding with multimodal large language models.
\newblock In \emph{Proceedings of the 42nd International Conference on Machine
  Learning}, 2025{\natexlab{a}}.

\bibitem[Lian et~al.(2025{\natexlab{b}})Lian, Liu, Xu, Liu, Liu, Zhang, Liu,
  Li, Cheng, Zuo, et~al.]{lian2025mer}
Zheng Lian, Rui Liu, Kele Xu, Bin Liu, Xuefei Liu, Yazhou Zhang, Xin Liu, Yong
  Li, Zebang Cheng, Haolin Zuo, et~al.
\newblock Mer 2025: When affective computing meets large language models.
\newblock In \emph{Proceedings of the 33th ACM International Conference on
  Multimedia}, 2025{\natexlab{b}}.

\bibitem[Lian et~al.(2025{\natexlab{c}})Lian, Sun, Sun, Chen, Chen, Gu, Wen,
  Chen, Siyuan, Yao, et~al.]{lian2025ov}
Zheng Lian, Haiyang Sun, Licai Sun, Haoyu Chen, Lan Chen, Hao Gu, Zhuofan Wen,
  Shun Chen, Zhang Siyuan, Hailiang Yao, et~al.
\newblock Ov-mer: Towards open-vocabulary multimodal emotion recognition.
\newblock In \emph{Proceedings of the 42nd International Conference on Machine
  Learning}, 2025{\natexlab{c}}.

\bibitem[Lin et~al.(2024)Lin, Ye, Zhu, Cui, Ning, Jin, and Yuan]{lin2024video}
Bin Lin, Yang Ye, Bin Zhu, Jiaxi Cui, Munan Ning, Peng Jin, and Li~Yuan.
\newblock Video-llava: Learning united visual representation by alignment
  before projection.
\newblock In \emph{Proceedings of the 2024 Conference on Empirical Methods in
  Natural Language Processing}, pp.\  5971--5984, 2024.

\bibitem[Lindquist \& Barrett(2008)Lindquist and
  Barrett]{lindquist2008emotional}
Kristen~A Lindquist and Lisa~Feldman Barrett.
\newblock Emotional complexity.
\newblock \emph{Handbook of emotions}, 4:\penalty0 513--530, 2008.

\bibitem[Maaz et~al.(2024)Maaz, Rasheed, Khan, and Khan]{maaz2024video}
Muhammad Maaz, Hanoona Rasheed, Salman Khan, and Fahad Khan.
\newblock Video-chatgpt: Towards detailed video understanding via large vision
  and language models.
\newblock In \emph{Proceedings of the 62nd Annual Meeting of the Association
  for Computational Linguistics (Volume 1: Long Papers)}, pp.\  12585--12602,
  2024.

\bibitem[Minsky(1986)]{minsky1986society}
Marvin Minsky.
\newblock \emph{Society of mind}.
\newblock Simon and Schuster, 1986.

\bibitem[Noroozi et~al.(2018)Noroozi, Corneanu, Kami{\'n}ska, Sapi{\'n}ski,
  Escalera, and Anbarjafari]{noroozi2018survey}
Fatemeh Noroozi, Ciprian~Adrian Corneanu, Dorota Kami{\'n}ska, Tomasz
  Sapi{\'n}ski, Sergio Escalera, and Gholamreza Anbarjafari.
\newblock Survey on emotional body gesture recognition.
\newblock \emph{IEEE Transactions on Affective Computing}, 12\penalty0
  (2):\penalty0 505--523, 2018.

\bibitem[OpenAI(2024)]{openai24gpt4o}
OpenAI.
\newblock Gpt-4o system card, 2024.
\newblock URL \url{https://cdn.openai.com/gpt-4o-system-card.pdf}.

\bibitem[Plutchik(1980)]{plutchik1980general}
Robert Plutchik.
\newblock A general psychoevolutionary theory of emotion.
\newblock \emph{Emotion: Theory, research, and experience}, 1, 1980.

\bibitem[Russell(1980)]{russell1980circumplex}
James~A Russell.
\newblock A circumplex model of affect.
\newblock \emph{Journal of personality and social psychology}, 39\penalty0
  (6):\penalty0 1161, 1980.

\bibitem[Russell \& Mehrabian(1977)Russell and Mehrabian]{russell1977evidence}
James~A Russell and Albert Mehrabian.
\newblock Evidence for a three-factor theory of emotions.
\newblock \emph{Journal of research in Personality}, 11\penalty0 (3):\penalty0
  273--294, 1977.

\bibitem[Tan et~al.(2025)Tan, Zhuang, Montgomery, Tang, Cuadron, Wang, Popa,
  and Stoica]{tan2024judgebench}
Sijun Tan, Siyuan Zhuang, Kyle Montgomery, William~Yuan Tang, Alejandro
  Cuadron, Chenguang Wang, Raluca~A. Popa, and Ion Stoica.
\newblock Judgebench: {A} benchmark for evaluating llm-based judges.
\newblock In \emph{Proceedings of the 13th International Conference on Learning
  Representations}, 2025.

\bibitem[Team et~al.(2023)Team, Anil, Borgeaud, Alayrac, Yu, Soricut,
  Schalkwyk, Dai, Hauth, Millican, et~al.]{team2023gemini}
Gemini Team, Rohan Anil, Sebastian Borgeaud, Jean-Baptiste Alayrac, Jiahui Yu,
  Radu Soricut, Johan Schalkwyk, Andrew~M Dai, Anja Hauth, Katie Millican,
  et~al.
\newblock Gemini: a family of highly capable multimodal models.
\newblock \emph{arXiv preprint arXiv:2312.11805}, 2023.

\bibitem[Team et~al.(2024)Team, Georgiev, Lei, Burnell, Bai, Gulati, Tanzer,
  Vincent, Pan, Wang, et~al.]{team2024gemini}
Gemini Team, Petko Georgiev, Ving~Ian Lei, Ryan Burnell, Libin Bai, Anmol
  Gulati, Garrett Tanzer, Damien Vincent, Zhufeng Pan, Shibo Wang, et~al.
\newblock Gemini 1.5: Unlocking multimodal understanding across millions of
  tokens of context.
\newblock \emph{arXiv preprint arXiv:2403.05530}, 2024.

\bibitem[Wang et~al.(2024{\natexlab{a}})Wang, Li, Chen, Cai, Zhu, Lin, Cao,
  Kong, Liu, Liu, et~al.]{wang2024large}
Peiyi Wang, Lei Li, Liang Chen, Zefan Cai, Dawei Zhu, Binghuai Lin, Yunbo Cao,
  Lingpeng Kong, Qi~Liu, Tianyu Liu, et~al.
\newblock Large language models are not fair evaluators.
\newblock In \emph{Proceedings of the 62nd Annual Meeting of the Association
  for Computational Linguistics (Volume 1: Long Papers)}, pp.\  9440--9450,
  2024{\natexlab{a}}.

\bibitem[Wang et~al.(2024{\natexlab{b}})Wang, Yu, Yao, Zeng, Yang, Wang, Chen,
  Jiang, Xie, Wang, Xie, Ye, Zhang, and Zhang]{wang2024pandalm}
Yidong Wang, Zhuohao Yu, Wenjin Yao, Zhengran Zeng, Linyi Yang, Cunxiang Wang,
  Hao Chen, Chaoya Jiang, Rui Xie, Jindong Wang, Xing Xie, Wei Ye, Shikun
  Zhang, and Yue Zhang.
\newblock Pandalm: An automatic evaluation benchmark for {LLM} instruction
  tuning optimization.
\newblock In \emph{Proceedings of the 12th International Conference on Learning
  Representations}, 2024{\natexlab{b}}.

\bibitem[Xu et~al.(2025)Xu, Guo, He, Hu, He, Bai, Chen, Wang, Fan, Dang,
  et~al.]{xu2025qwen2}
Jin Xu, Zhifang Guo, Jinzheng He, Hangrui Hu, Ting He, Shuai Bai, Keqin Chen,
  Jialin Wang, Yang Fan, Kai Dang, et~al.
\newblock Qwen2. 5-omni technical report.
\newblock \emph{arXiv preprint arXiv:2503.20215}, 2025.

\bibitem[Xu et~al.(2024)Xu, Zhao, Zhou, Lin, Ng, and Feng]{xu2024pllava}
Lin Xu, Yilin Zhao, Daquan Zhou, Zhijie Lin, See~Kiong Ng, and Jiashi Feng.
\newblock Pllava: Parameter-free llava extension from images to videos for
  video dense captioning.
\newblock \emph{arXiv preprint arXiv:2404.16994}, 2024.

\bibitem[Yang et~al.(2024)Yang, Yang, Zhang, Hui, Zheng, Yu, Li, Liu, Huang,
  Wei, Lin, Yang, Tu, Zhang, Yang, Yang, Zhou, Lin, Dang, Lu, Bao, Yang, Yu,
  Li, Xue, Zhang, Zhu, Men, Lin, Li, Tang, Xia, Ren, Ren, Fan, Su, Zhang, Wan,
  Liu, Cui, Zhang, and Qiu]{qwen2025qwen25technicalreport}
An~Yang, Baosong Yang, Beichen Zhang, Binyuan Hui, Bo~Zheng, Bowen Yu,
  Chengyuan Li, Dayiheng Liu, Fei Huang, Haoran Wei, Huan Lin, Jian Yang,
  Jianhong Tu, Jianwei Zhang, Jianxin Yang, Jiaxi Yang, Jingren Zhou, Junyang
  Lin, Kai Dang, Keming Lu, Keqin Bao, Kexin Yang, Le~Yu, Mei Li, Mingfeng Xue,
  Pei Zhang, Qin Zhu, Rui Men, Runji Lin, Tianhao Li, Tianyi Tang, Tingyu Xia,
  Xingzhang Ren, Xuancheng Ren, Yang Fan, Yang Su, Yichang Zhang, Yu~Wan,
  Yuqiong Liu, Zeyu Cui, Zhenru Zhang, and Zihan Qiu.
\newblock Qwen2.5 technical report.
\newblock \emph{arXiv preprint arXiv:2412.15115}, 2024.

\bibitem[Yang et~al.(2025)Yang, Li, Yang, Zhang, Hui, Zheng, Yu, Gao, Huang,
  Lv, et~al.]{yang2025qwen3}
An~Yang, Anfeng Li, Baosong Yang, Beichen Zhang, Binyuan Hui, Bo~Zheng, Bowen
  Yu, Chang Gao, Chengen Huang, Chenxu Lv, et~al.
\newblock Qwen3 technical report.
\newblock \emph{arXiv preprint arXiv:2505.09388}, 2025.

\bibitem[Yasunaga et~al.(2025)Yasunaga, Zettlemoyer, and
  Ghazvininejad]{yasunaga2025multimodal}
Michihiro Yasunaga, Luke Zettlemoyer, and Marjan Ghazvininejad.
\newblock Multimodal rewardbench: Holistic evaluation of reward models for
  vision language models.
\newblock \emph{arXiv preprint arXiv:2502.14191}, 2025.

\bibitem[Ye et~al.(2025)Ye, Wang, Huang, Chen, Zhang, Moniz, Gao, Geyer, Huang,
  Chen, Chawla, and Zhang]{ye2025justice}
Jiayi Ye, Yanbo Wang, Yue Huang, Dongping Chen, Qihui Zhang, Nuno Moniz, Tian
  Gao, Werner Geyer, Chao Huang, Pin{-}Yu Chen, Nitesh~V. Chawla, and
  Xiangliang Zhang.
\newblock Justice or prejudice? quantifying biases in llm-as-a-judge.
\newblock In \emph{Thirteenth International Conference on Learning
  Representations}, 2025.

\bibitem[Ye et~al.(2023)Ye, Xu, Xu, Ye, Yan, Zhou, Wang, Hu, Shi, Shi, Jiang,
  Li, Xu, Chen, Tian, Qi, Zhang, and Huang]{ye2023mplugowl}
Qinghao Ye, Haiyang Xu, Guohai Xu, Jiabo Ye, Ming Yan, Yiyang Zhou, Junyang
  Wang, Anwen Hu, Pengcheng Shi, Yaya Shi, Chaoya Jiang, Chenliang Li, Yuanhong
  Xu, Hehong Chen, Junfeng Tian, Qian Qi, Ji~Zhang, and Fei Huang.
\newblock mplug-owl: Modularization empowers large language models with
  multimodality, 2023.

\bibitem[Yin et~al.(2024)Yin, Fu, Zhao, Li, Sun, Xu, and Chen]{yin2024survey}
Shukang Yin, Chaoyou Fu, Sirui Zhao, Ke~Li, Xing Sun, Tong Xu, and Enhong Chen.
\newblock A survey on multimodal large language models.
\newblock \emph{National Science Review}, 11\penalty0 (12):\penalty0 nwae403,
  2024.

\bibitem[Zhang et~al.(2023)Zhang, Yu, Yu, Lv, Liu, Huang, Xu, and
  Li]{zhang2023wider}
Xinghua Zhang, Bowen Yu, Haiyang Yu, Yangyu Lv, Tingwen Liu, Fei Huang, Hongbo
  Xu, and Yongbin Li.
\newblock Wider and deeper llm networks are fairer llm evaluators.
\newblock \emph{arXiv preprint arXiv:2308.01862}, 2023.

\bibitem[Zhang et~al.(2025)Zhang, Yu, Tian, Fu, Li, Zeng, Xie, Shi, Zhang, Wu,
  et~al.]{zhang2025mm}
YiFan Zhang, Tao Yu, Haochen Tian, Chaoyou Fu, Peiyan Li, Jianshu Zeng, Wulin
  Xie, Yang Shi, Huanyu Zhang, Junkang Wu, et~al.
\newblock Mm-rlhf: The next step forward in multimodal llm alignment.
\newblock In \emph{Forty-second International Conference on Machine Learning},
  2025.

\bibitem[Zhao et~al.(2023)Zhao, Zhou, Li, Tang, Wang, Hou, Min, Zhang, Zhang,
  Dong, et~al.]{zhao2023survey}
Wayne~Xin Zhao, Kun Zhou, Junyi Li, Tianyi Tang, Xiaolei Wang, Yupeng Hou,
  Yingqian Min, Beichen Zhang, Junjie Zhang, Zican Dong, et~al.
\newblock A survey of large language models.
\newblock \emph{arXiv preprint arXiv:2303.18223}, 2023.

\bibitem[Zheng et~al.(2023)Zheng, Chiang, Sheng, Zhuang, Wu, Zhuang, Lin, Li,
  Li, Xing, et~al.]{zheng2023judging}
Lianmin Zheng, Wei-Lin Chiang, Ying Sheng, Siyuan Zhuang, Zhanghao Wu, Yonghao
  Zhuang, Zi~Lin, Zhuohan Li, Dacheng Li, Eric Xing, et~al.
\newblock Judging llm-as-a-judge with mt-bench and chatbot arena.
\newblock \emph{Proceedings of the Advances in Neural Information Processing
  Systems}, 36:\penalty0 46595--46623, 2023.

\end{thebibliography}
\bibliographystyle{iclr2026_conference}

\clearpage
\appendix

\section*{Appendix}  
\addcontentsline{toc}{chapter}{Appendix Contents} 
\etocsettocstyle{}{}  
\localtableofcontents 

\newpage


\section{LLM Usage}
In this paper, LLMs are merely used to polish language, such as improving the clarity, coherence, and fluency. We do not leverage LLMs in the generation of research ideas. Our research ideation involves deep domain expertise, creativity, and an understanding of unresolved scientific challenges, all of which rely on our insights into the field of affective computing. Therefore, LLMs serve as supportive tools for language enhancement rather than as drivers of idea breakthroughs.

\section{Related Works}
\label{appendix:related_works}

\subsection{Descriptive Emotion Representation}
Emotion is intrinsically linked to human cognition, decision-making, and behavior \citep{cacioppo1999emotion,lerner2015emotion,dolan2002emotion} and plays a pivotal role in artificial intelligence \citep{minsky1986society}. Emotion representation methods serve as the foundation, aiming to map complex human emotions into computable values \citep{gunes2011emotion}. Within the field of emotion representation, two predominant paradigms exist: categorical representation and dimensional representation.

\paragraph{Categorical Representation.}
Rooted in psychological theory, researchers classify human emotions into discrete categories. For instance, \cite{ekman1970universal} proposed that a set of universal emotional states, termed basic emotions, exists across all human cultures, including \emph{anger}, \emph{disgust}, \emph{fear}, \emph{happiness}, \emph{sadness}, and \emph{surprise}. Later, \cite{plutchik1980general} introduced eight primary emotions, organized in opposing pairs and arranged by intensity, which can combine to form secondary emotions. However, human emotions are far more complex than these six or eight labels \citep{lindquist2008emotional}. Restricting the emotional space to such limited categories inevitably overlooks nuanced emotional experiences. To address this, \cite{lian2025ov} proposed recognizing emotions in an open-vocabulary manner, enabling the prediction of arbitrary emotion categories. This approach extends beyond traditional categorical representation paradigms.

\paragraph{Dimensional Representation.}
Rather than discrete categories, dimensional representation theory models human emotions as points in a continuous multi-dimensional space. For example, \cite{russell1977evidence} proposed the Valence–Arousal–Dominance (VAD) model, where valence reflects the positivity of an emotion, arousal reflects its level of excitement, and dominance reflects the perceived sense of control in a situation. Later, \cite{russell1980circumplex} introduced a simpler Circumplex Model of Affect, reducing emotions to a two-dimensional space—valence and arousal. However, dimensional emotions are abstract and less intuitive than discrete labels (e.g., happy), limiting their practical use in downstream tasks where simpler categorical models are preferred.

\paragraph{Descriptive Representation.}
Neither categorical nor dimensional representations fully capture human cognition in emotion perception and understanding, rendering the emotion understanding process a black-box operation. To address this limitation, \cite{lian2023explainable,lian2025affectgpt} proposed a novel representation strategy called \emph{descriptive emotions}. They leverage free-form natural language to visualize the human emotion understanding process by incorporating multimodal clues and additional analysis. Within its internal analysis, this representation further accounts for emotions' temporal dynamics, intensity, and uncertainty, thereby providing more human-like emotion representations. Despite its advantages, evaluating the quality of such open-ended descriptions remains a non-trivial challenge. Previous studies rely on costly human annotators to identify human-preferred descriptions. In this paper, we propose \emph{EmoPrefer}, which leverages MLLMs to achieve more cost-efficient preference annotation, offering a meaningful complement to existing evaluation strategies for descriptive emotions. To the best of our knowledge, this is the first work to explore the capabilities of MLLMs in emotion preference decoding.

\subsection{LLM-based Judge}
LLM-as-a-Judge aims to leverage LLMs as evaluators to replace traditional human-driven assessments, offering a cost-effective evaluation solution \citep{gu2024survey}. Beyond evaluation, these LLM-based judges can also serve as reward models in reinforcement learning \citep{christiano2017deep} or act as verifiers to select the best-of-N responses from multiple candidates \citep{cobbe2021training}. To assess the effectiveness of LLM-based judges, it is necessary to measure the agreement between LLM-generated responses and human judgments. This necessitates a series of benchmark datasets with human annotations. For example,  MT-Bench \citep{zheng2023judging} consists of multi-turn questions designed to evaluate a chatbot’s conversational abilities and adherence to instructions. FairEval \citep{wang2024large} provides human-annotated preferences for responses generated by ChatGPT and Vicuna. JudgeBench \citep{tan2024judgebench} evaluates LLM-based judges on challenging response pairs requiring advanced reasoning skills, such as knowledge, math, and coding. These benchmarks primarily focus on text-only preference tasks. In contrast, MLLM-as-a-Judge \citep{chen2024mllm} extends such tasks to single-image or image-sequence preference evaluations, broadening the range of involved modalities. In this paper, we propose \emph{EmoPrefer-Data}, the first multimodal preference dataset specifically designed for human emotions. Unlike previous benchmarks that focus on text-only, single-image, or image-sequence inputs, human emotions are often conveyed through subtle behaviors embedded in multimodal cues. To achieve this, EmoPrefer-Data not only expands the supported modalities to full multimodal scenarios (including audio, text, and video) but also serves as the first dataset centered on human emotions. Therefore, our work provides a valuable data resource for current research on LLM-based judges.

\section{Details of EmoPrefer-Data}
\label{appendix:dataset_details}
To construct EmoPrefer-Data, we provide pairwise emotion descriptions for each video and then recruit multiple annotators to evaluate which description better aligns with the character's emotional state. This section provides further details about the construction and characteristics of our dataset.

\subsection{Data Visualization}
\label{appendix:dataset_data_visualization}
\begin{figure}[h]
	\centering
	\includegraphics[width=\linewidth]{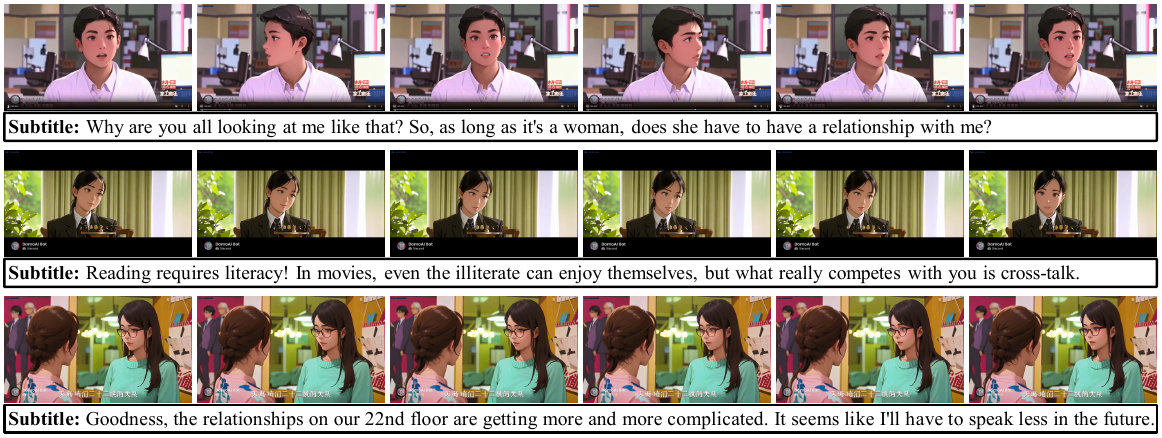}
	\caption{\textbf{Data visualization.} We cartoonize these video data to eliminate privacy concerns.}
	\label{fig:example}
\end{figure}

Figure~\ref{fig:example} presents three examples from EmoPrefer-Data. The original videos capture real people’s emotional expressions in uncontrolled settings. To preserve privacy, we apply a style-transfer technique to anonymize the individuals. The use of this dataset has been authorized by its owner. As shown in Figure~\ref{fig:example}, these videos primarily depict a single front-facing individual and include complete audio recordings, ensuring the full expression of emotions for the target speaker.

\subsection{Description Statistics}
\label{appendix:dataset_description_statistic}
To ensure the quality of description pairs, we leverage data in two benchmark datasets: MERR-Fine and MER-Caption+. To visualize their description richness, we extract emotion words and nouns from these descriptions using Qwen2.5-7B with the following prompts:

\textbf{Emotion Words:} \emph{\textcolor[rgb]{0.93,0.0,0.47}{{Please assume the role of an expert in the field of emotions. We provide clues that may be related to the emotions of the characters. Based on the provided clues, please identify the emotional states of the main characters. Please separate different emotional categories with commas and output only the clearly identifiable emotional categories in a list format. If none are identified, please output an empty list.}}} 

\textbf{Nouns:} \emph{\textcolor[rgb]{0.93,0.0,0.47}{{We provide clues that may be related to the emotions of the characters. Please extract all nouns from the provided clues. Please separate different words with commas and output in a list format. If none are identified, please output an empty list.}}}

\subsection{Annotation Platform Layout}
\label{appendix:dataset_annotation_platform}
\begin{figure}[h]
	\centering
	\includegraphics[width=\linewidth]{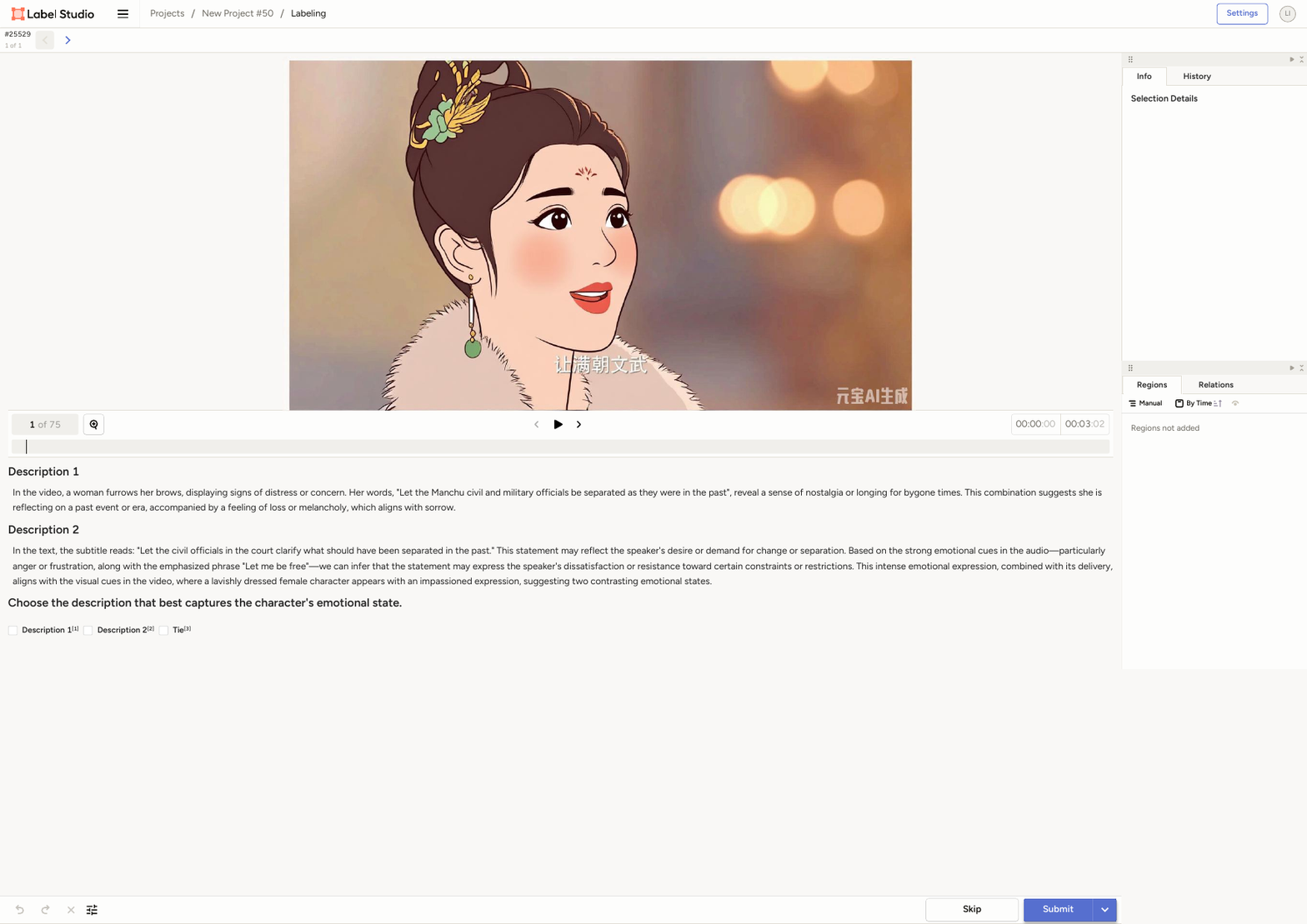}
	\caption{\textbf{Annotation platform layout.} We provide pairwise descriptions for each video and ask annotators to select the one that better reflects the character's emotional state. In this figure, we apply cartoonization to the video data to eliminate privacy concerns.}
	\label{fig:annotation_layout}
\end{figure}

Figure \ref{fig:annotation_layout} shows the layout of the annotation platform. During the annotation process, we provide two descriptions for each video. Annotators should select the preferred description, i.e., the one that better matches the character's emotional state, based on the video content. Annotators can choose from three options: (1) \emph{description 1 is better}; (2) \emph{description 2 is better}; (3) \emph{tie}.

\section{Details of MLLM-based Strategies}
\label{appendix:prompts}
Figures \ref{fig:s1}$\sim$\ref{fig:s4} illustrate the prompts for S1$\sim$S4. As shown in these figures, the length of the reasoning chains increases progressively from S1 to S4.

\subsection{Details of S1}
\label{appendix:prompts-s1}
\begin{figure}[h]
	\centering
	\includegraphics[width=0.92\linewidth]{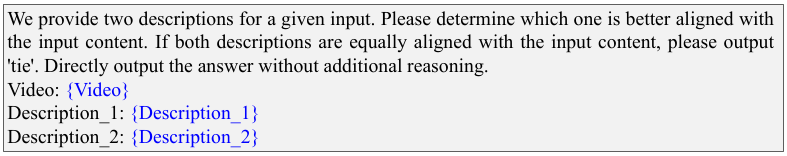}
	\caption{{Prompt for S1.}}
	\label{fig:s1}
\end{figure}

For S1, we provide the MLLM with a video and two descriptions, instructing it to identify the better one. Figure \ref{fig:s1} shows the prompt used in S1.

\newpage

\subsection{Details of S2 and S3}
\label{appendix:prompts-s2S3}
\begin{figure}[h]
	\centering
	\includegraphics[width=0.92\linewidth]{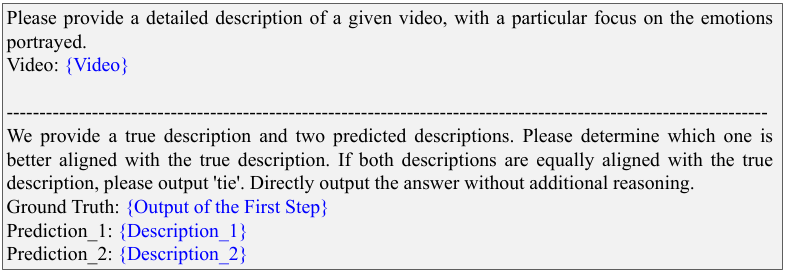}
	\caption{{Prompt for S2 and S3.}}
	\label{fig:s2_s3}
\end{figure}

S2 and S3 decompose S1 into two stages: description generation and preference prediction. Figure \ref{fig:s2_s3} illustrates the prompts for S2 and S3.

\subsection{Details of S4}
\label{appendix:prompts-s4}
\begin{figure}[!h]
	\centering
	\includegraphics[width=0.92\linewidth]{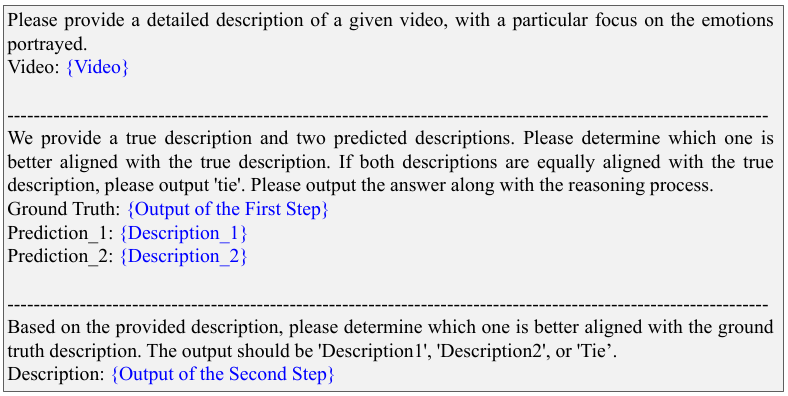}
	\caption{{Prompt for S4.}}
	\label{fig:s4}
\end{figure}

S4 decomposes S1 into three steps: description generation, preference reasoning, and answer extraction. Figure \ref{fig:s4} provides details of the prompt used in S4.

\newpage
\section{Prompt Template for Swap Consistency}
\label{appendix:metric_swap_consistency}
\begin{figure}[h]
	\centering
	\includegraphics[width=0.92\linewidth]{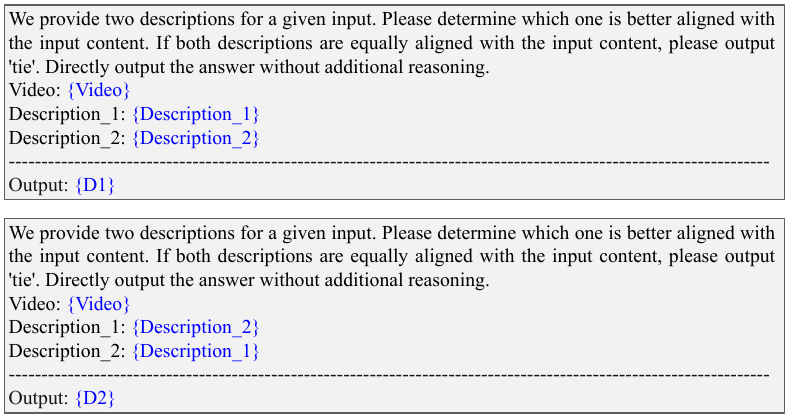}
	\caption{\textbf{Swap consistency.} This figure illustrates an ideal case, where the model prefers consistent descriptions regardless of input order.}
	\label{fig:swap_consistency}
\end{figure}

Figure \ref{fig:swap_consistency} visualizes the prompts for normal and swapped inputs. We require the model to be robust to input order changes.

\section{Details of Bradley-Terry Algorithm}
\label{appendix:bradley_terry}
Suppose there are $M$ models, each denoted as $m_i$. In this algorithm, each model $m_i$ is associated with a positive parameter $\theta_i$, which quantifies its relative advantage. The probabilities that model $m_i$ wins, loses, or ties against model $m_j$ are defined as follows:
\begin{equation}
    P(m_i>m_j) = \frac{\theta_i}{\theta_i+\theta_j+\beta},\;P(m_i<m_j) = \frac{\theta_j}{\theta_i+\theta_j+\beta},\;P(m_i \sim m_j) = \frac{\beta}{\theta_i+\theta_j+\beta},
\end{equation}
where $\beta$ controls the likelihood of ties. The parameters $\theta$ can then be estimated by maximizing the following likelihood function. Models with higher $\theta$ values are ranked higher.
\begin{equation}
    \max \mathcal{L}(\theta) = \prod_{i<j}\left(\frac{\theta_i}{\theta_i + \theta_j+\beta}\right)^{\mathbf{\text{win}}_{ij}}\left(\frac{\theta_j}{\theta_i + \theta_j+\beta}\right)^{\text{lose}_{ij}}\left(\frac{\beta}{\theta_i + \theta_j+\beta}\right)^{\text{tie}_{ij}},
\end{equation}
where $\text{win}_{ij}$, $\text{lose}_{ij}$, and $\text{tie}_{ij}$ denote the numbers of wins, losses, and ties between model $m_i$ and model $m_j$, respectively.

\section{MLLM Performance across Prompting Strategies}
\label{appendix:main_raw}
Table \ref{tab:main_raw} presents the performance of MLLMs across four prompting techniques. Our analysis reveals that different MLLMs favor distinct prompting strategies. Notably, high-performing MLLMs on multimodal tasks (such as Qwen2.5-Omni and Qwen2.5-VL) favor S1 or S2, whereas other MLLMs generally achieve better performance with S3 or S4.
\begin{table*}[!h]
	\centering
	\renewcommand\tabcolsep{10pt}
	\caption{Performance of MLLMs under four prompting strategies.}
	\label{tab:main_raw}
        \scalebox{0.8}{
		\begin{tabular}{c|c|cc|cc|c}
            \toprule
            \multirow{2}{*}{\textbf{Model}} & \multirow{2}{*}{\textbf{S.}} & \multicolumn{2}{c|}{\textbf{Rec. (2-Class)}} & \multicolumn{2}{c|}{\textbf{Rec. (3-Class)}} & \multirow{2}{*}{\begin{tabular}[c]{@{}c@{}}\textbf{Swap} \\ \textbf{Cons. ($\uparrow$)}\end{tabular}} \\
            & & \textbf{WAF($\uparrow$)} & \textbf{ACC($\uparrow$)} & \textbf{WAF($\uparrow$)} & \textbf{ACC($\uparrow$)} \\
            
            \midrule
            
\multirow{4}{*}{\begin{tabular}[c]{@{}c@{}}VideoChat \\ \citep{li2023videochat} \end{tabular}}& S1& 38.81$\pm$1.68& 50.18$\pm$1.51& 36.62$\pm$1.30& 43.55$\pm$0.70& 13.59$\pm$0.52\\
 & S2& 38.45$\pm$0.76& 49.82$\pm$0.27& 33.90$\pm$0.74& 38.07$\pm$0.78& 14.90$\pm$2.00\\
 & S3& 45.94$\pm$1.62& 51.60$\pm$0.62& 33.79$\pm$2.26& 27.70$\pm$1.92& 41.99$\pm$2.44\\
 & S4& 51.79$\pm$1.21& 52.31$\pm$0.98& 44.88$\pm$1.54& 40.77$\pm$1.57& 40.85$\pm$1.83\\

            \midrule

            \multirow{4}{*}{\begin{tabular}[c]{@{}c@{}}Video-ChatGPT \\ \citep{maaz2024video} \end{tabular}}& S1& 31.98$\pm$0.08& 48.67$\pm$0.18& 19.81$\pm$1.03& 18.29$\pm$0.87& 46.69$\pm$0.17\\
 & S2& 32.06$\pm$0.00& 48.85$\pm$0.00& 2.83$\pm$0.46& 3.40$\pm$0.26& 93.12$\pm$0.44\\
 & S3& 45.04$\pm$0.03& 47.87$\pm$0.27& 37.99$\pm$0.38& 35.71$\pm$0.52& 45.30$\pm$0.00\\
 & S4& 52.10$\pm$0.94& 52.13$\pm$0.98& 43.97$\pm$0.44& 40.42$\pm$0.35& 45.91$\pm$1.13\\

            \midrule

            \multirow{4}{*}{\begin{tabular}[c]{@{}c@{}}Otter \\ \citep{li2023otter} \end{tabular}}& S1& 32.06$\pm$0.00& 48.85$\pm$0.00& 0.07$\pm$0.00& 1.92$\pm$0.00& 100.00$\pm$0.00\\
 & S2& 32.06$\pm$0.00& 48.85$\pm$0.00& 0.07$\pm$0.00& 1.92$\pm$0.00& 98.95$\pm$0.00\\
 & S3& 42.21$\pm$0.13& 46.45$\pm$0.09& 34.51$\pm$0.10& 31.79$\pm$0.09& 46.95$\pm$0.44\\
 & S4& 52.12$\pm$1.10& 52.49$\pm$0.98& 43.41$\pm$1.67& 38.15$\pm$1.74& 46.08$\pm$0.96\\

            \midrule

            \multirow{4}{*}{\begin{tabular}[c]{@{}c@{}}VideoChat2 \\ \citep{li2024mvbench} \end{tabular}}& S1& 39.79$\pm$0.53& 42.63$\pm$0.18& 37.01$\pm$0.87& 37.63$\pm$0.87& 36.24$\pm$0.17\\
 & S2& 42.02$\pm$2.48& 49.64$\pm$2.04& 40.09$\pm$2.30& 45.12$\pm$1.57& 21.60$\pm$0.17\\
 & S3& 45.31$\pm$1.64& 48.31$\pm$1.60& 37.73$\pm$1.62& 34.84$\pm$1.05& 38.85$\pm$1.74\\
 & S4& 52.38$\pm$0.10& 52.49$\pm$0.09& 45.88$\pm$0.96& 43.82$\pm$0.44& 41.46$\pm$1.74\\

            \midrule

            \multirow{4}{*}{\begin{tabular}[c]{@{}c@{}}mPLUG-Owl \\ \citep{ye2023mplugowl} \end{tabular}}& S1& 39.13$\pm$2.32& 49.20$\pm$1.07& 33.32$\pm$3.08& 33.62$\pm$2.79& 22.13$\pm$2.61\\
 & S2& 37.59$\pm$0.32& 48.31$\pm$0.18& 31.71$\pm$0.83& 33.01$\pm$1.48& 18.47$\pm$0.70\\
 & S3& 48.41$\pm$0.24& 51.33$\pm$0.36& 42.87$\pm$0.40& 40.16$\pm$0.44& 42.25$\pm$5.31\\
 & S4& 53.76$\pm$0.17& 53.82$\pm$0.18& 46.90$\pm$0.24& 44.34$\pm$0.44& 42.33$\pm$1.39\\

            \midrule

            \multirow{4}{*}{\begin{tabular}[c]{@{}c@{}}Video-LLaVA \\ \citep{lin2024video} \end{tabular}}& S1& 32.06$\pm$0.00& 48.85$\pm$0.00& 31.04$\pm$0.00& 47.91$\pm$0.00& 0.00$\pm$0.00\\
 & S2& 32.06$\pm$0.00& 48.85$\pm$0.00& 31.04$\pm$0.00& 47.91$\pm$0.00& 0.09$\pm$0.09\\
 & S3& 48.10$\pm$0.29& 51.69$\pm$0.18& 39.61$\pm$0.80& 35.63$\pm$1.13& 39.55$\pm$0.35\\
 & S4& 54.53$\pm$0.81& 54.62$\pm$0.80& 43.61$\pm$0.82& 38.94$\pm$0.44& 42.86$\pm$3.31\\

            \midrule

            \multirow{4}{*}{\begin{tabular}[c]{@{}c@{}}Chat-UniVi \\ \citep{jin2024chat} \end{tabular}}& S1& 36.66$\pm$0.26& 48.49$\pm$0.18& 35.52$\pm$0.28& 47.47$\pm$0.09& 10.71$\pm$1.13\\
 & S2& 35.20$\pm$0.05& 49.29$\pm$0.09& 33.92$\pm$0.21& 47.30$\pm$0.26& 6.10$\pm$0.17\\
 & S3& 48.66$\pm$0.14& 51.15$\pm$0.18& 44.05$\pm$0.02& 42.25$\pm$0.09& 44.34$\pm$2.53\\
 & S4& 54.97$\pm$1.64& 55.06$\pm$1.60& 47.03$\pm$2.32& 44.69$\pm$2.00& 43.90$\pm$0.52\\

            \midrule

            \multirow{4}{*}{\begin{tabular}[c]{@{}c@{}}LLaVA-Next-Video \\ \citep{li2024llava} \end{tabular}}& S1& 35.13$\pm$0.00& 50.27$\pm$0.00& 34.04$\pm$0.00& 49.30$\pm$0.00& 2.79$\pm$0.00\\
 & S2& 33.23$\pm$0.00& 49.38$\pm$0.00& 32.18$\pm$0.00& 48.43$\pm$0.00& 2.26$\pm$0.00\\
 & S3& 49.25$\pm$0.03& 51.42$\pm$0.09& 44.93$\pm$0.23& 42.16$\pm$0.17& 51.74$\pm$0.52\\
 & S4& 56.41$\pm$0.98& 56.57$\pm$0.98& 52.84$\pm$1.25& 50.78$\pm$1.31& 53.92$\pm$2.00\\

            \midrule

            \multirow{4}{*}{\begin{tabular}[c]{@{}c@{}}LLaMA-VID \\ \citep{li2024llama} \end{tabular}}& S1& 33.03$\pm$0.19& 49.29$\pm$0.09& 31.99$\pm$0.19& 48.34$\pm$0.09& 0.61$\pm$0.26\\
 & S2& 32.25$\pm$0.20& 48.93$\pm$0.09& 31.23$\pm$0.19& 48.00$\pm$0.09& 0.09$\pm$0.09\\
 & S3& 54.30$\pm$1.31& 56.48$\pm$1.24& 47.76$\pm$1.37& 44.16$\pm$1.13& 44.16$\pm$1.31\\
 & S4& 57.10$\pm$0.63& 57.10$\pm$0.62& 50.42$\pm$1.17& 47.13$\pm$1.66& 45.12$\pm$0.70\\

            \midrule

            \multirow{4}{*}{\begin{tabular}[c]{@{}c@{}}PLLAVA \\ \citep{xu2024pllava} \end{tabular}}& S1& 35.79$\pm$0.00& 50.44$\pm$0.00& 34.69$\pm$0.00& 49.48$\pm$0.00& 5.05$\pm$0.00\\
 & S2& 32.84$\pm$0.00& 49.20$\pm$0.00& 31.80$\pm$0.00& 48.26$\pm$0.00& 1.39$\pm$0.00\\
 & S3& 51.13$\pm$0.44& 53.02$\pm$0.44& 47.76$\pm$0.01& 46.43$\pm$0.09& 51.22$\pm$0.87\\
 & S4& 57.29$\pm$0.18& 57.55$\pm$0.18& 54.02$\pm$0.05& 52.61$\pm$0.00& 53.83$\pm$1.22\\

            \midrule

            \multirow{4}{*}{\begin{tabular}[c]{@{}c@{}}VITA-1.5 \\ \citep{fu2025vita} \end{tabular}}& S1& 48.85$\pm$0.00& 48.85$\pm$0.00& 47.44$\pm$0.00& 47.91$\pm$0.00& 70.73$\pm$0.00\\
 & S2& 56.34$\pm$0.00& 60.04$\pm$0.00& 54.74$\pm$0.00& 58.89$\pm$0.00& 41.99$\pm$0.00\\
 & S3& 56.38$\pm$0.37& 57.28$\pm$0.27& 52.85$\pm$0.70& 51.83$\pm$0.78& 58.54$\pm$0.17\\
 & S4& 60.08$\pm$0.61& 60.12$\pm$0.62& 57.08$\pm$0.16& 56.01$\pm$0.09& 59.06$\pm$1.22\\

            \midrule

            \multirow{4}{*}{\begin{tabular}[c]{@{}c@{}}Qwen2-Audio \\ \citep{chu2024qwen2} \end{tabular}}& S1& 45.13$\pm$0.37& 54.62$\pm$0.27& 39.91$\pm$0.08& 50.96$\pm$0.09& 6.18$\pm$0.09\\
 & S2& 50.52$\pm$0.24& 54.44$\pm$0.44& 44.18$\pm$0.04& 48.17$\pm$0.44& 14.81$\pm$0.17\\
 & S3& 58.10$\pm$0.29& 58.44$\pm$0.36& 56.16$\pm$0.46& 56.36$\pm$0.61& 63.94$\pm$0.70\\
 & S4& 63.17$\pm$0.19& 63.23$\pm$0.18& 60.15$\pm$0.76& 59.32$\pm$0.78& 61.50$\pm$0.17\\

            \midrule

            \multirow{4}{*}{\begin{tabular}[c]{@{}c@{}}Qwen2.5-VL \\ \citep{bai2025qwen2} \end{tabular}}& S1& 64.43$\pm$0.87& 65.28$\pm$0.80& 62.60$\pm$0.84& 64.02$\pm$0.78& 62.02$\pm$0.35\\
 & S2& 62.42$\pm$0.99& 62.79$\pm$0.98& 60.63$\pm$0.97& 61.59$\pm$0.96& 68.90$\pm$0.61\\
 & S3& 61.79$\pm$1.41& 62.61$\pm$1.33& 58.38$\pm$1.62& 56.53$\pm$1.48& 58.71$\pm$1.05\\
 & S4& 63.37$\pm$3.47& 63.59$\pm$3.37& 60.11$\pm$2.78& 59.23$\pm$2.44& 58.19$\pm$1.92\\

            \midrule

            \multirow{4}{*}{\begin{tabular}[c]{@{}c@{}}Qwen2.5-Omni \\ \citep{xu2025qwen2} \end{tabular}}& S1& 63.36$\pm$0.00& 63.41$\pm$0.00& 61.54$\pm$0.00& 62.20$\pm$0.00& 73.87$\pm$0.00\\
 & S2& 67.21$\pm$0.00& 67.32$\pm$0.00& 65.29$\pm$0.00& 66.03$\pm$0.00& 79.09$\pm$0.00\\
 & S3& 61.54$\pm$0.07& 61.72$\pm$0.09& 59.05$\pm$0.06& 56.62$\pm$0.00& 72.65$\pm$0.52\\
 & S4& 65.69$\pm$0.29& 65.81$\pm$0.27& 63.48$\pm$0.29& 62.54$\pm$0.35& 67.94$\pm$0.52\\

\bottomrule
		\end{tabular}
        }
\end{table*}

\end{document}